\documentclass[twocolumn]{aastex701}

\usepackage{graphicx}
\usepackage{natbib}
\usepackage{amsmath}
\usepackage{calrsfs}        
\usepackage{mathrsfs}
\usepackage{makecell}    
\usepackage{multirow}    
\usepackage[table]{xcolor}
\usepackage[normalem]{ulem}  
\newcommand\hspchorhalf{\hspace{0.18cm}}
\newcommand\hspchorone{\hspace{0.9cm}}
\newcommand\hspchor{\hspace{2cm}}

\newcommand\rhoz{\ensuremath{\rho_\mathrm{0}}}
\newcommand\pz{\ensuremath{p_\mathrm{0}}}
\newcommand\Tz{\ensuremath{T_\mathrm{0}}}
\newcommand\vz{\ensuremath{\vec v_\mathrm{0}}}

\newcommand\cs{\ensuremath{c_\mathrm{s}}}
\newcommand\csi{\ensuremath{c_\mathrm{si}}}
\newcommand\cv{\ensuremath{c_\mathrm{v}}}

\newcommand\calL{\ensuremath{\mathcal{L}}}

\newcommand\Arho{\ensuremath{A}}
\newcommand\Ap{\ensuremath{\tilde{A}}}

\newcommand\Np{\ensuremath{N_\mathrm{p}}}
\newcommand\Nrho{\ensuremath{N_\rho}}
\newcommand\NS{\ensuremath{N_\mathrm{S}}}
\newcommand\Npz{\ensuremath{N_{\mathrm{p}0}}}
\newcommand\Nrhoz{\ensuremath{N_{\rho 0}}}
\newcommand\NSz{\ensuremath{N_{\mathrm{S}0}}}
\newcommand\kF{\ensuremath{k_\mathrm{F}}}
\newcommand\tc{\ensuremath{t_\mathrm{cool}}}
\newcommand\tsnd{\ensuremath{t_\mathrm{snd}}}
\newcommand\tcond{\ensuremath{t_\mathrm{cond}}}
\newcommand\lambdaisoch{\ensuremath{\lambda_\mathrm{isochoric}}}

\newcommand\kB{\ensuremath{k_\mathrm{B}}}

\newcommand\nR{\ensuremath{n_\mathrm{R}}}
\newcommand\nI{\ensuremath{n_\mathrm{I}}}
\newcommand\nRmax{\ensuremath{n_\mathrm{R,max}}}

\newcommand\lambdaF{\ensuremath{\lambda_\mathrm{F}}}
\newcommand\lambdath{\ensuremath{\lambda_\mathrm{th}}}
\newcommand\lambdam{\ensuremath{\lambda_\mathrm{m}}}
\newcommand\km{\ensuremath{k_\mathrm{m}}}

\newcommand\kacoustic{\ensuremath{k_\mathrm{ac}}}
\newcommand\kcond{\ensuremath{k_\mathrm{therm}}}
\newcommand\lambdacond{\ensuremath{\lambda_\mathrm{therm}}}
\newcommand\lambdaacoustic{\ensuremath{\lambda_\mathrm{ac}}}

\newcommand\kc{\ensuremath{k_\mathrm{c}}}
\newcommand\lambdac{\ensuremath{\lambda_\mathrm{c}}}

\newcommand\Ncrit{\ensuremath{N_\mathrm{crit}}}

\newcommand\cond{thermal}
\newcommand\bfcond{\bf thermal}
\newcommand\itacoustic{\it acoustic}

\newcommand{\nHz}{n_{\rm H0}}
\newcommand{\aeff}{a_{\rm eff}}

\makeatletter\newcommand\figurehere{\def\@captype{figure}}\makeatother

\shorttitle{Reevaluating thermal instability}
\begin{document}

\title{Reevaluating Thermal Instability in a Uniform Plasma: An Extended Analysis of Instability Domains}


\author[orcid=0000-0003-4162-7240]{Ramon Oliver}
\affiliation{Departament de Física, Universitat de les Illes Balears, 07122 Palma de Mallorca, Spain}
\affiliation{Institute of Applied Computing \& Community Code (IAC3), Universitat de les Illes Balears, Spain}
\email[show]{ramon.oliver@uib.es}

\author[orcid=0009-0005-0187-6308]{Varsha Felsy}
\affiliation{Departament de Física, Universitat de les Illes Balears, 07122 Palma de Mallorca, Spain}
\affiliation{Institute of Applied Computing \& Community Code (IAC3), Universitat de les Illes Balears, Spain}
\email[show]{varsha.felsy@uib.es}
\correspondingauthor{Varsha Felsy}

\author[orcid=0000-0002-7013-1933]{Jaume Terradas}
\affiliation{Departament de Física, Universitat de les Illes Balears, 07122 Palma de Mallorca, Spain}
\affiliation{Institute of Applied Computing \& Community Code (IAC3), Universitat de les Illes Balears, Spain}
\email{jaume.terradas@uib.es}

\author[orcid=0000-0001-7751-8885]{Amanda Stricklan}
\affiliation{X-Computational Physics Division, Los Alamos National Laboratory, Los Alamos, NM, USA}
\affiliation{Department of Astronomy, New Mexico State University, Las Cruces, NM, USA}
\email{astricklan@lanl.gov}

\author[orcid=0000-0002-5205-9472]{Timothy Waters}
\affiliation{X-Computational Physics Division, Los Alamos National Laboratory, Los Alamos, NM, USA}
\affiliation{Center for Theoretical Astrophysics, Los Alamos National Laboratory, Los Alamos, NM, USA}
\email{waters@lanl.gov}

\author[orcid=0000-0003-2255-0305]{James A. Klimchuk}
\affiliation{NASA Goddard Space Flight Center, Heliophysics Science Division, Greenbelt, MD 20771, USA}
\email{james.a.klimchuk@nasa.gov}


\begin{abstract}
Thermal instability plays a crucial role in the dynamics of astrophysical plasmas. Building upon the work of \citet{field1965thermal} and the subsequent analysis by \citet{waters2019non}, this study revisits thermal instability in a uniform, non-magnetic medium, aiming to classify thermal and acoustic modes and their instability domains. Throughout the linear analysis, heating, cooling, and thermal conduction are kept unspecified to ensure generality, and are given specific forms only for applications to particular media. Additionally, we investigate isobaric and isochoric thermal modes at very short, intermediate, and very long wavelengths; we also address a misconception regarding the thermal mode at very short wavelengths in the presence of thermal conduction, and show that the catastrophic cooling solution of \citet{waters_stricklan2025} is the long-wavelength limit of the thermal mode. Our findings confirm that the classification of instability regions outlined by \citet{waters2019non} holds when the Field length is smaller than or comparable to the thermal length, $\lambdath = \cs \tc$, and ceases to hold for larger Field lengths. We show that the ratio of these lengths is independent of plasma density, and evaluate it for five environments, from the solar corona to the intracluster medium: all satisfy the condition for the classification to apply. We also show that the instability domains accessible to a plasma are restricted by the density dependence of its heating rate, and that these environments occupy only two of the seven domains. Furthermore, we discuss the implications of the catastrophic cooling solution for coronal rain formation.
\end{abstract}


\section{Introduction} \label{sect:intro}

Thermal instability is a fundamental mechanism that affects the physical states and evolutionary pathways of astrophysical plasmas across a wide range of cosmic environments. The concept was first introduced by \citet{parker1953instability}, who proposed that, in an infinite homogeneous medium in thermal equilibrium, instability can arise when radiative losses increase with decreasing temperature, leading to runaway cooling and the formation of cool condensations---early insights that laid the groundwork for subsequent theoretical developments. Later, in 1965, Field established a rigorous formalism and derived the local criteria for thermal instability, providing a foundational analytical framework \citep{field1965thermal}. His formulation elucidated the conditions under which small perturbations in temperature and density grow, forming the basis for understanding multi-phase structure formation in astrophysical plasmas. This seminal contribution was further generalized by \citet{balbus1986local,balbus1995thermal} to include the effects of dynamical systems and complex physical processes. Additionally, foundational studies by \citet{defouw_thermal-convective_1970} and \citet{mathews1978radiative} expanded the understanding of thermal instability in different astrophysical contexts, emphasizing the role of radiative cooling in various plasma environments.

Since these pioneering works, a vast body of research has sought to understand the complex interplay of radiative cooling, heating processes, and additional physical factors that contribute to the formation of multi-phase structures such as cold clouds and filaments in astrophysical plasmas. Many studies have extended the foundational analyses by exploring the behavior of thermal instability in more realistic configurations, including self-gravity, non-uniform backgrounds, magnetic fields, and external perturbations \citep[][to name just a few works]{gomezpelaez_thermal_2002,sharma_etal2010,mccourt_etal2012,ChoudhurySharma2016,fielding_etal2017,falle_etal2020,claes_etal2020}. These investigations have revealed how various physical mechanisms influence the onset, growth, and saturation of thermal instability, often moving beyond the idealized, homogeneous models to better reflect conditions encountered in actual astrophysical environments.

Among these processes, thermal conduction occupies a special position in the theory of thermal instability, since it is what suppresses the instability at short wavelengths and thereby sets a lower bound on the size of the condensations that can grow. The scale at which it does so is the Field length, $\lambdaF$, defined in Section~\ref{sect:dispersion_relation}, and much of the present work is concerned with the consequences of including it in the classification of the instability domains. In a magnetized plasma the conductivity is strongly anisotropic, and along the field it takes the classical Spitzer value \citep{Spitzer1962}. Whether that value is attained has been the subject of considerable discussion over the last decade, particularly for the intracluster medium. The tangling of the magnetic field lines reduces the effective conductivity by a factor, $f$, whose value is itself disputed: \citet{ChandranCowley1998} obtained $10^{-3}-10^{-2}$, whereas \citet{NarayanMedvedev2001} found values of the order of 0.2 for a chaotic field with a broad range of scales. Furthermore, kinetic calculations indicate that the parallel heat flux is itself limited by the whistler instability, and is then suppressed by a factor that depends on the electron plasma beta rather than being a constant \citep{Roberg-Clark_etal2018,Komarov_etal2018}. A further limit is imposed when the electron mean free path becomes comparable to the temperature scale height, in which case the flux saturates and cannot be described by a conductivity at all \citep{CowieMcKee1977}. We do not adopt any particular prescription in what follows, but we return to the question in Section~\ref{sect:environments}, where we show that the classification of the instability domains is affected by the suppression of the conductivity only through the combination $\lambdaF \propto f^{1/2}$, and that any suppression therefore widens the range of validity of that classification.

Over recent decades, a large body of research on thermal instability in more complex astrophysical settings has been amassed. While a comprehensive review is beyond the scope of this Introduction, interested readers are referred to extensive reviews \citep[e.g.,][]{antolin2020,HennebelleInutsuka2019,DonahueVoit2022,Faucher-GiguereOh2023,GronkeSchneider2026,ThompsonHeckman2024,RuszkowskiPfrommer2023} that collectively cover a variety of environments and spatial scales, ranging from the solar corona to the intracluster medium, including molecular clouds and the circumgalactic medium. Furthermore, thermal instability in uniform media also continues to attract significant attention from researchers nowadays \citep[e.g.,][]{Molevich_etal2024,FetschFisch2024,JainSharma2025,QianChiang2025,SarkarBora2026}. Two recent contributions are of particular relevance to what follows: \citet{waters_stricklan2025} identified a mode of runaway cooling in an optically thin plasma, which they termed the catastrophic cooling mode and proposed as the agent responsible for coronal rain condensations, and \citet{stricklan_etal2025} gave a detailed treatment of the instability criteria at finite wavelength and introduced the parameterization of the heating rate that we adopt in Appendix~\ref{sect:app_powerlaw}. Given this sustained interest and ongoing exploration across various astrophysical regimes, the physical insights derived from our work are particularly timely.

Thermal instability fundamentally influences the morphology and dynamics of the interstellar medium, galaxy clusters, and accretion environments. It underpins the formation of cold clouds from hot, diffuse media and plays a key role in the feedback mechanisms that regulate star formation and galaxy evolution \citep{mccourt_etal2012,voit_etal2015}. These phenomena are often observed as multi-phase systems, where cold, dense filaments coexist within hotter plasma, driven by the destabilization of thermal equilibrium \citep{KoyamaInutsuka2002,sharma_etal2010}. Understanding the precise conditions under which the thermal instability develops and saturates remains a scientific challenge, requiring advanced theoretical models and detailed numerical simulations.

Beyond extragalactic and galactic contexts, thermal instability has significant applications in solar physics. In the solar atmosphere, particularly within the chromosphere and corona, thermal instability mechanisms are believed to contribute to the formation of prominences, coronal rain, and other fine-structured cool plasmas embedded within hot surroundings \citep{vanderlinden1991a,vanderlinden1991b,li_etal2022}. These structures often originate from localized cooling regions triggered by radiative losses, with magnetic fields playing an essential role in structuring these condensations. Investigating thermal instability within this context not only enhances our understanding of solar activity and energetic events, but also provides fundamental insights into plasma processes that are applicable across astrophysics.

This study aims to provide a comprehensive understanding of the different regimes of thermal instability by exploring all possible modes within the same homogeneous hydrodynamical system initially investigated by \citet{field1965thermal}. Recently, \citet{falle_etal2020} applied the theory of wave hierarchies \citep{Whitham1974} to repeat Field's linear analysis from a different perspective and gave more physical insight into the terms of the dispersion relation. However, the aim of this work was not to characterize the domains of thermal instability. On the other hand, \citet{waters2019non} laid the ground for a detailed classification of the domains of thermal instability in terms of a single equilibrium parameter. In this paper we correct and extend the classification of \citet{waters2019non} with the inclusion of thermal conduction and provide diagrams of the growth rate and frequency as a function of wavelength for each instability domain; we also establish the range of validity of that classification and determine where a number of astrophysical plasmas fall within it. The theoretical framework is set out in Section~\ref{sect:main_equations}, where we discuss the advantage of the classification of the instability domains of \citet{waters2019non} over that of \citet{field1965thermal}, examine the time scales that govern the linear evolution of the thermal mode, and give the amplitudes of the linear perturbations. Next, we describe the solutions in the short- and long-wavelength limits in Section~\ref{sect:short_long_wavelength}. In Section~\ref{sect:environments} we then evaluate the relevant parameters for five astrophysical environments in which thermal instability is thought to operate. The instability domains and the wavelength dependence of the thermal and acoustic modes are investigated in Sections~\ref{sect:WP_fig2} and \ref{sect:nRnI_vs_lambda}. The presence and characteristics of truly isobaric and isochoric thermal modes are discussed in Section~\ref{sect:isobaric_perturbations}. Our discussion and conclusions are given in Section~\ref{sect:conclusions}, which collects the time scales and wavelengths used throughout the paper, examines the character of the thermal mode, discusses coronal rain and the catastrophic cooling instability, and summarizes our results. Closed-form expressions for the case in which the radiative losses, the heating rate, and the thermal conductivity are power laws are collected in Appendix~\ref{sect:app_powerlaw}, and the prescriptions adopted for each astrophysical environment in Appendix~\ref{sect:app_environments}.

\section{Theoretical framework} \label{sect:main_equations}

We study thermal instability in an infinite, homogeneous, non-magnetized plasma in mechanical and thermal equilibrium.  The equilibrium density,  pressure, temperature, and velocity are $\rhoz$, $\pz$, $\Tz$, $\vz = \vec 0$. In a plasma element, heat exchange is governed by the heat-loss function, $\calL$, and the thermal conduction heat flux, $-\nabla\cdot(\kappa\nabla T)$, where $\calL = \Lambda - H$ is equal to the heat losses, $\Lambda$, minus heat gains, $H$, per unit mass and unit time, and $\kappa$ is usually a function of $T$; neither $\calL$ nor $\kappa$ is specified in this work, except in Section~\ref{sect:environments} and in the appendices, where particular forms are adopted for each of the astrophysical environments considered. The definition of the non-adiabatic terms follows the convention adopted by \citet{field1965thermal}.

Perturbations of a quantity $f$ about its equilibrium state, $f_0$, are expressed as

\begin{equation} \label{eq:perturbation}
    f(\vec{r}, t) = f_0 + \delta f(x, t),
\end{equation}

\noindent where plane-wave solutions propagating in the $x$-direction are assumed,

\begin{equation} \label{eq:delta_f}
    \delta f(x, t) = A_f \exp(n t + i k x).
\end{equation}

\noindent Following \citet{field1965thermal}, we will refer to $n$ as the growth rate; sometimes we also refer to the real part of $n$ as the growth rate. Small amplitude perturbations are considered in what follows.

\subsection{Dispersion relation} \label{sect:dispersion_relation}

The dispersion relation in the form given by \citet{field1965thermal} is

\begin{equation} \label{eq:dr_Field}
    n^3 + n^2 \cs \left(k_T + \frac{k^2}{k_\kappa}\right) + n \cs^2 k^2 + \frac{\cs^3 k^2}{\gamma}\left(k_T - k_\rho + \frac{k^2}{k_\kappa}\right) = 0,
\end{equation}

\noindent where $\gamma$ is the ratio of specific heats and $\cs^2 = \gamma\pz/\rhoz$ is the square of the adiabatic sound speed. Furthermore, the following three wavenumbers are defined

\begin{equation} \label{eq:ks_Field}
    k_T = \frac{\calL_T}{\cv \cs}, \hspchorone k_\rho = \frac{\rhoz}{\Tz} \frac{\calL_\rho}{\cv \cs}, \hspchorone k_\kappa = \rhoz \frac{\cv \cs}{\kappa},
\end{equation}

\noindent with $\cv$ the specific heat at constant volume\footnote{In an ideal gas $\cv = \kB/[(\gamma-1)\mu m_{\rm H}]$, with $m_{\rm H}$ the hydrogen mass and $\mu$ the mean molecular weight, that is, the mean mass per particle in units of $m_{\rm H}$. For a fully ionized plasma of solar composition $\mu \simeq 0.6$.}. The derivatives of the heat-loss function,

\begin{equation} \label{eq:L_T_rho}
    \calL_T = \left(\frac{\partial \calL}{\partial T}\right)_\rho, \hspchor \calL_\rho = \left(\frac{\partial \calL}{\partial \rho}\right)_T,
\end{equation}

\noindent are evaluated in the equilibrium state.

It is customary to express the quantity $n$ as $n = \nR + i \nI$, where $\nR, \nI \in \mathbb{R}$. Given that Equation~(\ref{eq:dr_Field}) is a third degree polynomial with real coefficients, it can either have three real solutions ($\nI = 0$), or a real solution and a pair of complex conjugate solutions. Real solutions correspond to exponentially growing ($\nR > 0$) or damped ($\nR < 0$), non-propagating thermal modes, while the pair of complex conjugate solutions corresponds to two overstable ($\nR > 0$) or stable ($\nR \leq 0$) acoustic waves with frequency $\nI$.

\citet{waters2019non} chose to cast the dispersion relation as follows

\begin{equation} \label{eq:dr_WP}
    n^3 + \frac{\Nrho}{\tc} n^2 + k^2 \cs^2 n + \frac{\Np}{\gamma \tc} k^2 \cs^2 = 0.
\end{equation}

\noindent The dimensionless quantities \Np\ and \Nrho\ are given by

\begin{equation} \label{eq:Nrho_Np}
    \Nrho = \Nrhoz + \left(\frac{\lambdaF}{\lambda}\right)^2, \hspchor \Np = \Npz + \left(\frac{\lambdaF}{\lambda}\right)^2,
\end{equation}

\noindent with

\begin{equation} \label{eq:Nrho0_Np0}
    \Nrhoz = \frac{\Tz}{\Lambda_0} \calL_T, \hspchorhalf \Npz = \frac{\Tz}{\Lambda_0} \left(\frac{\partial \calL}{\partial T}\right)_p \equiv \frac{\Tz}{\Lambda_0} \left(\calL_T - \frac{\rhoz}{\Tz}\calL_\rho\right),
\end{equation}

\noindent and $\Lambda_0$ the heat losses in the equilibrium. In addition, $\lambdaF$ is the Field length as defined by \citet{waters2019non}, based on \citet{begelman_global_1990},

\begin{equation} \label{eq:lambdaF}
    \lambdaF = 2 \pi \sqrt\frac{\kappa \Tz}{\rhoz \Lambda_0}, \hspchor k_F = \frac{2 \pi}{\lambdaF}.
\end{equation}

\noindent In this definition, the Field length is the characteristic length scale for which cooling and conduction balance. We note that this is not the only definition of this quantity. In some works, e.g., \citet{KoyamaInutsuka2004}, the factor $2\pi$ in $\lambdaF$ is omitted. In other works, $\lambdaF$ is given as the critical wavelength below which the thermal mode is stabilized by heat conduction \citep{field1965thermal,mccourt_etal2012,falle_etal2020}. Here we refer to this length scale as $\lambdacond$, see Equation~(\ref{eq:kcond}).

The cooling time scale is defined as follows,
\begin{equation} \label{eq:tcool}
    \tc = \frac{\cv \Tz}{\Lambda_0}.
\end{equation}

\noindent Furthermore, \citet{waters2019non} defined the parameter

\begin{equation} \label{eq:R}
    R = \frac{\Npz}{\gamma \Nrhoz},
\end{equation}

\noindent  whose relevance will be highlighted later. Although Equation~(8) of \citet{waters2019non} does not include the subscript ``0'' in both $\Npz$ and $\Nrhoz$, it must be understood that the parameter $R$ is independent of $k$ and that the definition of $R$ must be properly written as in our Equation~(\ref{eq:R}).

A comparison of the two forms of the dispersion relation leads to the following identities:

\begin{equation} \label{eq:equivalence_ks_Field}
    k_T = \frac{\Nrhoz}{\tc \cs}, \hspchorhalf k_\rho = \frac{\Nrhoz-\Npz}{\tc \cs}, \hspchorhalf k_\kappa = \tc \cs \kF^2.
\end{equation}

\noindent The analytical solutions of cubic polynomials~(\ref{eq:dr_Field}) and (\ref{eq:dr_WP}) are well known, but they are rather long and not very useful to describe the nature of solutions and the instability domains of a given plasma. As we discuss later, Figure~1 of \citet{field1965thermal} provides a complex discussion of this point, while \citet{waters2019non} gave a simpler description in their Figure~2. One of the purposes of this work is to complete and amend this figure (see Table~\ref{table:R_and_lambda}).

\subsection{Instability criteria} \label{sect:instability_criteria}

From the physical analysis of the energy exchange between a plasma element and its surroundings in a uniform medium, \citet{field1965thermal} obtained the isochoric, isobaric, and isentropic instability criteria \citep[for an alternative derivation, see][]{falle_etal2020}:

\begin{equation}
\label{eq:inst_criteria1}
\begin{aligned}
&\calL_T < 0, \hspchorhalf
\left(\frac{\partial \calL}{\partial T}\right)_p < 0, \hspchorhalf \left(\frac{\partial \calL}{\partial T}\right)_S
\equiv \calL_T + \frac{1}{\gamma-1}\frac{\rhoz}{\Tz}\calL_\rho < 0 .
\end{aligned}
\end{equation}

\noindent These three criteria, which strictly apply in the absence of thermal conduction, can be rewritten as

\begin{equation}
\label{eq:inst_criteria2}
    \Nrhoz < 0, \hspchor \Npz < 0,  \hspchor \NSz < 0,
\end{equation}

\noindent where the new dimensionless parameter $\NSz$ has been defined as

\begin{equation}
\label{eq:NSz}
    \NSz = \frac{1}{\gamma -1}\left(\gamma \Nrhoz - \Npz\right) \equiv -\frac{\gamma}{\gamma -1} (R-1) \Nrhoz.
\end{equation}

\noindent Regarding the regions of instability of the acoustic modes, see the discussion in Section~3.4 of \citet{waters2019non}.

It is now obvious that one reason for using the quantities $\Npz$ and $\Nrhoz$ for the mathematical formulation of the thermal instability problem is the simplicity of the isochoric and isobaric instability criteria; something similar can be said about $\NSz$ and the isentropic criterion.

When conduction is taken into account, Equations~(\ref{eq:inst_criteria1}) or (\ref{eq:inst_criteria2}) no longer provide the instability criteria and must be substituted by \citep[see Equations~(25) in][]{field1965thermal}

\begin{equation}
\label{eq:inst_criteria3}
    \Nrhoz + \left(\frac{\lambdaF}{\lambda}\right)^2 < 0, \hspchorhalf \Npz + \left(\frac{\lambdaF}{\lambda}\right)^2 < 0,  \hspchorhalf \NSz + \left(\frac{\lambdaF}{\lambda}\right)^2 < 0.
\end{equation}

\noindent In view of these formulas, statements about a plasma satisfying some instability criterion should, in principle, reflect whether they refer to infinite (Equations~(\ref{eq:inst_criteria2})) or finite wavelength (Equations~(\ref{eq:inst_criteria3})); see \citet{stricklan_etal2025}, where a detailed treatment is given. Anyway, a relevant question is whether a system that is isobarically unstable for some wavelength has a  mode with this wavelength that evolves at constant pressure in the linear phase. We address this question in Section~\ref{sect:timescales}, on the basis of a comparison of time scales, and in Section~\ref{sect:isobaric_perturbations}, where the prediction is tested against the amplitudes of the linear perturbations.

\citet{field1965thermal} acknowledged the stabilizing effect of thermal conduction at short wavelengths. From the isobaric and isentropic instability criteria, the wavenumber for which the thermal (acoustic) mode transitions from unstable (overstable) to stable is

\begin{equation}\label{eq:kcond}
\begin{split}
&\kcond = \sqrt{-\Npz}\,\kF, \hspchorhalf \lambdacond = \frac{2\pi}{\kcond} \equiv \frac{\lambdaF}{\sqrt{-\Npz}}, \\
& \hspchor\text{(thermal mode)}
\end{split}
\end{equation}

\begin{equation}\label{eq:kacoustic}
\begin{split}
&\kacoustic = \sqrt{-\NSz}\,\kF, \hspchorone \lambdaacoustic = \frac{2\pi}{\kacoustic} \equiv \frac{\lambdaF}{\sqrt{-\NSz}}, \\
& \hspchor \text{(acoustic modes)}
\end{split}
\end{equation}

\noindent see Equations~(26) of \citet{field1965thermal}. The third of the criteria in Equation~(\ref{eq:inst_criteria3}) yields in the same way the wavelength $\lambdaF / \sqrt{-\Nrhoz}$, which we do not use in what follows: the isochoric criterion governs the thermal mode in the limit $\lambda \rightarrow \infty$, in which thermal conduction plays no role, whereas the transitions between stable and unstable behavior at finite wavelength are those of Equations~(\ref{eq:kcond}) and (\ref{eq:kacoustic}).

\subsection{Comparison between Field's and Waters \& Proga's formulations} \label{sect:comparison}

We have just seen that the parameters used in the formulation of \citet{waters2019non} have the benefit over those of \citet{field1965thermal} of expressing the isochoric, isobaric, and isentropic instability criteria as $\Nrho < 0$, $\Np < 0$, and $\NS < 0$, respectively, or $\Nrhoz < 0$, $\Npz < 0$, and $\NSz < 0$ if $\lambda$ is sufficiently large or heat conduction is negligible. Aside from this, it may appear that casting the dispersion relation as in Equation~(\ref{eq:dr_WP}) does not provide any advantage over Equation~(\ref{eq:dr_Field}): the former contains the five equilibrium parameters $\tc$, $\Nrhoz$, $\Npz$, $\lambdaF$, and $\cs$, while the latter only depends on four parameters: $k_T$, $k_\rho$, $k_\kappa$, and $\cs$.

Nevertheless, when it comes to the classification of the instability domains, the parameters used by \citet{waters2019non} provide a simpler description than those of \citet{field1965thermal}. To endorse this statement, let us consider a system given by its equilibrium temperature and density, $\Tz$ and $\rhoz$, together with the thermal conductivity, $\kappa(\Tz)$, and the derivatives $\calL_T$ and $\calL_\rho$. Now the constants $k_T$, $k_\rho$, and $k_\kappa$ are defined and one may wonder which modes are supported by this system and what are their stability features. Are there three thermal modes for some wavelength range? For which wavelengths are the modes unstable? The answer lies in Figure~1 of \citet{field1965thermal}: when $k$ is varied between 0 and $\infty$, our system is represented by a curve whose trajectory on the plane of this figure allows one to determine the nature of the modes for all wavelengths and the transition between stable and unstable domains and between one and three thermal modes. Then, what is the shape of this curve? Does it intersect the shaded regions, where there are three thermal modes? Does it cross the regions with none, 1, 2, or 3 positive roots? At which wavenumbers do these changes of domain take place? These questions are not easy to answer because the wavenumber enters the horizontal axis in a complex manner: while the variable in the vertical axis is $\sigma_\rho = k_\rho/k$, the one in the horizontal axis is $\sigma_T' = \sigma_T+\sigma_\kappa = k_T/k + k/k_\kappa$. Thus, for a given system, its exact trajectory in Figure~1 of \citet{field1965thermal} cannot be easily predicted from the values of the equilibrium parameters.

On the contrary, \citet{waters2019non} showed that the type of mode solutions supported by a given system depends exclusively on the parameter~$R$ and the signs of $\Npz$ and $\Nrhoz$. This information is conveyed in their Figure~2, which is updated in our Table~\ref{table:R_and_lambda}. To use this table it is enough to calculate $\Npz$ and $\Nrhoz$ and then use Equation~(\ref{eq:R}) to compute $R$ and determine the row of Table~\ref{table:R_and_lambda} the system belongs to, after which one straightforwardly knows the relevant wavelength ranges and the properties of modes in each of them. Read in this way, Table~\ref{table:R_and_lambda} is a map of the $(R, \lambda)$ plane, with the rows covering the possible values of $R$ and the columns the wavelength ranges that $R$ itself delimits. One can, for example, appreciate the change of modes between stable and unstable character (regular text versus bold and italic text) and the transition between thermal and acoustic solutions. The dispersion diagrams of Figures~\ref{fig:row1}--\ref{fig:row7} exemplify the wavelength-dependence of the mode characteristics summarized in Table~\ref{table:R_and_lambda}. The conclusion is that quantitative information about the modes of a system is easier to gain with Table~\ref{table:R_and_lambda} than with Figure~1 of \citet{field1965thermal}.

\begin{deluxetable*}{cccccc}
\tablecaption{Summary of mode properties in relevant ranges of $R$. The table is a map of the $(R, \lambda)$ plane: each row is a range of $R$, each column a range of wavelength, and each entry gives the type of the modes present and whether they are damped, unstable, or overstable. \label{table:R_and_lambda}}
\tablehead{
\colhead{Instability domain} & \multicolumn{5}{c}{Wavelength dependence} \\
\colhead{($R$ and signs of $\Npz$, $\Nrhoz$)} & \multicolumn{5}{c}{(Mode type and stability)}
}
\startdata
\multirow{4}{*}{\makecell{$R < 0$\\$\Npz < 0$\\$\Nrhoz > 0$}} & \multicolumn{2}{c}{$\lambda < \lambdacond$} & \multicolumn{2}{c}{$\lambdacond < \lambda < \lambdac$} & $\lambda > \lambdac$ \\
& \multicolumn{2}{c}{\cond} & \multicolumn{2}{c}{\bfcond} & \bfcond \\
& \multicolumn{2}{c}{acoustic} & \multicolumn{2}{c}{acoustic} & \cond \\
& \multicolumn{2}{c}{acoustic} & \multicolumn{2}{c}{acoustic} & \cond \\
\tableline
\multirow{4}{*}{\makecell{$R < 0$\\$\Npz > 0$\\$\Nrhoz < 0$}} & \multicolumn{2}{c}{$\lambda < \lambdaacoustic$} & \multicolumn{2}{c}{$\lambdaacoustic < \lambda < \lambdac$} & $\lambda > \lambdac$ \\
& \multicolumn{2}{c}{\cond} & \multicolumn{2}{c}{\cond} & \cond \\
& \multicolumn{2}{c}{acoustic} & \multicolumn{2}{c}{\itacoustic} & \bfcond \\
& \multicolumn{2}{c}{acoustic} & \multicolumn{2}{c}{\itacoustic} & \bfcond \\
\tableline
\tableline
\multirow{4}{*}{\makecell{$0 < R < 1/9$\\$\Npz, \Nrhoz < 0$}} & $\lambda < \lambdaacoustic$ & $\lambdaacoustic < \lambda < \lambdacond$ & $\lambdacond < \lambda < \lambda_-$ & $\lambda_- < \lambda < \lambda_+$ & $\lambda > \lambda_+$ \\
& \cond & \cond & \bfcond & \bfcond & \itacoustic \\
& acoustic & \itacoustic & \itacoustic & \bfcond & \itacoustic \\
& acoustic & \itacoustic & \itacoustic & \bfcond & \bfcond \\
\tableline
\multirow{4}{*}{\makecell{$1/9 < R < 1/\gamma$\\$\Npz, \Nrhoz < 0$}} & \multicolumn{2}{c}{$\lambda < \lambdaacoustic$} & \multicolumn{2}{c}{$\lambdaacoustic < \lambda < \lambdacond$} & $\lambda > \lambdacond$ \\
& \multicolumn{2}{c}{acoustic} & \multicolumn{2}{c}{\itacoustic} & \itacoustic \\
& \multicolumn{2}{c}{acoustic} & \multicolumn{2}{c}{\itacoustic} & \itacoustic \\
& \multicolumn{2}{c}{\cond} & \multicolumn{2}{c}{\cond} & \bfcond \\
\tableline
\multirow{4}{*}{\makecell{$1/\gamma < R < 1$\\$\Npz, \Nrhoz < 0$}} & \multicolumn{2}{c}{$\lambda < \lambdacond$} & \multicolumn{2}{c}{$\lambdacond < \lambda < \lambdaacoustic$} & $\lambda > \lambdaacoustic$ \\
& \multicolumn{2}{c}{acoustic} & \multicolumn{2}{c}{acoustic} & \itacoustic \\
& \multicolumn{2}{c}{acoustic} & \multicolumn{2}{c}{acoustic} & \itacoustic \\
& \multicolumn{2}{c}{\cond} & \multicolumn{2}{c}{\bfcond} & \bfcond \\
\tableline
\tableline
\multirow{4}{*}{\makecell{$R > 1$\\$\Npz, \Nrhoz < 0$}} & \multicolumn{2}{c}{$\lambda < \lambdacond$} & \multicolumn{3}{c}{$\lambda > \lambdacond$} \\
& \multicolumn{2}{c}{acoustic} & \multicolumn{3}{c}{acoustic} \\
& \multicolumn{2}{c}{acoustic} & \multicolumn{3}{c}{acoustic} \\
& \multicolumn{2}{c}{\cond} & \multicolumn{3}{c}{\bfcond} \\
\tableline
\multirow{4}{*}{\makecell{$R > 1$\\$\Npz, \Nrhoz > 0$}} & \multicolumn{2}{c}{$\lambda < \lambdaacoustic$} & \multicolumn{3}{c}{$\lambda > \lambdaacoustic$} \\
& \multicolumn{2}{c}{acoustic} & \multicolumn{3}{c}{\itacoustic} \\
& \multicolumn{2}{c}{acoustic} & \multicolumn{3}{c}{\itacoustic} \\
& \multicolumn{2}{c}{\cond} & \multicolumn{3}{c}{\cond} \\
\enddata
\tablecomments{The first column contains the value of $R$ together with the signs of $\Npz$ and $\Nrhoz$. The other columns display the wavelength range and the character of thermal and acoustic modes: regular text is used for stable modes, bold text for unstable thermal modes, and italic text for overstable acoustic modes. Meaning of wavelengths: $\lambdacond$ and $\lambdaacoustic$, given in Equations~(\ref{eq:kcond}) and~(\ref{eq:kacoustic}), mark the change of the thermal and the two acoustic modes from damped to exponentially growing. The wavelengths $\lambdac$ (case $R < 0$) and $\lambda_\pm$ (case $0 < R < 1/9$) signal the transition between a pair of acoustic modes and two thermal modes. They are obtained by solving Equation~(\ref{eq:WP2019_eq13}). This table is adapted from Figure~2 of \citet{waters2019non}, where approximate formulas for $\lambdac$ and $\lambda_\pm$  can be found (they are presented here in Equations~(\ref{eq:lambdac}) and (\ref{eq:lambdapm})). The classification of stability domains presented in this table is valid for small Field length ($\lambdaF \lesssim \lambdath$) or negligible thermal conduction ($\lambdaF \rightarrow 0$); see Section~\ref{sect:arbitrary_lambdaF}.}
\end{deluxetable*}

We emphasize that the number of wavenumbers at which the modes change character between thermal and acoustic --- the critical wavenumbers of Section~\ref{sect:WP_fig2_finite_kF} --- reveals the main features of the spectrum. If a system has no critical wavenumbers, then it will support a pair of acoustic modes that transition from damped to overstable at $\lambda = \lambdaacoustic$ and a thermal mode that transitions from stable to unstable at $\lambda = \lambdacond$; see rows 4 and 5 of Table~\ref{table:R_and_lambda}. The cases when $\lambdaacoustic$ or $\lambdacond$ do not exist correspond to rows 6 and 7, respectively. In the case of one critical wavenumber there is a single wavelength at which the two acoustic modes transform into two thermal modes (rows 1 and 2). Finally, for two critical wavelengths this conversion takes place at a wavelength $\lambda_-$ and is reversed at a wavelength $\lambda_+ > \lambda_-$.

Table~\ref{table:R_and_lambda} has a graphical counterpart in the $(\Npz, \Nrhoz)$ plane; see the top left panel of Figure~\ref{fig:Nzeros}. The first two rows of the table correspond to the second and fourth quadrants, respectively. Rows 3 to 6 completely fill the third quadrant and are in clockwise order from the negative $\Nrhoz$-axis to the negative $\Npz$-axis. Finally, row 7 lies between the positive $\Npz$-axis and the solid line in the first quadrant.  We see that Table~\ref{table:R_and_lambda} does not cover the whole of the $(\Npz, \Nrhoz)$ plane, because the region comprised between the line $R = 1$ in the first quadrant and the positive $\Nrhoz$-axis is not addressed in our table. When both $\Npz$ and $\Nrhoz$ are positive the plasma is isobarically and isochorically stable, and one might expect all of its modes to be damped at all wavelengths.  This is indeed so if $R < 1$, and we omit such systems from the table because they present no instability whatsoever.  It is not so if $R > 1$, because $\Nrhoz > 0$ then implies $\NSz < 0$ by Equation~(\ref{eq:NSz}): the isentropic instability criterion is satisfied, the two acoustic modes are overstable for $\lambda > \lambdaacoustic$, and this case is retained as the seventh row.  We also stress that the absence of instability does not imply the absence of structure in the spectrum.  A plasma with $\Npz, \Nrhoz > 0$ and $0 < R < 1/9$ still possesses two critical wavelengths, at which its pair of damped acoustic modes is converted into two damped thermal modes and back again, just as happens in the third row of Table~\ref{table:R_and_lambda} with overstable and unstable modes.  Since every mode is damped, this case is of no interest for thermal instability and is not tabulated.

Among the boundaries that separate the rows of Table~\ref{table:R_and_lambda}, the value $R = 1/\gamma$ deserves a comment, because it plays a larger role than its origin suggests.  It is defined by the coincidence of $\lambdacond$ and $\lambdaacoustic$, that is, by the thermal and the two acoustic modes becoming unstable at one and the same wavelength, and it separates the fourth row from the fifth.  We shall meet it again in Section~\ref{sect:arbitrary_lambdaF}, where it bounds the region of Figure~\ref{fig:Nzeros} that is accessible to real plasmas, and in Appendix~\ref{sect:app_powerlaw}, where it is shown to be the value taken by $R$ when the heating rate per unit volume shares the density scaling of the radiative losses.

One last comment is in order: to generate their Figure~2, \citet{waters2019non} realized that the critical wavelengths at which solutions change from three thermal modes to one thermal mode and two acoustic modes, and vice versa, depend solely on~$R $. These authors did not include in their Figure~2 the effect of conduction and in Section~\ref{sect:arbitrary_lambdaF} we show that Table~\ref{table:R_and_lambda} is still valid when conduction is included as long as $\lambdaF \lesssim \lambdath$.

\subsection{Time scales and the isobaric or isochoric character of the thermal mode} \label{sect:timescales}

It is well-known that, if the radiative cooling time scale is much shorter than the time taken by the pressure gradient to influence the dynamics of the plasma via acoustic waves, then the linear evolution of the thermal mode is isochoric. This time-scale comparison is described by the condition $\tc \ll \tsnd$, where $\tsnd$ is the sound crossing time,

\begin{equation} \label{eq:tsnd}
     \tsnd = \frac{\lambda}{\cs}.
\end{equation}

\noindent Given that $\tsnd$ is proportional to $\lambda$, it is clear that the thermal mode behaves isochorically for long wavelengths; to be more precise, this happens if $\lambda \gg \lambdath$, where

\begin{equation}\label{eq:lambdath}
\tc = \tsnd \hspchorone \Rightarrow \hspchorone \lambda = \lambdath = \cs \tc.
\end{equation}

In addition, if $\lambda \ll \lambdath$ then acoustic waves can restore the pressure balance across one wavelength before cooling has time to act. Then, the thermal mode can, {\it in principle}, evolve isobarically. We say ``in principle'' because the last statement does not take into account the role of thermal conduction and, as we will now see, the thermal mode is not isobaric if the wavelength is sufficiently small. Thermal conduction acts on the time scale

\begin{equation} \label{eq:tcond}
     \tcond = \frac{\pz}{(\gamma - 1) \kappa \Tz} \frac{1}{k^2},
\end{equation}

\noindent where $\tcond$ is the conduction time across a length $k^{-1}$ rather than across a wavelength, which is the convention under which $\lambdaF$ carries the factor $2\pi$ of Equation~(\ref{eq:lambdaF}). If $\tcond \ll \tsnd$, then the heat lost by a plasma element can be quickly compensated by heat carried to this element by thermal conduction. Thus, the linear dynamics of the plasma is governed by thermal conduction without the need of mass motions and this allows the plasma to evolve isochorically, such as happens for long wavelengths. The particular wavelength below which the thermal mode turns isochoric, i.e., at which the conduction and sound time scales balance, is

\begin{equation}\label{eq:lambda_isochoric}
\tcond = \tsnd \hspchorone \Rightarrow \hspchorone \lambdaisoch = \frac{\lambdaF^2}{\lambdath}.
\end{equation}

Given the large number of wavelengths and time scales that arise in this study, we compile them in Table~\ref{table:scales}. To summarize our findings from the time-scale analysis it is very relevant to note that $\tc$ does not depend on $\lambda$, whereas $\tsnd$ and $\tcond$ are linear and quadratic in $\lambda$, respectively. The wavelength regions in which the thermal mode is isochoric or isobaric then depend on whether $\lambdaF < \lambdath$ or $\lambdath < \lambdaF$. If $\lambdaF < \lambdath$, then we have the ordering $\lambdaisoch < \lambdaF < \lambdath$ and the thermal mode is isochoric for $\lambda \ll \lambdaisoch$ and $\lambda \gg \lambdath$; it is isobaric otherwise, that is, in the range $\lambdaisoch \ll \lambda \ll \lambdath$. On  the other hand, if $\lambdaF > \lambdath$, then the wavelength ordering becomes $\lambdath < \lambdaF < \lambdaisoch$ and, since $\lambdath < \lambdaisoch$, there {\it may be} no wavelength range in which the thermal mode behaves isobarically. Of course, the limits of the wavelength ranges presented in this paragraph are only approximate and they are put to the test in Section~\ref{sect:isobaric_perturbations} for two particular values of $\lambdaF/ \lambdath$.

It is worth being explicit about why the thermal mode is isochoric at both ends of the spectrum, since the two ranges have different physical origins. In both of them the condition is the same, namely that the temperature perturbation is removed faster than a sound wave can cross the perturbation and restore pressure balance, so that the density has no time to respond; what differs is the process that removes it. At long wavelengths it is radiation, and the comparison $\tc < \tsnd$ gives $\lambda \gg \lambdath$. At short wavelengths it is conduction, and the comparison $\tcond < \tsnd$ gives $\lambda \ll \lambdaisoch$. Since $\tcond \propto \lambda^2$ and $\tsnd \propto \lambda$, conduction necessarily wins below some wavelength, so the second range always exists when conduction is present. The short-wavelength isochoric range is therefore an effect of thermal conduction and exists only because conduction is retained: if $\kappa = 0$ then $\lambdaF = 0$, $\lambdaisoch = \lambdaF^2/\lambdath$ vanishes with it, and the thermal mode is isobaric at short wavelengths, as in the classical treatment. Our result does not contradict that one; it reduces to it.

An important message is that, regardless of whether $\lambdaF < \lambdath$ or $\lambdath < \lambdaF$, thermal conduction prevents the thermal mode from being isobaric at small wavelengths, it is isochoric instead. Failing to appreciate this fact can lead to incorrect results; see our discussion of \citet{seno_etal2026} in Section~\ref{sect:discussion_character}. Another major outcome is that the ratio $\lambdaF/ \lambdath$ determines whether the thermal mode can lead to isobaric perturbations. Our analysis in Section~\ref{sect:environments} of several astrophysical environments in which thermal instability may be at work reveals that all these environments satisfy $\lambdaF/ \lambdath < 1$, so that the thermal mode is isochoric for small and large wavelengths and is isobaric at intermediate wavelengths.

\begin{deluxetable}{lcl}
\tablecaption{Time scales and wavelengths used in this work.\label{table:scales}}
\tablehead{\colhead{Symbol} & \colhead{Equation} & \colhead{Meaning}}
\startdata
\multicolumn{3}{c}{\textit{Time scales}} \\
\hline
$\tc$                & (\ref{eq:tcool})      & radiative cooling time; independent of $\lambda$ \\
$t_{\rm snd}$        & (\ref{eq:tsnd})       & sound crossing time; $\propto \lambda$ \\
$t_{\rm cond}$       & (\ref{eq:tcond})      & thermal conduction time; $\propto \lambda^2$ \\
\hline
\multicolumn{3}{c}{\textit{Wavelengths at which two time scales balance}} \\
\hline
$\lambdath$          & (\ref{eq:lambdath})   & thermal length, $\tc = t_{\rm snd}$; the thermal mode is isochoric for $\lambda \gg \lambdath$ \\
$\lambdaF$           & (\ref{eq:lambdaF}), (\ref{eq:lambda_F2})    & Field length, $t_{\rm cond} = \tc$; conduction and cooling balance \\
$\lambdaisoch$       & (\ref{eq:lambda_isochoric})& $t_{\rm cond} = t_{\rm snd}$; the thermal mode is isochoric for $\lambda \ll \lambdaisoch$ \\
\hline
\multicolumn{3}{c}{\textit{Wavelengths at which the modes change character}} \\
\hline
$\lambdacond$        & (\ref{eq:kcond})      & the thermal mode changes from damped to unstable \\
$\lambdaacoustic$    & (\ref{eq:kacoustic})  & the two acoustic modes change from damped to overstable \\
$\lambda_{\rm c}$    & (\ref{eq:lambdac})    & two acoustic modes are converted into two thermal modes (rows 1 and 2) \\
$\lambda_\pm$        & (\ref{eq:lambdapm})   & the same conversion, and its reversal (row 3) \\
$\lambda_{\rm m}$    & (\ref{eq:km})    & the growth rate of the thermal mode is largest \\
\enddata
\tablecomments{The row numbers refer to Table~\ref{table:R_and_lambda}.  The critical wavelengths $\lambda_{\rm c}$ and $\lambda_\pm$ are obtained by solving Equation~(\ref{eq:WP2019_eq13}); the approximate expressions of \citet{waters2019non} are given in Equations~(\ref{eq:lambdac}) and (\ref{eq:lambdapm}).}
\end{deluxetable}

There is yet another relevant wavelength, which is related to the stabilizing role of thermal conduction. This takes place if the time scale of conduction is shorter than that of cooling, because the former can transport heat to a cooling element faster than the latter removes it. The wavelength at which the two time scales are balanced is

\begin{equation} \label{eq:lambda_F2}
     \tcond = \tc \hspchorone \Rightarrow \hspchorone \lambda = \lambdaF.
\end{equation}

\noindent Not surprisingly, we have recovered the wavelength $\lambdaF$, below which radiative cooling cannot sustain the instability of the thermal mode \citep{field1965thermal}. Now, let us put together this result with our previous finding that the thermal mode is isochoric for $\lambda \ll \lambdaisoch$. Regardless of the wavelength ordering prescribed above, i.e., regardless of whether $\lambdaisoch < \lambdaF$ or not, the thermal mode is isochoric and stable for sufficiently short wavelength. This conclusion agrees with the results of Section~\ref{sect:short_condensation}, based on the analysis of the equations of linear perturbations. Note that the thermal mode is damped throughout this short-wavelength range, since $\lambda < \lambdacond$ there, so that whether it is isobaric or isochoric has no bearing on the existence of the instability. It has a direct bearing on the growth rate, however, because the asymptotic behaviour of $\nR$ for $\lambda \rightarrow 0$ is sometimes derived by assuming the mode to be isobaric; Section~\ref{sect:discussion_character} gives a published example in which that assumption leads to an incorrect result.

\subsection{Linear theory perturbations} \label{sect:perturbations}

We express the density, pressure, and temperature perturbations as a fraction of their equilibrium value. Denoting the initial amplitude of the density perturbation by $A_\rho = \Arho \rhoz$ for some dimensionless constant $A$, we have, upon plugging Equation~(\ref{eq:delta_f}) into the linearized equations of gas dynamics \citep[see][]{field1965thermal},

\begin{align}
    \frac{\delta \rho(x,t)}{\rhoz} &= \Arho \exp(n t + i k x), \label{eq:perturb_rho} \\
    \delta v(x,t) &= i \Arho \frac{n}{k} \exp(n t + i k x), \label{eq:perturb_v} \\
    \frac{\delta p(x,t)}{\pz} &= - \Arho \frac{1}{\csi^2} \left(\frac{n}{k}\right)^2 \exp(n t + i k x), \label{eq:perturb_p}
\end{align}

\noindent with $\csi^2 = \pz/\rhoz$ the square of the isothermal sound speed.

To investigate the limit $k \rightarrow 0$ one needs to write these equations in terms of the initial amplitude of the pressure perturbation, $A_\mathrm{p} = \Ap \pz$:

\begin{align}
    \frac{\delta \rho(x,t)}{\rhoz} &= - \Ap \csi^2 \left(\frac{k}{n}\right)^2 \exp(n t + i k x), \label{eq:perturb_rho2} \\
    \delta v(x,t) &= -i \Ap \frac{k}{n} \csi^2 \exp(n t + i k x), \label{eq:perturb_v2} \\
    \frac{\delta p(x,t)}{\pz} &= \Ap \exp(n t + i k x), \label{eq:perturb_p2}
\end{align}

\noindent so that $k$ no longer appears in the denominator of $\delta v$ and $\delta p$, but in the numerator of $\delta \rho$ and $\delta v$.

In both formulations, the temperature perturbation is

\begin{equation} \label{eq:perturb_T}
    \frac{\delta T}{\Tz} = \frac{\delta p}{\pz} - \frac{\delta \rho}{\rhoz}.
\end{equation}

\section{The short- and long-wavelength limits} \label{sect:short_long_wavelength}

\subsection{Short wavelength solutions ($\lambda \rightarrow 0$)} \label{sect:short_wavelength}

Regardless of the value of $R$, when $\lambda \rightarrow 0$ the system supports one thermal mode and a pair of acoustic modes. All these modes are stable, and the dominant terms in their growth rates are independent of $\Nrhoz$, $\Npz$, and $\NSz$ and depend only on $\lambdaF$, which is an indication of their stability being determined by thermal conduction alone, as expected. The growth rates given here and in Section~\ref{sect:long_wavelength} have been obtained analytically or using symbolic manipulation in Wolfram Mathematica.

\subsubsection{Thermal mode} \label{sect:short_condensation}

 The growth rate of the thermal mode has real and imaginary parts given by
\begin{equation} \label{eq:lambda0_condens_n}
    \nR = -\frac{\Nrhoz}{\tc} + \frac{\gamma-1}{\gamma} \tc \cs^2 \kF^2 -\frac{1}{\tc} \left(\frac{k}{\kF}\right)^2, \hspchorhalf \nI = 0,
\end{equation}

\noindent where the third term in $\nR$ is the dominant one in the short wavelength limit. Due to the thermal conduction term, the factor $n/k$ in Equations~(\ref{eq:perturb_v}) and (\ref{eq:perturb_p}) is proportional to $k$, so it diverges as $\lambda \rightarrow 0$. Therefore, to study this solution, we must use Equations~(\ref{eq:perturb_rho2})--(\ref{eq:perturb_T}) rather than Equations~(\ref{eq:perturb_rho})--(\ref{eq:perturb_p}); in the limit $k \rightarrow \infty$ we obtain,

\begin{equation} \label{eq:lambda0_condens_perturb}
    \delta \rho = 0, \hspchorone \delta p \neq 0, \hspchorone \delta v = 0, \hspchorone \delta T \neq 0.
\end{equation}

\noindent We thus see that this is an isochoric mode with no mass motions and whose pressure and temperature perturbations damp in time. It is worth emphasizing that this mode is attenuated by the effect of thermal conduction alone, yet it has no density perturbations regardless of whether the system is isochorically unstable for $\lambda \rightarrow \infty$ ($\Nrhoz < 0$) or not ($\Nrhoz > 0$), that is, regardless of the sign of $\Nrhoz$. The point we want to make here is that the isochoric instability criterion formulated by \citet{field1965thermal} has no implications about the stability of the short-wavelength (isochoric) thermal mode. Another interesting feature of this mode is its change from isobaric character in a system without conduction \citep[e.g.,][]{Das2021} to isochoric in the present case with thermal conduction; also see the time-scale discussion of Section~\ref{sect:timescales}.

Further insight into the properties of this mode comes from the analysis of the perturbed mass continuity and momentum equations, i.e., Equations~(11) and (12) of \citet{field1965thermal}. Imposing the dominant behavior $n \propto k^2$ from Equation~(\ref{eq:lambda0_condens_n}) we obtain for $k \rightarrow \infty$:

\begin{equation}
    k \delta \rho = \mathcal{O}(\delta v), \hspchorone k \delta v = \mathcal{O}(\delta p) \hspchorhalf \Rightarrow \hspchorhalf k^2 \delta \rho = \mathcal{O}(\delta p).
\end{equation}

\noindent It is then clear that, for $\lambda \rightarrow 0$, the thermal mode has vanishing density and velocity perturbations, in agreement with Equation~(\ref{eq:lambda0_condens_perturb}). Obviously, plugging $\delta \rho = 0$, $\delta v = 0$ into the perturbed energy equation and perfect gas law, that is, Equations~(13) and (14) of \citet{field1965thermal}, allows us to retrieve the dominant $n \propto k^2$ part of Equation~(\ref{eq:lambda0_condens_n}).

\subsubsection{Acoustic modes} \label{sect:short_acoustic}

For very short wavelengths, the acoustic modes have

\begin{equation} \label{eq:lambda0_acoustic_n}
    \nR = -\frac{\gamma-1}{2\gamma} \tc \cs^2 \kF^2, \hspchor \nI = \pm k \csi.
\end{equation}

\noindent After inserting this solution into Equations~(\ref{eq:perturb_rho})--(\ref{eq:perturb_p}) or (\ref{eq:perturb_rho2})--(\ref{eq:perturb_p2}) and Equation~(\ref{eq:perturb_T}) we obtain

\begin{equation} \label{eq:lambda0_acoustic_perturb}
    \delta \rho \neq 0, \hspchorone \delta p \neq 0, \hspchorone \delta v \neq 0, \hspchorone \delta T = 0.
\end{equation}

\noindent For $\gamma > 1$, this is a pair of damped isothermal acoustic waves.

\subsection{Long wavelength solutions ($\lambda \rightarrow \infty$)} \label{sect:long_wavelength}

When $\lambda$ is sufficiently large, the system supports three thermal modes if $R < 0$ and one thermal mode and a pair of acoustic modes if $R > 0$.

\subsubsection{Thermal modes} \label{sect:long_condensation}

For all values of $R$ one thermal mode has

\begin{equation} \label{eq:k0_condens_n}
    \nR = -\frac{\Nrhoz}{\tc} -\left[\frac{1}{\kF^2\tc} +\frac{\tc}{\Nrhoz} (R-1) \cs^2\right] k^2, \hspchorhalf \nI = 0,
\end{equation}

\noindent where the first term on the right-hand side is the dominant one for $\lambda \rightarrow \infty$. When dealing with perturbations, here we face the same situation we found in Section~\ref{sect:short_condensation}: for this growth rate, the factor $n/k$ in Equations~(\ref{eq:perturb_v}) and (\ref{eq:perturb_p}) diverges as $\lambda \rightarrow \infty$ and we are forced to use Equations~(\ref{eq:perturb_rho2})--(\ref{eq:perturb_T}) to study the perturbations. We get

\begin{equation} \label{eq:k0_condens_perturb}
    \delta \rho = 0, \hspchorone \delta p \neq 0, \hspchorone \delta v = 0, \hspchorone \delta T \neq 0.
\end{equation}

\noindent Just like the thermal mode for $\lambda \rightarrow 0$, this is an isochoric solution with no mass motions. In this case, Field's isochoric instability criterion determines the stability of the thermal mode: it is unstable if the isochoric instability criterion is satisfied ($\Nrhoz < 0$) and is stable otherwise.

\citet{waters_stricklan2025} derived this solution directly from the internal energy equation by assuming a spatially constant temperature perturbation with zero density and velocity perturbations. Furthermore, the growth rate of this solution is that of Equation~(\ref{eq:k0_condens_n}). These authors stated that this solution is not governed by thermal instability and that it is ``the mode of a separate instability''; this led them to term it ``catastrophic cooling mode''. It is now clear that thermal instability theory recovers this solution in the $k = 0$ limit as an isochoric thermal mode whose properties can be well addressed by writing the perturbation amplitudes in terms of that of the pressure perturbation, as in Equations~(\ref{eq:perturb_rho2})--(\ref{eq:perturb_p2}).

The other two thermal modes ($R < 0$) have

\begin{equation} \label{eq:k0_condens_n_Rlt0}
    \nR = \pm \sqrt{-R} k \cs +\frac{\tc}{2\Nrhoz} (R-1) k^2 \cs^2, \hspchorhalf \nI = 0, \hspchorhalf \mathrm{if } \; R < 0,
\end{equation}

\noindent hence, one of these modes is unstable and the other is stable, their growth/damping rate tending to zero as $\lambda \rightarrow \infty$. In this respect, one must bear in mind that $R < 0$ means that the system is either isochorically ($\Nrhoz < 0$) or isobarically ($\Npz < 0$) unstable, but not both. In addition, these linear theory solutions have non-vanishing perturbations:

\begin{equation} \label{eq:k0_condens_perturb_Rlt0}
    \delta \rho \neq 0, \hspchorone \delta p \neq 0, \hspchorone \delta v \neq 0, \hspchorone \delta T \neq 0.
\end{equation}

\noindent These two thermal modes are neither isobaric nor isochoric.

\subsubsection{Acoustic modes} \label{sect:long_acoustic}

The two acoustic modes, which are present when $R > 0$, have

\begin{equation} \label{eq:k0_acoustic_n}
    \nR = \frac{\tc}{2\Nrhoz} (R-1) k^2 \cs^2, \hspchorhalf \nI = \pm \sqrt{R} k \cs, \hspchorhalf \mathrm{if } \; R > 0.
\end{equation}

\noindent As noted in \citet{waters2019non}, these solutions are overstable if either $R < 1$ and $\Nrhoz < 0$ or $R > 1$ and $\Nrhoz > 0$; and they are stable otherwise. Unlike adiabatic sound waves, their phase speed is not $\cs$ but $\sqrt{R} \cs$. Furthermore, in this case, all perturbations are non-zero:

\begin{equation} \label{eq:k0_acoustic_perturb}
    \delta \rho \neq 0, \hspchorone \delta p \neq 0, \hspchorone \delta v \neq 0, \hspchorone \delta T \neq 0.
\end{equation}


\section{Application to representative astrophysical environments} \label{sect:environments}
 
The classification of Table~\ref{table:R_and_lambda} is expressed in terms of $R$, $\Npz$, and $\Nrhoz$, and its validity is controlled by the ratio $\lambdaF/\lambdath$.  It is therefore natural to ask where real plasmas sit in this classification.  In this section we evaluate these parameters, together with the characteristic scales of Section~\ref{sect:timescales}, for five environments in which thermal instability is invoked as the mechanism that produces multi-phase structure. It is worth being clear about the division of labour here. The parameter $R$, together with the signs of $\Npz$ and $\Nrhoz$, fixes the sequence of regimes that a plasma passes through as the wavelength is increased, and the stability of each mode in each of them; it does not fix the wavelengths at which one regime gives way to the next, which are dimensional quantities and require the equilibrium temperature and density of the medium. Table~\ref{table:R_and_lambda} supplies the first and Table~\ref{table:environments_output} the second.

Our selection of environments is guided by two criteria.  First, we retain only media in which thermal instability is thought to operate, and not the structures that result from it: solar prominences and coronal rain, for instance, are believed to be products of thermal instability in the corona and are themselves thermally stable, so they are not included.  Second, for each environment we adopt a cooling function, a thermal conductivity, and a heating prescription that are in common use in that field, rather than constructing our own. The five media considered are the two-phase atomic interstellar medium (ISM), the solar corona, in an active-region loop in equilibrium and in a loop late in a thermal non-equilibrium (TNE) cycle, the circumgalactic medium (CGM), the cool-core intracluster medium (ICM), and the X-ray photoionized outflow of an active galactic nucleus. The prescriptions adopted for each of them, and the reasons for adopting them, are set out in Appendix~\ref{sect:app_environments}; the inputs are listed in Table~\ref{table:environments_input} and the resulting parameters in Table~\ref{table:environments_output}.

We use the closed-form expressions collected in Appendix~\ref{sect:app_powerlaw}. In most of the environments considered here the radiative losses, the heating rate, and the thermal conductivity are, to a good approximation, power laws of the temperature and density. We write the radiative losses per unit volume as $\rho\Lambda = x\, n_{\rm H}^{2} Q(T)$ with $Q(T) = Q_0 T^{-\beta}$ (Equation~(\ref{eq:appQ})), where $Q(T)$ is the radiative loss function of the plasma, tabulated for each medium from atomic physics, and $\beta$ measures how steeply the losses grow as the plasma cools.  The heating rate per unit volume is written as $\rho H = H_\ast \rho^{\,a}T^{\,b}$ (Equation~(\ref{eq:appH})), where the exponents $a$ and $b$ give its dependence on density and temperature. Finally, the thermal conductivity is $\kappa = f\kappa_0 T^{\,\nu}$ (Equation~(\ref{eq:appkappa})), with $\nu = 5/2$ for Spitzer conduction in a fully ionized plasma, $\nu = 1/2$ for conduction by neutral atoms, and $f \le 1$ a suppression factor. Values of $f$ well below unity are expected in the intracluster and circumgalactic media, for the reasons recalled in Section~\ref{sect:intro}; we adopt $f = 0.1$ for both, which is conservative in the sense that a stronger suppression would only reduce $\lambdaF$ further.

Each environment is thus specified by the exponents $\beta$, $a$, $b$, and $\nu$, the suppression factor $f$, the normalizations $Q(\Tz)$ and $\kappa_0$, the composition constants $\mu_{\rm H}$, $\mu$, and $x$, and the equilibrium temperature $\Tz$ and hydrogen number density $\nHz$; Appendix~\ref{sect:app_powerlaw} turns these into $\Nrhoz$, $\Npz$, $R$, and the characteristic time and length scales. Wherever the radiative losses, the heating rate, and the thermal conductivity are not power laws of the temperature and density, as happens for the X-ray photoionized outflow, $\Npz$ and $\Nrhoz$ are evaluated directly from Equations~(\ref{eq:Nrho0_Np0}), which is all that is needed to place the medium in Table~\ref{table:R_and_lambda}; the restriction on the rows accessible to it is then expressed through its effective heating exponent, Equation~(\ref{eq:aeff}).

\begin{deluxetable*}{lllcccccccc}
\tablecaption{Input parameters for the five environments, the solar corona being represented by two states of a coronal loop.\label{table:environments_input}}
\tablehead{
\colhead{Environment} & \colhead{Heat-loss function} & \colhead{$\kappa_0$} &
\colhead{$\nu$} & \colhead{$f$} & \colhead{$Q(\Tz)$} & \colhead{$\beta$} & \colhead{$a$} & \colhead{$b$} &
\colhead{$\Tz$ (K)} & \colhead{$\nHz$ (cm$^{-3}$)}}
\startdata
Atomic ISM & KI02, VS07          & $2.5\times10^{3}$ & $1/2$ & 1 & $8.1\times10^{-27}$ & $-0.59$ & 1 & 0 & $1.0\times10^{3}$ & $2.5$ \\
Equilibrium loop             & CHIANTI             & $9.2\times 10^{-7}$         & $5/2$ & 1 & $2.1\times10^{-22}$ & 2.1 & 0 & 0 & $2.0\times10^{6}$ & $1.0\times10^{9}$ \\
TNE loop   & CHIANTI             & $1.0\times 10^{-6}$         & $5/2$ & 1 & $3.5\times10^{-22}$ & 0.11 & 0 & 0 & $1.0\times10^{6}$ & $1.0\times10^{10}$ \\
CGM       & SD93 / TN01 & $5.6\times 10^{-7}$ & $5/2$ & 0.1 & $6.4\times10^{-23}$ & $1.47$ & 0, 1 & 0 & $1.0\times10^{6}$ & $1.0\times10^{-3}$ \\
Cool-core ICM               & SD93 / TN01 & $5.4\times 10^{-7}$ & $5/2$ & 0.1 & $1.5\times10^{-23}$ & $-0.25$ & 0, 1 & 0 & $2.3\times10^{7}$ & $4.2\times10^{-2}$ \\
X-ray outflow & B94 & $1.0\times 10^{-6}$ & $5/2$ & 1 & $9.5\times10^{-24}$ & \multicolumn{3}{c}{not power laws} & $4.5\times10^{5}$ & $1.0\times10^{9}$ \\
\enddata
\tablecomments{KI02: \citet{KoyamaInutsuka2002}; VS07: \citet{vazquez-semadeni_etal2007}; SD93: \citet{SutherlandDopita1993}; TN01: \citet{TozziNorman2001}; B94: \citet{Blondin1994}. The thermal conductivity is $\kappa = f \kappa_0 T^\nu$ and has units erg~cm$^{-1}$~s$^{-1}$~K$^{-1}$; 
the Spitzer coefficient, used in all rows but the first one, is $\kappa_0 = 1.84\times 10^{-5} / \ln\Lambda_{\rm C}$ erg~cm$^{-1}$~s$^{-1}$~K$^{-1-\nu}$, with $\ln\Lambda_{\rm C} =29.7 + \ln[(T/10^6~{\rm K})\,n_{\rm e}^{-1/2}]$ and $n_{\rm e} = 1.2\,n_{\rm H}$ the electron number density in cm$^{-3}$. The units of $Q(\Tz)$ are erg~cm$^3$~s$^{-1}$. The composition is $\mu_{\rm H} = 1.4$, $\mu = 1.27$ for the atomic interstellar medium and $\mu_{\rm H} = 1.4$, $\mu = 0.6$ for the four ionized media. We take $x = 1.2$ for the corona, whose loss function is tabulated as $n_{\rm e}n_{\rm H}Q$; $x = 1.32$ for the circumgalactic and the intracluster media, tabulated as $n_{\rm e}n_{\rm i}Q$; and $x = 1$ for the atomic interstellar medium and for the X-ray photoionized outflow, whose rates are given per $n_{\rm H}^{2}$. The remaining parameters that specify each medium are given in Appendix~\ref{sect:app_environments}.}
\end{deluxetable*}
 
\begin{deluxetable*}{lcccccccccc}
\tabletypesize{\footnotesize}
\tablecaption{Derived parameters and characteristic scales.\label{table:environments_output}}
\tablehead{
\colhead{Environment} & \colhead{$\Nrhoz$} & \colhead{$\Npz$} & \colhead{$R$} &
\colhead{Row} & \colhead{$\tc$} & \colhead{$\lambdaF$} & \colhead{$\lambdath$} &
\colhead{$\lambdaF/\lambdath$} & \colhead{$\lambdacond/L$} & \colhead{$\lambdath/L$}}
\startdata
Atomic ISM        & $+0.59$ & $-0.41$ & $-0.41$ & 1 & $0.36$ Myr & $0.081$ pc & $1.2$ pc & $0.067$ & $0.13$ & 1.2 \\
Equilibrium loop   & $-2.1$ & $-4.1$ & $+1.2$ & 6 & $1.1$ h    & $400$ Mm   & $820$ Mm & $0.49$  & $2.0$ & 8.2 \\
TNE loop & $-0.11$ & $-2.1$ & $+12$ & 6 & $110$ s     & $9.7$ Mm   & $17$ Mm & $0.56$  & $0.067$ & 0.17 \\
CGM, $a = 0$      & $-1.5$ & $-3.5$ & $+1.4$ & 6 & $180$ Myr & $1.7$ kpc & $28$ kpc  & $0.059$ & $0.090$ & 2.8 \\
\quad $a = 1$     &         & $-2.5$ & $+1.0$ & 6 &           &           &           &         & $0.11$ &  \\
Cool core, $a = 0$ & $+0.25$ & $-1.8$ & $-4.3$ & 1 & $430$ Myr & $19$ kpc  & $320$ kpc & $0.061$ & $0.29$ & 6.4 \\
\quad $a = 1$      &         & $-0.75$ & $-1.8$ & 1 &           &           &           &         & $0.45$ & \\
X-ray outflow     & $+0.47$ & $-0.47$ & $-0.61$ & 1 & $6.4$ h    & $1.6\times10^{10}$ cm & $2.3\times10^{11}$ cm & $0.069$ & $1.2\times10^{-6}$ & $1.2\times10^{-5}$ \\
\enddata
\tablecomments{``Row'' is the row of Table~\ref{table:R_and_lambda} to which the system belongs and $L$ is its characteristic size: 1~pc for the ISM, 100~Mm for the coronal loops, 10~kpc for the CGM, 50~kpc for the ICM, and $2\times10^{16}$~cm for the X-ray photoionized outflow; the environments are those of Table~\ref{table:environments_input}, in the same order, with two rows each for the circumgalactic and intracluster media. A blank entry means that the quantity is equal to that in the row above it. The thermal mode is unstable at some wavelength that fits inside the system when $\lambdacond/L < 1$, and the isochoric regime is accessible when $\lambdath \lesssim L$.  All values of $\lambdaF$ include the factor $2\pi$ of Equation~(\ref{eq:lambdaF}). All entries are quoted to two significant figures. Where the loss function is nearly flat, as for the TNE loop, $R$ is very sensitive to the slope $\beta$ and should be taken as indicative.}
\end{deluxetable*}
 
\subsection{What the numbers show}

Three conclusions follow from Table~\ref{table:environments_output}.

The first, and the most important for the present work, is that every environment has $\lambdaF/\lambdath$ between $0.059$ and $0.56$, comfortably inside the range for which Table~\ref{table:R_and_lambda} is valid (Section~\ref{sect:arbitrary_lambdaF}). This is not an accident of our choice of fiducial values. Equation~(\ref{eq:appratio}) shows that the ratio depends weakly on the density of the plasma through $\ln \Lambda_{\rm C}$ and varies slowly with its temperature, and below we confirm that it remains well below unity throughout the range of temperatures over which each of the five loss functions applies. Conversely, any suppression of the conductivity moves a system further into the domain of validity of Table~\ref{table:R_and_lambda}, since $\lambdaF \propto f^{1/2}$. Our fiducial $f = 0.1$ for the circumgalactic and intracluster media is thus a conservative choice in this respect: with $f = 1$ the ratio $\lambdaF/\lambdath$ would rise only to $0.19$ in both cases. We are therefore not aware of an environment in which thermal instability operates and the Field length exceeds the thermal length: the classification of Table~\ref{table:R_and_lambda} applies wherever thermal instability is invoked, and the regime $\lambdaF \gtrsim \lambdath$ studied in Section~\ref{sect:arbitrary_lambdaF} is a remote corner of parameter space.
 
The second is that the eight entries populate only the first and the sixth rows of Table~\ref{table:R_and_lambda}, exactly as anticipated by Equation~(\ref{eq:appclassification}).  All of them have $a \le 2$, including the photoionized outflow, for which the radiative losses do not scale as $\rho^2$, because the ionization parameter varies with density, and for which Equation~(\ref{eq:aeff}) gives $\aeff = 1.06$; the second, third, fourth, and seventh rows are therefore excluded.  The exclusion of the second row has a simple physical reading: that row requires isochoric instability together with isobaric stability, but Equation~(\ref{eq:appsep}) with $a \leq 2$, or $\aeff \le 2$ for the X-ray photoionized outflow, gives $\Npz \le \Nrhoz$, so that a medium that is isochorically unstable is always isobarically unstable as well.  The exclusion of the third row matters for a different reason: it is the only row whose spectrum contains the pair of critical wavelengths $\lambda_\pm$, and hence the only one in which the two acoustic modes turn into thermal modes and back again.  That spectrum, the most elaborate of the seven, is not realized in any of the media considered here. Finally, for the four environments with $b = 0$ the distinction between the first and the sixth rows is simply the sign of $\Nrhoz$, that is, whether the cooling function rises or falls with temperature: the interstellar medium at $10^{3}$~K and the cluster cool core sit on rising branches and are isochorically stable, whereas the corona and the circumgalactic medium sit on falling branches and are unstable by both criteria. The photoionized outflow also falls in the first row, but there $\Nrhoz > 0$ arises from the temperature dependence of the heating rather than from the slope of the losses.
 
The third concerns the sensitivity to the heating prescription. For the circumgalactic and intracluster media, where both values of $a$ are tabulated, a change from $a = 1$ to $a = 0$ leaves the row of Table~\ref{table:R_and_lambda} unchanged, and the same is true of the remaining environments. It is worth noting, nevertheless, that the circumgalactic medium is close to a boundary: its slope $\beta = 1.47$ lies just below the critical value $1/(\gamma-1) = 3/2$ at which $R = 1$ for $a = 1$, $b = 0$, so that $R = 1.01$ in that case and a slightly steeper cooling function would place it in the fifth row instead of the sixth. The corona is the one case in which the choice of $a$ itself decides the row: $\beta \simeq 2.1$ lies between the values 3 and 3/2 at which $R = 1$ for $a = 0$ and $a = 1$, respectively, given by the last of Equations~(\ref{eq:appcrit}), so that $R$ falls from $1.20$ for $a=0$ to $0.89$ for $a=1$, crossing the value $R=1$ and moving the system from the sixth row to the fifth.  As discussed in Section~\ref{sect:discussion_CR_CC}, this is precisely the boundary that decides whether the thermal mode reaches its largest growth rate at a finite wavelength or as $\lambda\to\infty$.  We adopt $a=0$ because it is the prescription used in the loop models that motivate the comparison, but the conclusion should be understood as conditional on that choice.
 
\subsection{Which wavelengths fit inside the system} \label{sect:fit_inside}
 
The classification of Table~\ref{table:R_and_lambda} describes an infinite medium, and the wavelengths it involves must be compared with the size of the system before they can be given physical meaning.  Two comparisons are informative.
 
The thermal mode is unstable only for $\lambda > \lambdacond$, so a system is liable to thermal instability only if $\lambdacond < L$. The values of $\lambdacond / L$ in Table~\ref{table:environments_output} show that this is comfortably satisfied by the interstellar medium, the circumgalactic medium, and the photoionized outflow, and marginally by the cluster cool core with suppressed conduction; with $f=1$ and $a=1$ the cool core would instead have $\lambdacond = 1.4\,L$ and no unstable wavelength would fit inside it.  The equilibrium loop is the clearest case of stabilization: $\lambdacond = 2.0\,L$, so that thermal conduction damps the thermal mode at every wavelength that fits along the loop, in agreement with \citet[][their Section~2.3]{stricklan_etal2025}.  Since $\lambdacond \propto T^{7/4}/n_{\rm H}$, however, the higher density and lower temperature of the loop late in a TNE cycle reduce $\lambdacond$ by a factor of about $30$, to $0.067\,L$.  Thermal instability in a coronal loop is therefore not a property of the loop but of the state it has been driven into.
 
The second comparison concerns the isochoric regime.  Section~\ref{sect:timescales} shows that the thermal mode evolves isochorically for $\lambda \gg \lambdath$, which is also the limit in which the catastrophic cooling solution of \citet{waters_stricklan2025} is recovered (Section~\ref{sect:discussion_CR_CC}).  This regime is accessible only if $\lambdath \lesssim L$. Table~\ref{table:environments_output} shows that it is comfortably reached in the photoionized outflow, where $\lambdath/L \sim 10^{-5}$, and in the TNE loop, where $\lambdath = 0.17\,L$.  It is marginal in the interstellar medium, whose $\lambdath = 1.2$~pc is comparable to the size of a small cloud.  And it is out of reach in the circumgalactic medium, where $\lambdath = 2.8\,L$, in the cluster cool core, where $\lambdath = 6.4\,L$, and in the equilibrium loop, where $\lambdath = 8.2\,L$.
 
\subsection{Robustness of the numbers} \label{sect:robustness}

The entries of Table~\ref{table:environments_output} inherit the uncertainties of their inputs, but not uniformly.  For all environments but the X-ray photoionized outflow, the dimensionless parameters $\Nrhoz$, $\Npz$, and $R$ depend only on the logarithmic slope of the radiative loss function and on $a$ and $b$ (Equations~(\ref{eq:appN}) and~(\ref{eq:appR})); they are unaffected by the normalization of $Q(T)$ and by the chemical composition. They are, on the other hand, quite sensitive to the slope itself, and the loss functions in use for the same medium differ in it considerably: at $2$~MK the CHIANTI and SPEX curves give $\beta \simeq 2.1$ and $2.0$ respectively, the piecewise fit of \citet{Klimchuk_etal2008} gives $3/2$, and the expression of \citet{KlimchukLuna2019} gives $1/2$. In the intracluster and circumgalactic media the slope depends in addition on the metallicity. The characteristic scales carry the opposite sensitivity: $\tc$ and $\lambdath$ are proportional to $Q_0^{-1}$ and $\lambdaF$ to $Q_0^{-1/2}$, so that they respond directly to the normalization of the loss function, as do $\lambdacond$, $\lambdaacoustic$, and $\lambdaisoch$, which are built from $\lambdaF$ and $\lambdath$; the composition, by contrast, changes them by no more than about ten per cent. Finally, we have treated the suppression factor $f$ as a constant, which is the usual practice; if instead the conductivity is limited by the whistler instability, $f$ depends on the electron plasma beta and on the background temperature gradient \citep{Roberg-Clark_etal2018,Komarov_etal2018}, and the ratio then acquires a dependence on density and on magnetic field strength that Equation~(\ref{eq:appratio}) does not contain.

The values of $\lambdaF/\lambdath$ quoted above are those at the fiducial equilibrium of each medium, and it is worth asking how far they are representative. We have therefore evaluated the ratio over the whole range of temperatures in which each of the five loss functions is meant to apply. For the atomic interstellar medium, between $50$~K and $10^{4}$~K, it lies between $0.031$ and $0.27$, the smallest value occurring at $5.3\times10^{3}$~K; for the solar corona, between $10^{5}$~K and $10^{7}$~K, between $0.31$ and $0.57$, with the largest value at $1.3\times10^{6}$~K; and for the circumgalactic and intracluster media, between $10^{5}$~K and $10^{8}$~K, between $0.039$ and $0.23$. For the X-ray photoionized outflow the free parameter is the ionization parameter rather than the temperature, since the equilibrium temperature follows from $\xi$; over the interval $122 \lesssim \xi \lesssim 615$ in which the medium is isobarically unstable, corresponding to $1.1\times10^{5}~{\rm K} \lesssim \Tz \lesssim 1.4\times10^{6}$~K, the ratio lies between $0.063$ and $0.075$. Taken together, the five environments give $0.031 \lesssim \lambdaF/\lambdath \lesssim 0.57$ over the whole of the range $50~{\rm K} \lesssim T \lesssim 10^{8}$~K. The figures for the circumgalactic and intracluster media include the suppression factor $f = 0.1$; without it they would reach $0.72$ and $0.74$, still below unity.
 
That the ratio never approaches unity has a simple reason. By Equation~(\ref{eq:appratio}) with the Spitzer conductivity, $\lambdaF/\lambdath \propto [\kappa_0 T^{1/2}Q(T)]^{1/2}$, so it grows only as $T^{1/2}$ once the losses are dominated by bremsstrahlung and $Q \propto T^{1/2}$. Taking the bremsstrahlung term of \citet{TozziNorman2001} and a fully ionized plasma of solar composition, the equality $\lambdaF = \lambdath$ is not reached until $T \simeq 1.3\times10^{9}$~K, or about $110$~keV, a value that changes by less than $6\%$ over two decades in density. At the temperature of the hottest intracluster plasmas, $k_{\rm B}T \simeq 20$~keV, the ratio is still only $\simeq 0.4$. Plasmas hot enough for $\lambdaF$ to approach $\lambdath$ are in any case dominated by processes that violate the assumption of two-body cooling on which Equation~(\ref{eq:appratio}) rests, since Compton and synchrotron losses do not scale as $n^{2}$.

\section{Classification of instability domains} \label{sect:WP_fig2}

In his Figure~1, \citet{field1965thermal} presented a diagram representing the instability domains and mode properties as a function of $k$. As discussed in Section~\ref{sect:comparison}, we suggest that the classification provided in Table~\ref{table:R_and_lambda} offers a more direct and simplified representation of these domains. This table summarizes the properties of the entire spectrum according to the value of $R$ and the sign of $\Nrhoz$ or $\Npz$. In addition, the values $R$~=~0, 1/9, $1/\gamma$, and~1 separate the different regimes of thermal instability and determine in which wavelength ranges there are two acoustic modes plus one thermal mode or three thermal modes, together with their stable/unstable character. Table~\ref{table:R_and_lambda} differs from Figure~2 of \citet{waters2019non} in several respects that are discussed in Sections~\ref{sect:WP_fig2_finite_kF}--\ref{sect:WP_fig2_kF}. First, \citet{waters2019non} obtained expressions for the critical wavelengths at which the acoustic modes convert into thermal modes using the assumption $\lambdaF = 0$, which is equivalent to neglecting thermal conduction. Second, the range $1/9 < R < 1/3$ was included as a separate one, but it is actually part of the larger range $1/9 < R < 1$. And, third, the stabilization of the thermal and acoustic modes at small wavelengths due to thermal conduction was not dealt with. In other words, the wavelengths that correspond to the thermal and acoustic wavenumbers of Equations~(\ref{eq:kcond}) and (\ref{eq:kacoustic}) were not included in their table. In Sections~\ref{sect:WP_fig2_finite_kF}--\ref{sect:WP_fig2_kF} we consider small values of $\lambdaF$, while in Section~\ref{sect:arbitrary_lambdaF} we study the domains of instability when this assumption is removed.

\subsection{Critical wavelengths and the assumption of small $\lambdaF$} \label{sect:WP_fig2_finite_kF}

To gather the information of Table~\ref{table:R_and_lambda}, we follow \citet{waters2019non} and study the critical wavelengths at which the acoustic modes turn into thermal modes, or vice versa. This involves obtaining the wavelengths at which the dispersion relation~(\ref{eq:dr_WP}) transitions from having only one real solution to having three real solutions, or vice versa. This condition leads to

\begin{equation} \label{eq:WP2019_eq13}
    \left(\frac{\Nrho^2}{9} - \frac{1}{3}\kc^2\lambdath^2\right)^3 = \left[\frac{\Nrho}{2}\kc^2\lambdath^2\right(\frac{\Np}{\gamma\Nrho}-\frac{1}{3}\left) + \frac{\Nrho^3}{27}\right]^2,
\end{equation}

\noindent where $\kc$ is a critical wavenumber and the thermal length has been defined in Equation~(\ref{eq:lambdath}). Equation~(\ref{eq:WP2019_eq13}) is a quartic polynomial in $\kc^2$, hence the critical wavenumbers come in pairs $\pm\kc$, of which only the real, positive values are of interest. \citet{waters2019non} obtained the following formulas for the corresponding critical wavelengths:

\begin{equation} \label{eq:lambdac}
    \lambdac \approx \frac{2\pi}{\Nrhoz} \lambdath \sqrt{\frac{-B}{R}}, \hspchorone R < 0,
\end{equation}

\begin{equation} \label{eq:lambdapm}
    \lambda_\pm^2 = \frac{8\pi^2}{\Nrhoz^2} \frac{\lambdath^2}{B} \left[-1 \pm \sqrt{1-\frac{4R}{B^2}} \right]^{-1}, \hspchorone 0 < R < 1/9,
\end{equation}

\noindent where

\begin{equation} \label{eq:B}
    B = \frac{27}{4} \left(R - \frac{1}{3}\right)^2 -1.
\end{equation}

\noindent Equation~(\ref{eq:lambdac}) applies to the first two rows of Table~\ref{table:R_and_lambda}, in which the two acoustic modes are transformed into two thermal modes at $\lambda = \lambdac$. Moreover, Equation~(\ref{eq:lambdapm}) applies to the third row of Table~\ref{table:R_and_lambda}, in which the two acoustic modes turn into two thermal modes at $\lambda = \lambda_-$. This conversion is reversed at $\lambda = \lambda_+$. For $R > 1/9$ no mode conversions take place (rows 4--7).

\begin{figure*}
    \centering
    \includegraphics[width=0.47\linewidth]{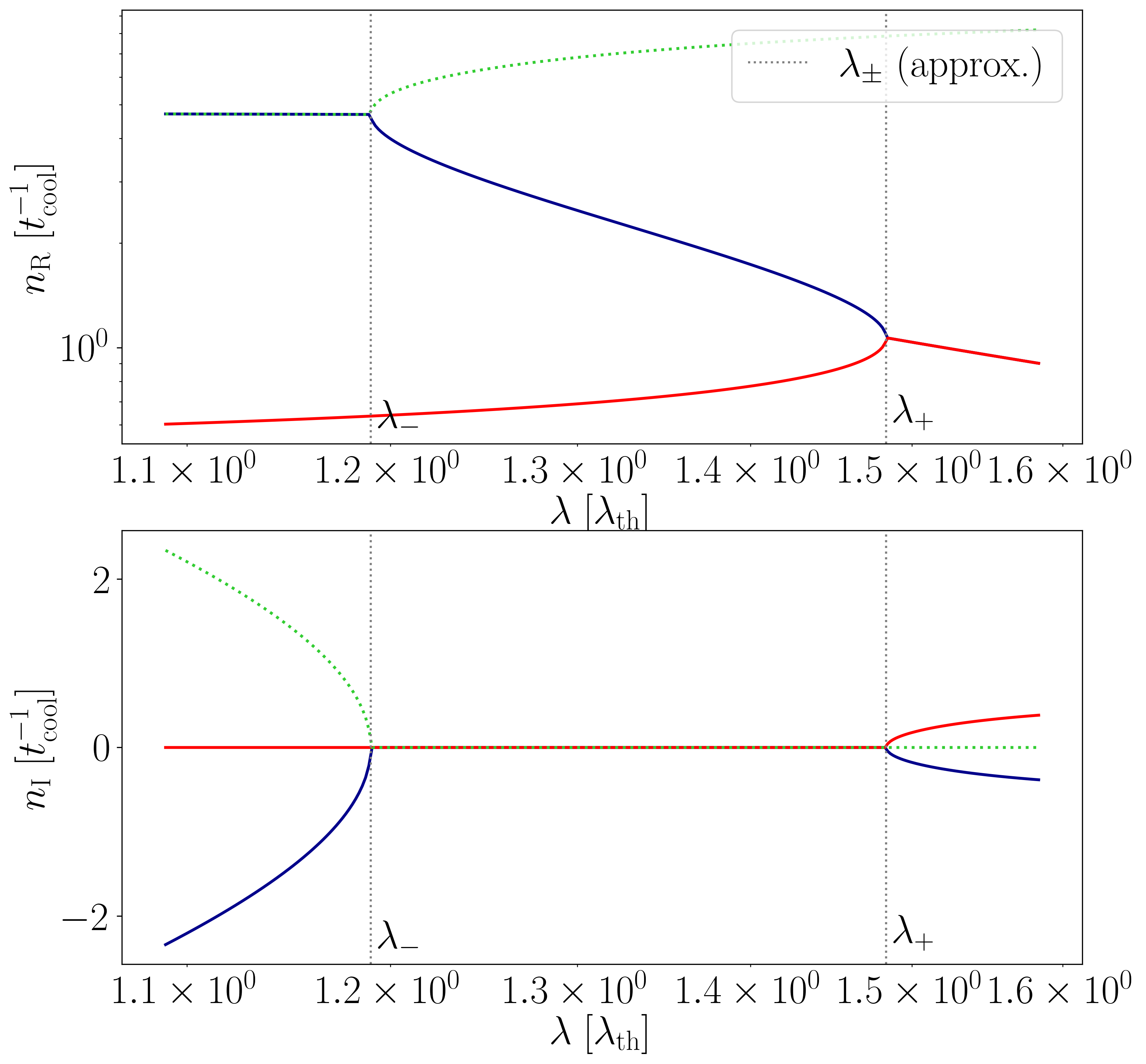}
    \includegraphics[width=0.47\linewidth]{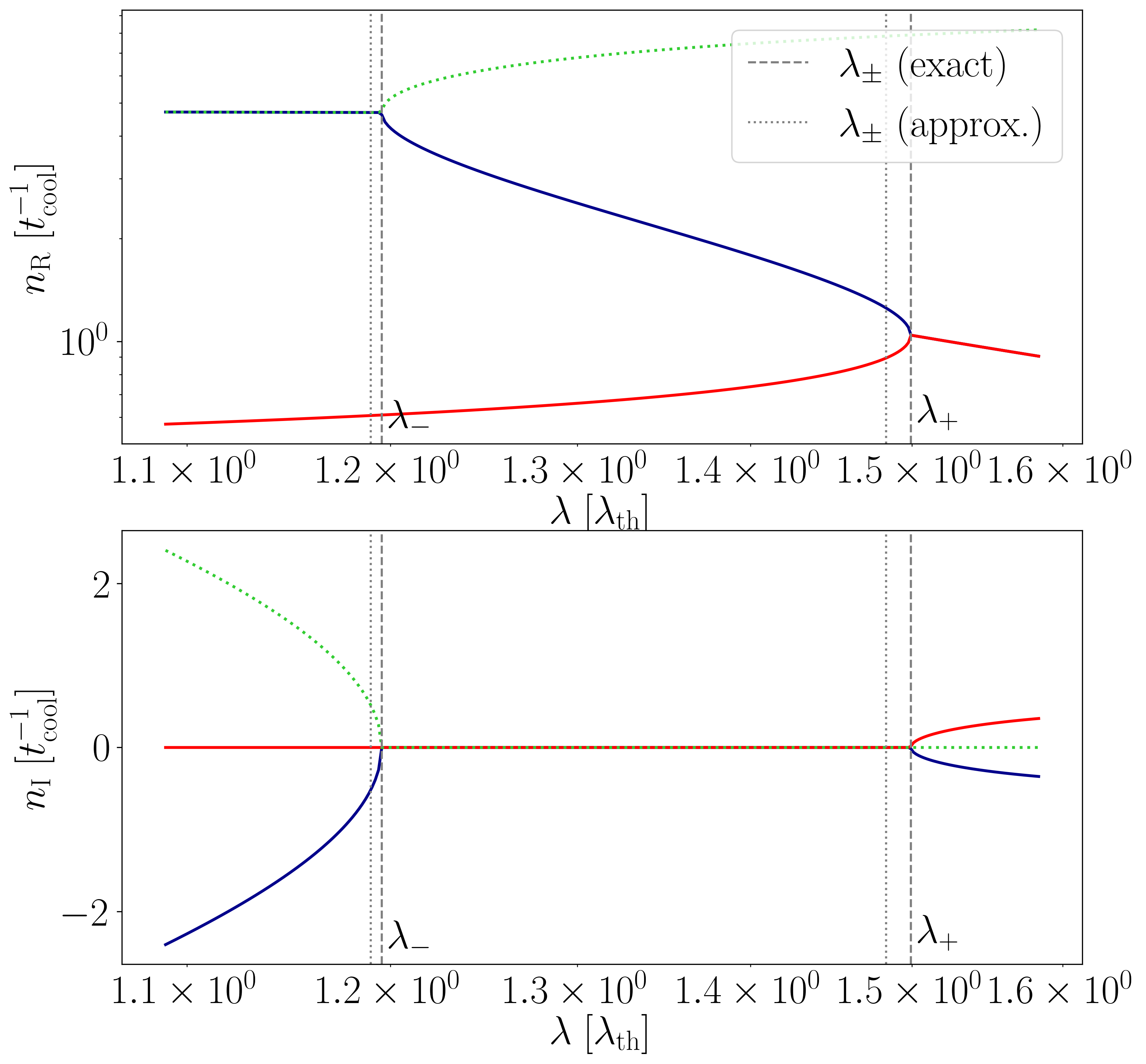}
    \caption{Left column, top and bottom panels: real and imaginary parts of the growth rate versus wavelength around the critical wavelengths $\lambda_\pm$ for $\lambdaF = 0$. The solid blue, solid red, and dotted green lines are the three solutions of the dispersion relation. The vertical gray dotted lines labeled approx. are the values given by Equation~(\ref{eq:lambdapm}), while those labeled exact come from the solutions to Equation~(\ref{eq:WP2019_eq13}). Right column: same as left column for $\lambdaF = 0.2 \lambdath$; this is approximately the value used by \citet{waters2019non} in their Figure~1. The other parameter values are $R = 0.05$ and $\Nrhoz = -10$ (hence $\Npz = -2.5/3$).}
    \label{fig:lambda_pm}
\end{figure*}

In Equation~(\ref{eq:WP2019_eq13}) we have  corrected a typographical error in Equation~(13) of \citet{waters2019non}, namely the substitution of the term $-1$ by $-1/3$ on the right-hand side. Furthermore, when writing their Equation~(13) these authors assumed $\lambdaF \ll \lambda$ and substituted $\Np/\gamma\Nrho$ by $R$; this assumption was also used to derive Equations~(\ref{eq:lambdac}) and (\ref{eq:lambdapm}), which are only valid if $\lambdaF$ is small. To illustrate the importance of $\lambdaF$ we consider a system with $R = 0.05$, i.e., which belongs to the range $0 < R < 1/9$. Figure~\ref{fig:lambda_pm} shows $\nR$ and $\nI$ in a range of wavelengths around [$\lambda_-$, $\lambda_+$]. When $\lambdaF = 0$ the analytical equations provide a perfect match to the critical wavelengths. When $\lambdaF \neq 0$, however, there is no such an excellent agreement, although the analytical formulas still give a very close estimation of the actual critical wavelengths.

\begin{figure*}
    \centering
    \includegraphics[width=0.55\linewidth]{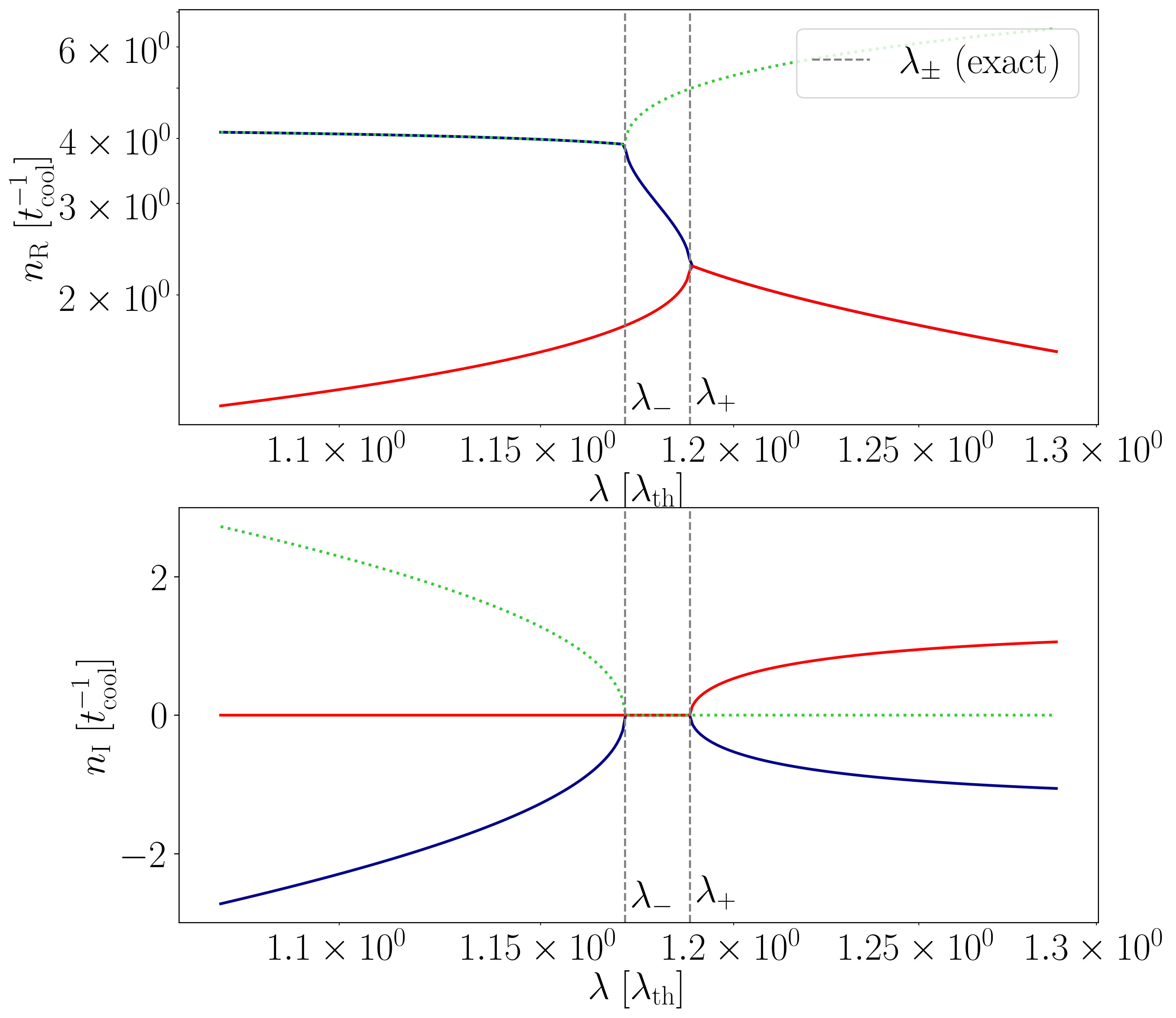}
    
    \caption{Same as Figure~\ref{fig:lambda_pm} for $\lambdaF=0.8 \lambdath$, $\Nrhoz=-10$, and $R=0.12$. This value of $R$ is slightly larger than 1/9 and so, according to Table~\ref{table:R_and_lambda}, there should be no critical wavelengths instead of two. This disparity is caused by $\lambdaF$ not being zero and will be explored in Section~\ref{sect:arbitrary_lambdaF}.}
    \label{fig:lambda_pm_fail}
\end{figure*}

The conclusion of this subsection is that the analytical formulas for the critical wavelengths $\lambdac$ and $\lambda_\pm$ provided by \citet{waters2019non} are good approximations to the exact ones as long as $\lambdaF$ is small, where ``small'' means $\lambdaF \lesssim \lambdath$, as explained in Section~\ref{sect:arbitrary_lambdaF}. And, what is more important, the classification of the modes supported by the system given in terms of $R$ is {\it essentially} correct. We use the word ``essentially'' because values of $R$ near 0, 1/9, or 1 may behave differently to what is exposed in Table~\ref{table:R_and_lambda}; Figure~\ref{fig:lambda_pm_fail} shows an example of this.

\subsection{The claimed range $1/9 < R < 1/3$} \label{sect:WP_fig2_row5}
In their Section~3.3, \citet{waters2019non} state that for $1/9 < R < 1/3$, instead of the two critical wavelengths that are present for $0 < R < 1/9$ (see Figure~\ref{fig:lambda_pm}), there is just one. We now explain why this claim is incorrect. This single critical wavelength should come from Equation~(13) of \citet{waters2019non}. An examination of the term $1-4R/B^2$, with $B$ given by Equation~(\ref{eq:B}), shows that it is negative for $R$ in the range $1/9 < R < 1$. Thus, there is no critical wavelength for $1/9 < R < 1$ and this implies that rows 4 and 5 of Figure~2 of \citet{waters2019non} must be merged.

\subsection{Mode stabilization at small wavelengths due to thermal conduction} \label{sect:WP_fig2_kF}

At wavelengths shorter than those included in Figure~2 of \citet{waters2019non}, thermal conduction stabilizes unstable thermal modes and overstable acoustic modes. Just to give two examples, this is what happens with the unstable thermal mode (bold text) in the first row and the overstable acoustic modes (italic text) in the second row of Table~\ref{table:R_and_lambda}. Our aim in this subsection is to incorporate the wavelengths $\lambdacond$ and $\lambdaacoustic$ defined in Equations~(\ref{eq:kcond}) and~(\ref{eq:kacoustic}).

We note that $\lambdacond$ exists provided that the system is isobarically unstable for $\lambda \rightarrow \infty$ ($\Npz < 0$). This happens in rows 1 and 3--6 of Table~\ref{table:R_and_lambda}. In addition, $\lambdaacoustic$ exists if $R - 1$ and $\Nrhoz$ have the same sign, i.e., in rows 2--5 and 7 of the same table. Hence, both wavenumbers exist in rows 3--5 ($0 < R < 1$). It can be easily proved that the wavelengths $\lambdacond$ and $\lambdaacoustic$ satisfy  $\lambdaacoustic < \lambdacond$ if $0 < R < 1/\gamma$ and $\Npz, \Nrhoz < 0$. We see that $\lambdaacoustic < \lambdacond$ in rows 3 and 4, whereas in row 5 the opposite applies, namely $\lambdacond < \lambdaacoustic$. Of course, for $R = 1/\gamma$ we have $\lambdaacoustic = \lambdacond$ and all three modes transition between stable and unstable at the same wavelength.

The summary of this subsection is that, to extend Figure~2 of \citet{waters2019non} all the way to $\lambda = 0$, it is necessary to include the possible mode stabilization that takes place at $\lambdaacoustic$ and $\lambdacond$. Table~\ref{table:R_and_lambda} corrects and expands Figure~2 of \citet{waters2019non} according to the issues discussed in this section. Section~\ref{sect:nRnI_vs_lambda} contains plots of $\nR$ and $\nI$ versus $\lambda$ for the values of $R$ in each row of Table~\ref{table:R_and_lambda}.

\subsection{Instability domains for arbitrary  $\lambdaF$} \label{sect:arbitrary_lambdaF}

\begin{figure*}
    \begin{interactive}{animation}{Nzeros_vs_lambdaF.mov}
    \centering
    \includegraphics[width=0.47\linewidth,trim={0 22ex 0 0},clip]{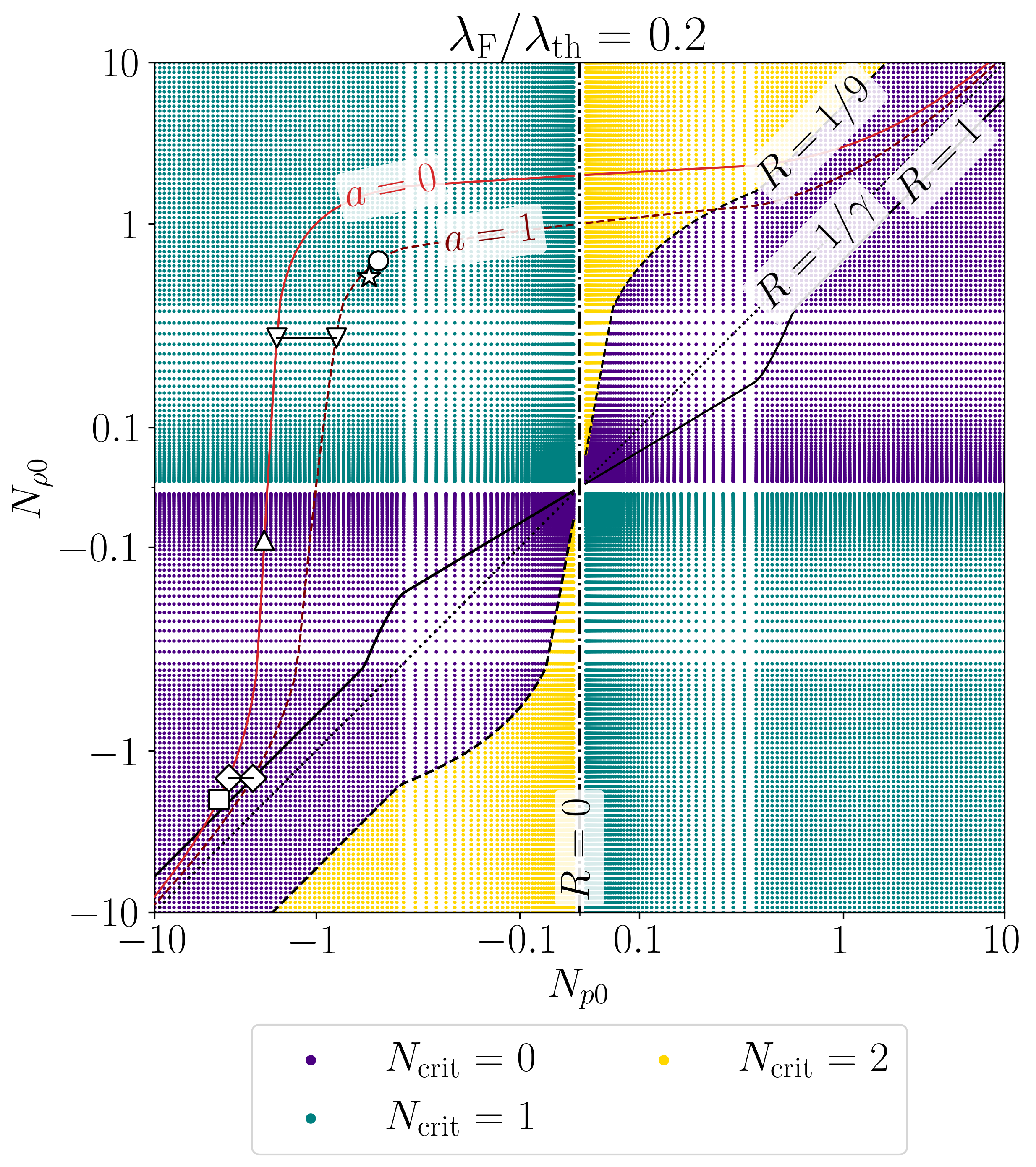}\includegraphics[width=0.47\linewidth,trim={0 22ex 0 0},clip]{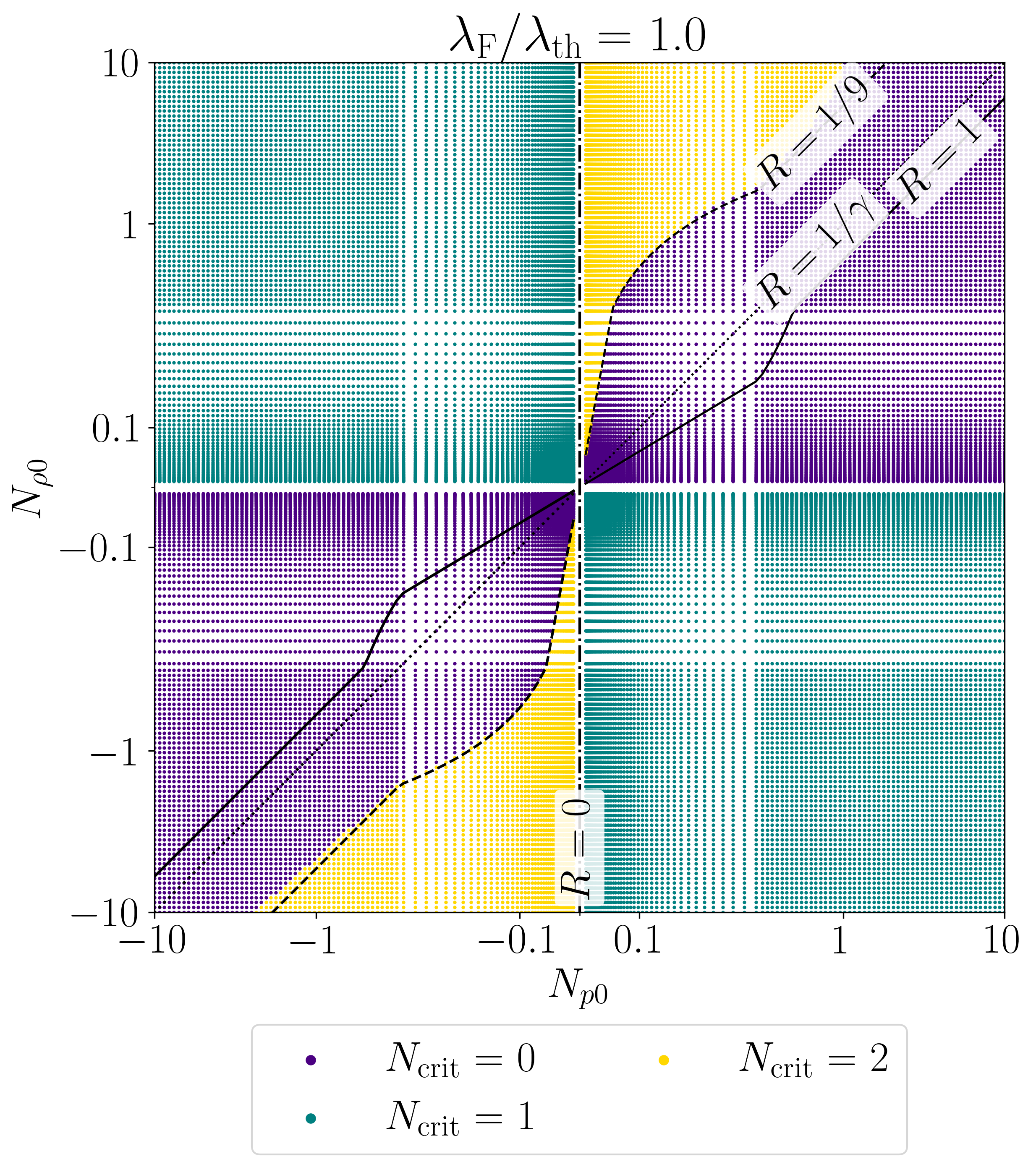}
    \includegraphics[width=0.47\linewidth,trim={0 22ex 0 0},clip]{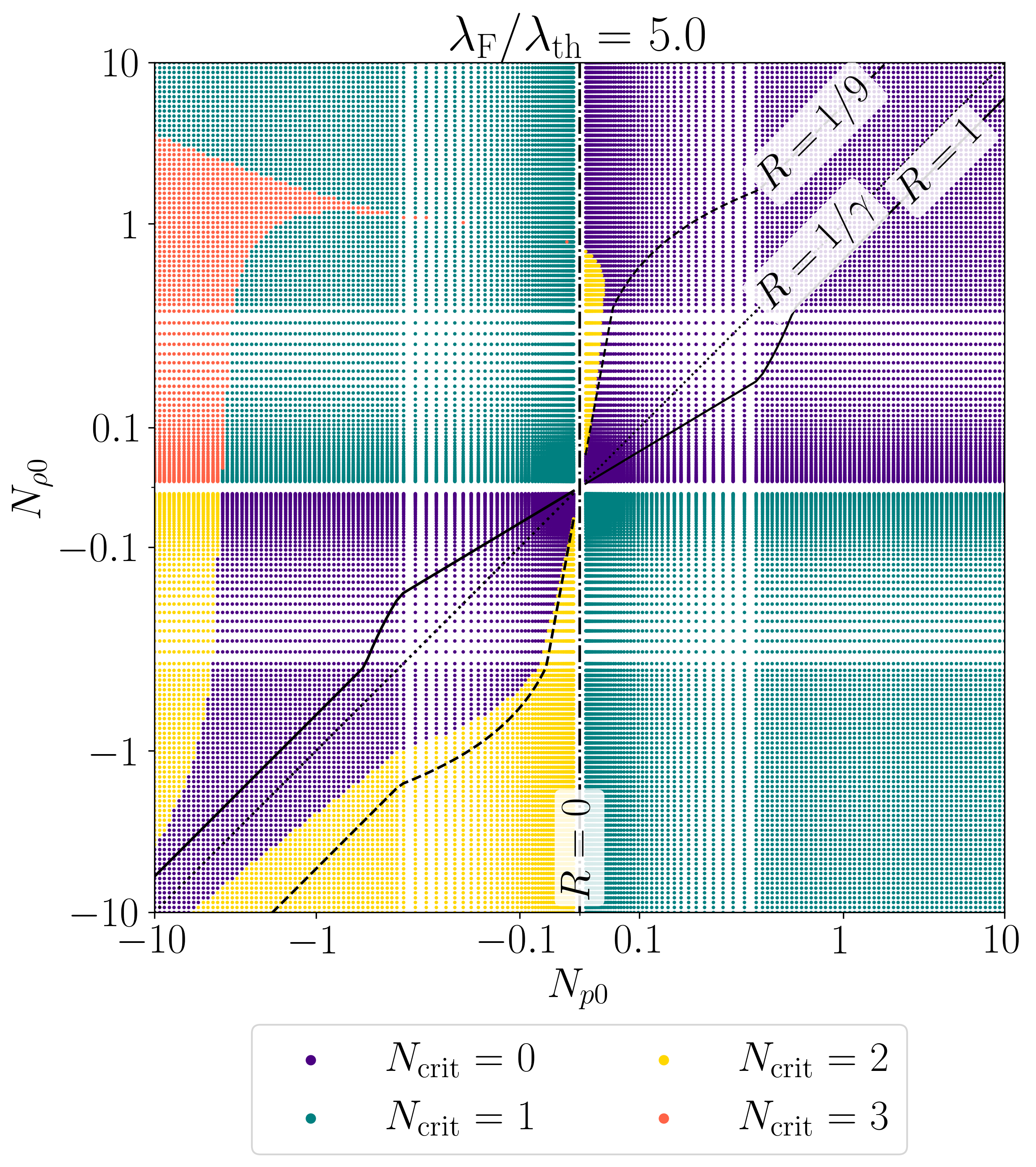}\includegraphics[width=0.47\linewidth,trim={0 22ex 0 0},clip]{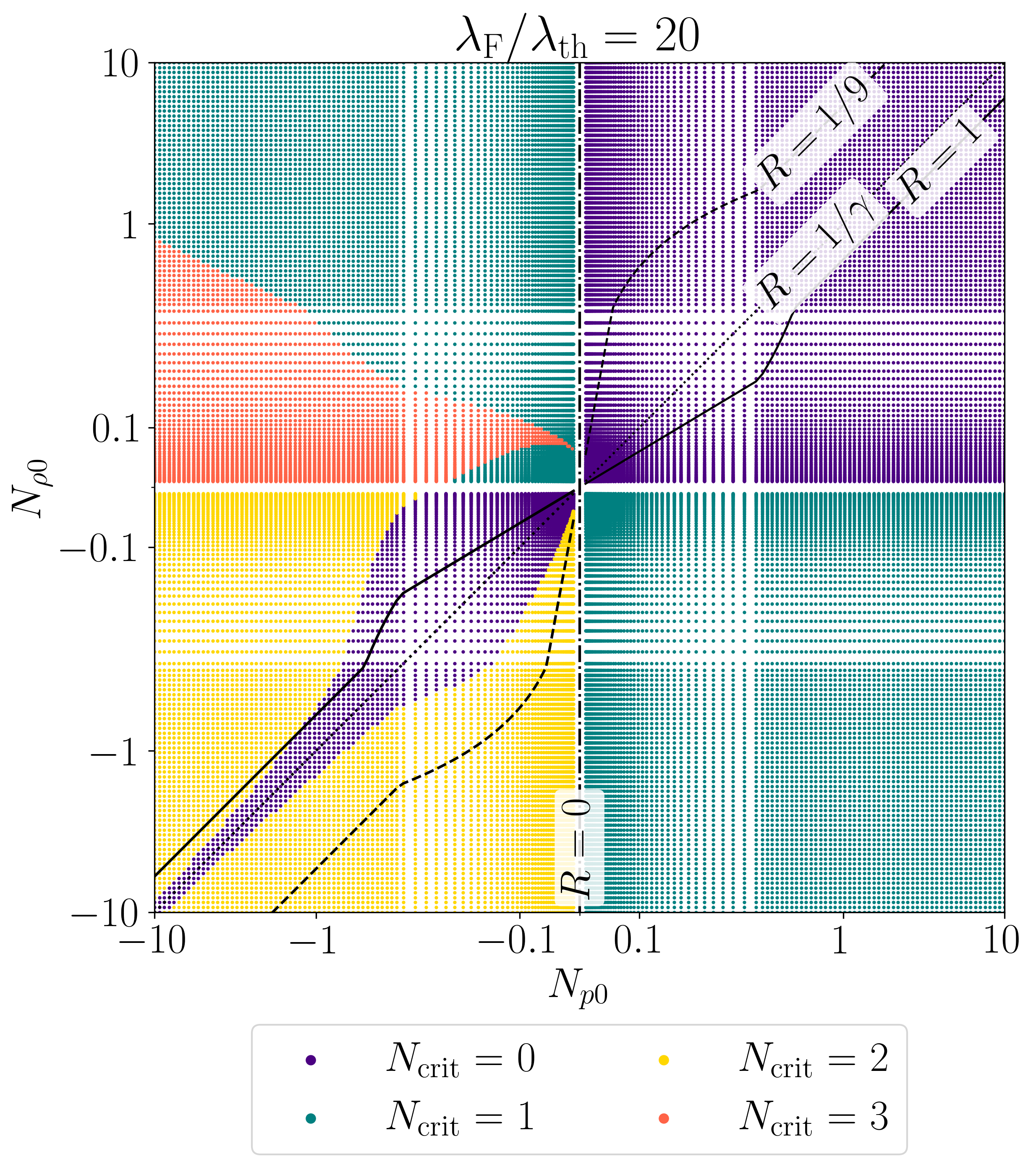}
    {\centering\includegraphics[width=0.47\linewidth,trim={0 0 0 150ex},clip]{nzeros_20.0.png}}
    \end{interactive}
    \caption{Number of critical wavenumbers as a function of $\Npz$ and $\Nrhoz$ for several values of the ratio $\lambdaF / \lambdath$. This figure shows the cases $\lambdaF / \lambdath = 0.2$, 1, 5, 20, with other values of this ratio shown in the accompanying movie. The four panels correspond to snapshots from the movie at the respective values of $\lambdaF / \lambdath$. The movie has a playback duration of 8.5 s and shows the evolution of the distribution as $\lambdaF / \lambdath$ increases from 0.2 to 500, with intermediate values connecting the four snapshots. Between 00~s and 01~s the ratio $\lambdaF / \lambdath$ is smaller than or equal to 1 and the distribution of critical wavenumbers resembles that of $\lambdaF = 0$. At 1.5~s ($\lambdaF / \lambdath = 2$) a region with 3 critical wavenumbers appears for the first time and its size increases until roughly 05~s ($\lambdaF / \lambdath = 10$). After 07~s this region almost disappears and the distribution of critical wavenumbers tends to a simple one with no critical wavenumbers in the first quadrant, one in the second and fourth quadrants, and two in the third quadrant. The varying density of the points in certain regions is a result of the discrete sampling mesh applied over a symmetric logarithmic scale and does not carry additional physical significance. The line $R = 0$ coincides with the vertical axis $\Npz = 0$. The $R = 1/9, 1/\gamma, 1$ lines and the lines $\Nrhoz - \Npz = 2-a$ (Equation~(\ref{eq:appsep})), on which lie the plasmas whose heating rate per unit volume scales as $\rho^a$, with $a = 0$ and 1, do not appear straight because of the symmetric logarithmic scale used in both axes. The symbols displayed for $\lambdaF/\lambdath = 0.2$, and also for $\lambdaF/\lambdath = 0.5$ in the accompanying movie, correspond to the astrophysical environments analyzed in Section~\ref{sect:environments}: atomic interstellar medium (circle), active-region coronal loop in equilibrium and late in a TNE cycle (square and triangle), circumgalactic medium (diamond), cool-core intracluster medium (inverted triangle), and X-ray photoionized outflow (star). For the circumgalactic and intracluster media two symbols are shown, joined by a horizontal segment, corresponding to the two heating prescriptions $a = 0$ and $a = 1$ in use for those media. (An animation of this figure is available.)}
    \label{fig:Nzeros}
\end{figure*}

In Section~\ref{sect:WP_fig2_finite_kF} we have concluded that Table~\ref{table:R_and_lambda} is correct as long as $\lambdaF$ is small. But what does small $\lambdaF$ mean? To give a quantitative answer to this question, we solve Equation~(\ref{eq:WP2019_eq13}) on a grid of values of $\Npz$ and $\Nrhoz$ for a set of values of $\lambdaF/\lambdath$. The results are presented in Figure~\ref{fig:Nzeros} and in its accompanying movie. This figure only considers the number of critical wavenumbers ($\Ncrit$), hence it reveals whether the spectrum of modes is analogous to that of rows 4--7 of Table~\ref{table:R_and_lambda} ($\Ncrit = 0$), rows~1--2 ($\Ncrit = 1$), or row 3 ($\Ncrit = 2$). We note that the horizontal and vertical axes respectively correspond to the isobaric and isochoric instability criteria for $\lambda \rightarrow \infty$; see Equation~(\ref{eq:inst_criteria2}). Therefore, the system is isobarically stable (unstable) in the first and fourth quadrants (second and third quadrants), where the quadrants of each panel of Figure~\ref{fig:Nzeros} are numbered counterclockwise starting at the upper right. Moreover, the system is isochorically stable (unstable) in the first and second quadrants (third and fourth quadrants). The wavelength dependence of the spectrum is not included in Figure~\ref{fig:Nzeros}; instead, this figure maps the parameter space of the stability criteria. Consequently, a system is represented in Figure~\ref{fig:Nzeros} by a single point rather than by a curve, and a direct comparison via over-plotting with the curves of \citet{field1965thermal}, which are parameterized by the wavenumber, is not applicable.  This is precisely the advantage noted in Section~\ref{sect:comparison}: the wavelength dependence has been absorbed into Table~\ref{table:R_and_lambda}, so that the position of a plasma in this plane can be computed once and for all from its equilibrium parameters.

We start inspecting the top left panel ($\lambdaF/\lambdath = 0.2$). The second and fourth quadrants of this figure correspond to rows~1 and 2 of Table~\ref{table:R_and_lambda}, respectively, for which there is only one critical wavenumber (whose wavelength is $\lambdac$). In the third quadrant, the yellow dots are found in the region between the lines $R = 0$ and $R = 1/9$: this is the third row of Table~\ref{table:R_and_lambda}, for which there are two critical wavenumbers (whose wavelengths are $\lambda_-$ and $\lambda_+$). In addition, the purple dots in the third quadrant show that for $R > 1/9$ there are no critical wavenumbers, as shown in rows 4--6 of Table~\ref{table:R_and_lambda}. It is worth recalling that $\lambdacond$ and $\lambdaacoustic$ give the wavelengths at which the thermal and the two acoustic modes transition from stable to unstable, respectively; hence, they are not critical wavelengths in the sense given in this paper. Finally, in the first quadrant both $\Npz$ and $\Nrhoz$ are positive. Table~\ref{table:R_and_lambda} covers this case only for $R > 1$, its seventh row, in which the acoustic modes are overstable for $\lambda > \lambdaacoustic$ and there are no critical wavelengths; for $R < 1$ every mode is damped at every wavelength and the case is omitted, as explained in Section~\ref{sect:comparison}. The points with $\Ncrit = 2$ that lie above the $R = 1/9$ line belong to this omitted class: their two damped acoustic modes are converted into two damped thermal modes at $\lambda = \lambda_-$ and back at $\lambda = \lambda_+$, with Equation~(\ref{eq:lambdapm}) still giving an excellent approximation to both wavelengths.

It is worth asking which parts of this plane are actually occupied by astrophysical plasmas.  If the heating and the radiative losses are power laws of the density and the temperature (Equations~(\ref{eq:appQ}) and (\ref{eq:appH})), then Equation~(\ref{eq:appsep}) shows that a system lies on the straight line $\Nrhoz = \Npz + (2-a)$, of unit slope, whose intercept is fixed by the density scaling of the heating alone.  The two values of $a$ that cover most of the literature, $a = 0$ and $a = 1$, therefore select two such lines, and the position of a system along its line is set by $\beta + b$, where $\beta$ and $b$ give the temperature scaling of the radiative cooling and the heating.  The $\lambdaF/\lambdath = 0.2$ panel of Figure~\ref{fig:Nzeros} shows the $a = 0$ and $a = 1$ lines and the positions of the five astrophysical environments in the $(\Npz, \Nrhoz)$ plane. Three consequences are apparent in this panel.  First, a plasma with $a \le 2$ can never lie in the fourth quadrant, since $\Nrhoz \ge \Npz$ excludes the combination of isochoric instability with isobaric stability; the second row of Table~\ref{table:R_and_lambda}, which occupies that quadrant, is thus not realized by such a plasma.  Second, the same restriction excludes the region of the third quadrant lying between the lines $R = 0$ and $R = 1/\gamma$, and with it the third and fourth rows of Table~\ref{table:R_and_lambda}.  Since the third row is the only one whose spectrum contains the pair of critical wavelengths $\lambda_\pm$, the yellow region of the third quadrant of Figure~\ref{fig:Nzeros} is, in this sense, inaccessible. Third, the same restriction empties the part of the first quadrant lying below the diagonal, which is where the seventh row of Table~\ref{table:R_and_lambda} is found: when $\Npz$ and $\Nrhoz$ are both positive and $\Nrhoz \ge \Npz$, one has $R \le 1/\gamma < 1$ and hence $\NSz > 0$ by Equation~(\ref{eq:NSz}), so that the plasma is stable by all three criteria. A plasma with $a \le 2$ that lies in the first quadrant is therefore always a member of the class described above, which is omitted from Table~\ref{table:R_and_lambda} because every one of its modes is damped at every wavelength; the seventh row, like the second, third and fourth, requires $a > 2$. The five environments examined in Section~\ref{sect:environments} bear this out: all of them fall in the second or the third quadrant, and none in the fourth. Those whose heating is a power law lie exactly on one or other of the two diagonals, the circumgalactic and intracluster media appearing on both because both heating prescriptions are in use for them; the horizontal segments that join each such pair are a direct illustration of Equation~(\ref{eq:appsep}), since a change in the heating exponent displaces a system along $\Npz$ while leaving $\Nrhoz$ untouched. Only the X-ray photoionized outflow departs slightly from the $a = 1$ line, because its heating is not a power law of $\rho$ and $T$ and its effective exponent, Equation~(\ref{eq:aeff}), is $\aeff = 1.06$.

As we increase $\lambdaF / \lambdath$, the distribution of the number of critical wavenumbers of Figure~\ref{fig:Nzeros} remains essentially unchanged up to $\lambdaF / \lambdath = 1$ (see top right panel of Figure~\ref{fig:Nzeros}). For this reason, we identify ``small $\lambdaF$'' with $\lambdaF / \lambdath \lesssim 1$. For $\lambdaF / \lambdath = 2$ the deviation from the plot for $\lambdaF / \lambdath = 0.2$ is notable in the first quadrant and, moreover, many red dots appear in the second quadrant, meaning that, for some values of $\Npz$ and $\Nrhoz$, there are mode conversions between acoustic and thermal modes at three critical wavelengths. It is not the purpose of this work to analyze in detail the stability of modes in these cases; it is enough to state that Table~\ref{table:R_and_lambda} is no longer valid for $\lambdaF / \lambdath \gtrsim 1$ because the various regions in Figure~\ref{fig:Nzeros} are not limited by $R = \mathrm{const}$ lines. To understand the presence of these three critical wavenumbers, we recall that Equation~(\ref{eq:WP2019_eq13}) is a quartic polynomial in $\kc^2$, hence it admits up to four pairs of real solutions.

Continuing to even larger values of the Field length (see bottom panels of Figure~\ref{fig:Nzeros} and the accompanying movie), we notice that the area with three critical wavenumbers in the second quadrant becomes the largest for $\lambdaF / \lambdath \simeq 7-20$ and starts to disappear for $\lambdaF / \lambdath \simeq 100$, after which the first, second, third, and fourth quadrants respectively contain mostly systems with 0, 1, 2, and 0 critical wavenumbers.

We close this subsection by recalling a result of Section~\ref{sect:environments}.  There we evaluate $\lambdaF/\lambdath$ for five environments in which thermal instability is believed to operate and obtain values between $0.059$ and $0.56$, all of them well inside the range of validity of Table~\ref{table:R_and_lambda} established here. Moreover, Equation~(\ref{eq:appratio}) shows that this ratio is independent of the density of the plasma, apart from the weak logarithmic dependence noted in Section~\ref{app:ratio}, and varies slowly with its temperature, so that these values are not peculiar to the fiducial models adopted in that section.  The two upper panels of Figure~\ref{fig:Nzeros}, computed for $\lambdaF/\lambdath = 0.2$ and 1, therefore bracket every system we have examined.  The complications described in the present subsection, which arise only above $\lambdaF/\lambdath \simeq 1$ and become conspicuous at 2, are consequently of formal rather than practical interest: their role is to delimit the range of validity of Table~\ref{table:R_and_lambda} rather than to describe a regime that is commonly realized.

\section{Growth rate and dispersion diagrams} \label{sect:nRnI_vs_lambda}

The purpose of this section is to present plots of $\nR$ and $\nI$ versus $\lambda$ for values of $\Npz$ and $\Nrhoz$ representative of each row of Table~\ref{table:R_and_lambda}. In all plots except Figure~\ref{fig:row3} the thermal mode is represented by a red line for all wavelengths. A symmetric logarithmic scale is used in the vertical axis and for this reason the curves show a kink when $\nR \tc$ or $\nI \tc$ enter the range $[-10,10]$. No assumptions are made on the heat-loss function or the thermal conductivity. Both $\nR$ and $\nI$ are non-dimensionalized against $1/\tc$ and all wavelengths are non-dimensionalized against $\lambdath = \cs \tc$. The asymptotic formulas for $\lambda \rightarrow 0$ (Equations~(\ref{eq:lambda0_condens_n}) and~(\ref{eq:lambda0_acoustic_n})) and $\lambda \rightarrow \infty$ (Equations~(\ref{eq:k0_condens_n}), (\ref{eq:k0_condens_n_Rlt0}), and~(\ref{eq:k0_acoustic_n})) included in the plots are presented in Sections~\ref{sect:short_wavelength} and \ref{sect:long_wavelength}. In our plots we have only used the dominant terms of the thermal mode formulas, although considering all terms yields an agreement with the exact values over a larger wavelength range. Finally, we have taken $\lambdaF / \lambdath = 0.2$ and $\gamma = 5/3$ in all calculations.

Other significant wavelengths are $\lambdacond$ and $\lambdaacoustic$, which mark the change of the thermal and acoustic modes from damped to exponentially growing, and $\lambdac$ (case $R < 0$) and $\lambda_\pm$ (case $0 < R < 1/9$), which signal the transition between a pair of acoustic modes and two thermal modes; they are obtained by solving Equation~(\ref{eq:WP2019_eq13}).

\begin{figure*}
    \centering
    \begin{minipage}[t]{0.48\textwidth}
        \centering
    \includegraphics[width=\linewidth]{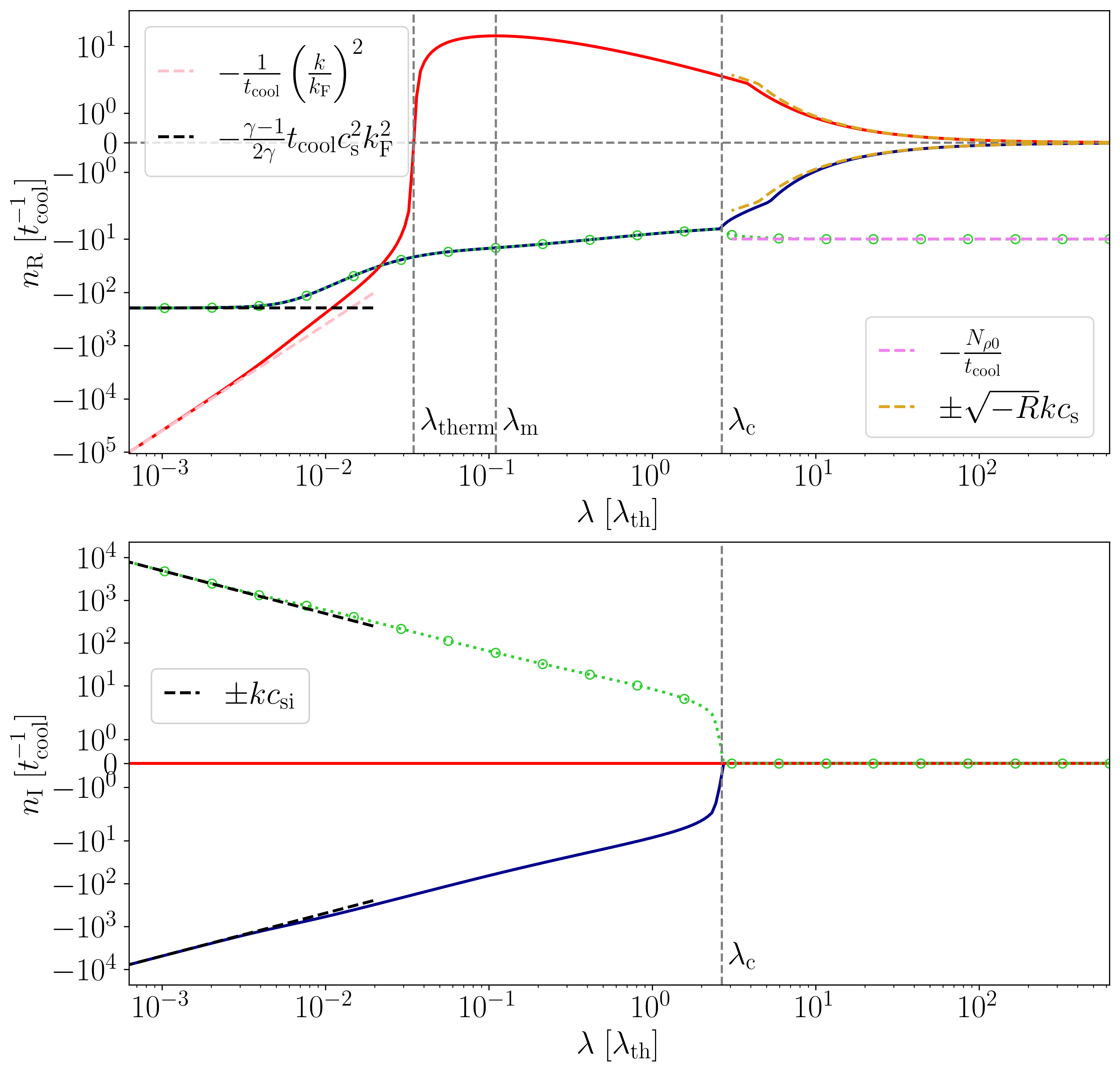}
    
    \caption{Case $R < 0$, $\Nrhoz > 0$, $\Npz < 0$. Top and bottom panels: real and imaginary parts of the growth rate versus wavelength. The solid blue, solid red, and dotted green lines are the three solutions of the dispersion relation. The legends refer to the asymptotic curves in the limits $\lambda \rightarrow 0$ (dominant term of Equation~(\ref{eq:lambda0_condens_n}) and Equation~(\ref{eq:lambda0_acoustic_n})) and $\lambda \rightarrow \infty$ (dominant terms of Equations~(\ref{eq:k0_condens_n}) and~(\ref{eq:k0_condens_n_Rlt0})). The vertical gray dotted lines mark the position of the relevant wavelengths (Equations~(\ref{eq:kcond}), (\ref{eq:kacoustic}), (\ref{eq:WP2019_eq13}), and (\ref{eq:km})), the horizontal grey dotted line in the top panel corresponds to $\nR = 0$. The parameter values are $\Nrhoz = 10$, $R = -2$, and $\lambdaF = 0.2 \lambdath$.}
    \label{fig:row1}
    \end{minipage}\hfill
    \begin{minipage}[t]{0.48\textwidth}
        \centering
        \includegraphics[width=\linewidth]{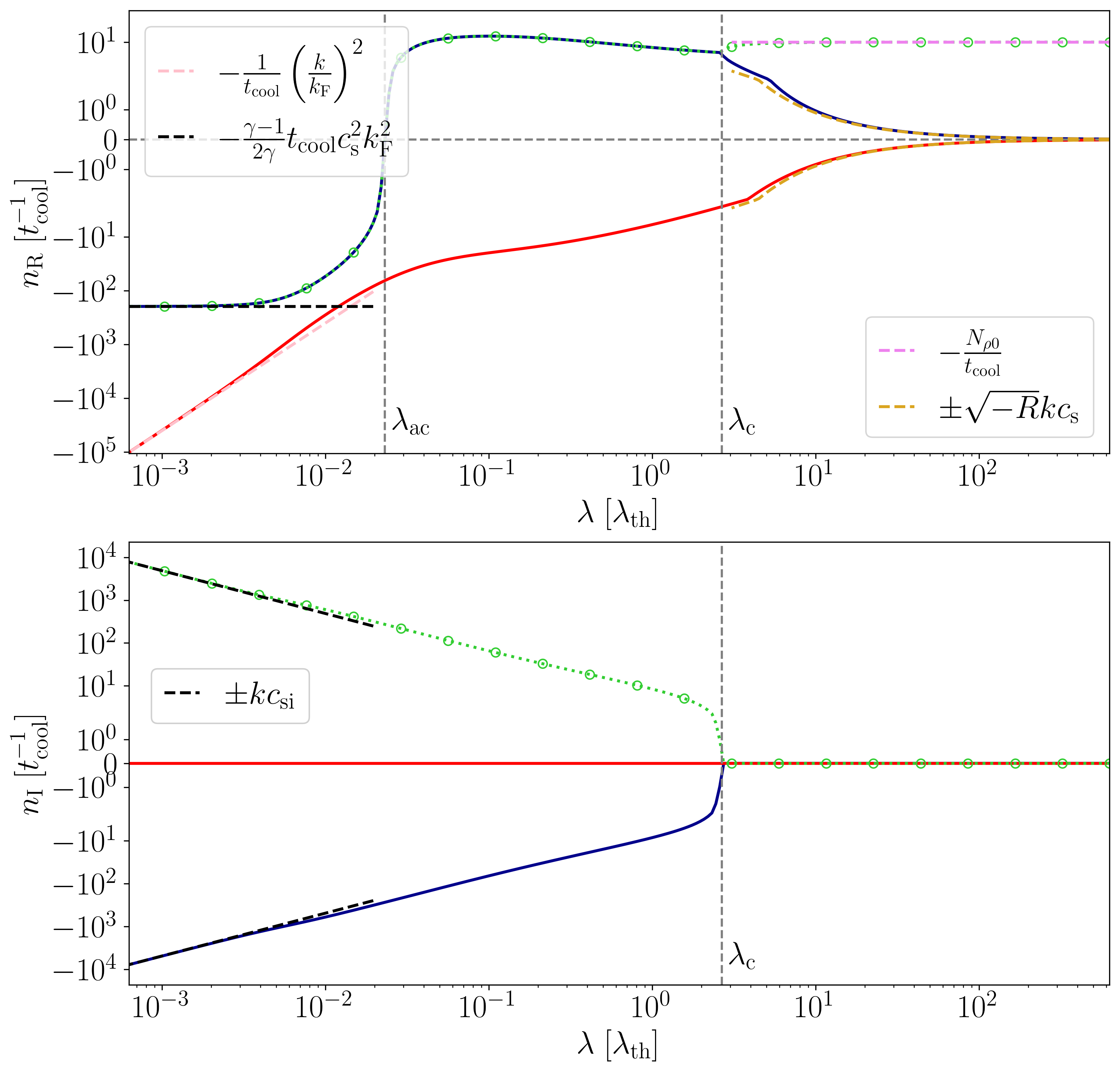}
    
    \caption{As in Figure~\ref{fig:row1} but for case $R < 0$, $\Nrhoz < 0$, $\Npz > 0$. The legends refer to the asymptotic curves in the limits $\lambda \rightarrow 0$ (dominant term of Equation~(\ref{eq:lambda0_condens_n}) and Equation~(\ref{eq:lambda0_acoustic_n})) and $\lambda \rightarrow \infty$ (dominant terms of Equations~(\ref{eq:k0_condens_n}) and~(\ref{eq:k0_condens_n_Rlt0})). The parameter values are $\Nrhoz = -10$, $R = -2$, and $\lambdaF = 0.2 \lambdath$.}
    \label{fig:row2}
    \end{minipage}
\end{figure*}

\begin{figure*}
    \centering
    \begin{minipage}[t]{0.48\textwidth}
        \centering
    \includegraphics[width=\linewidth]{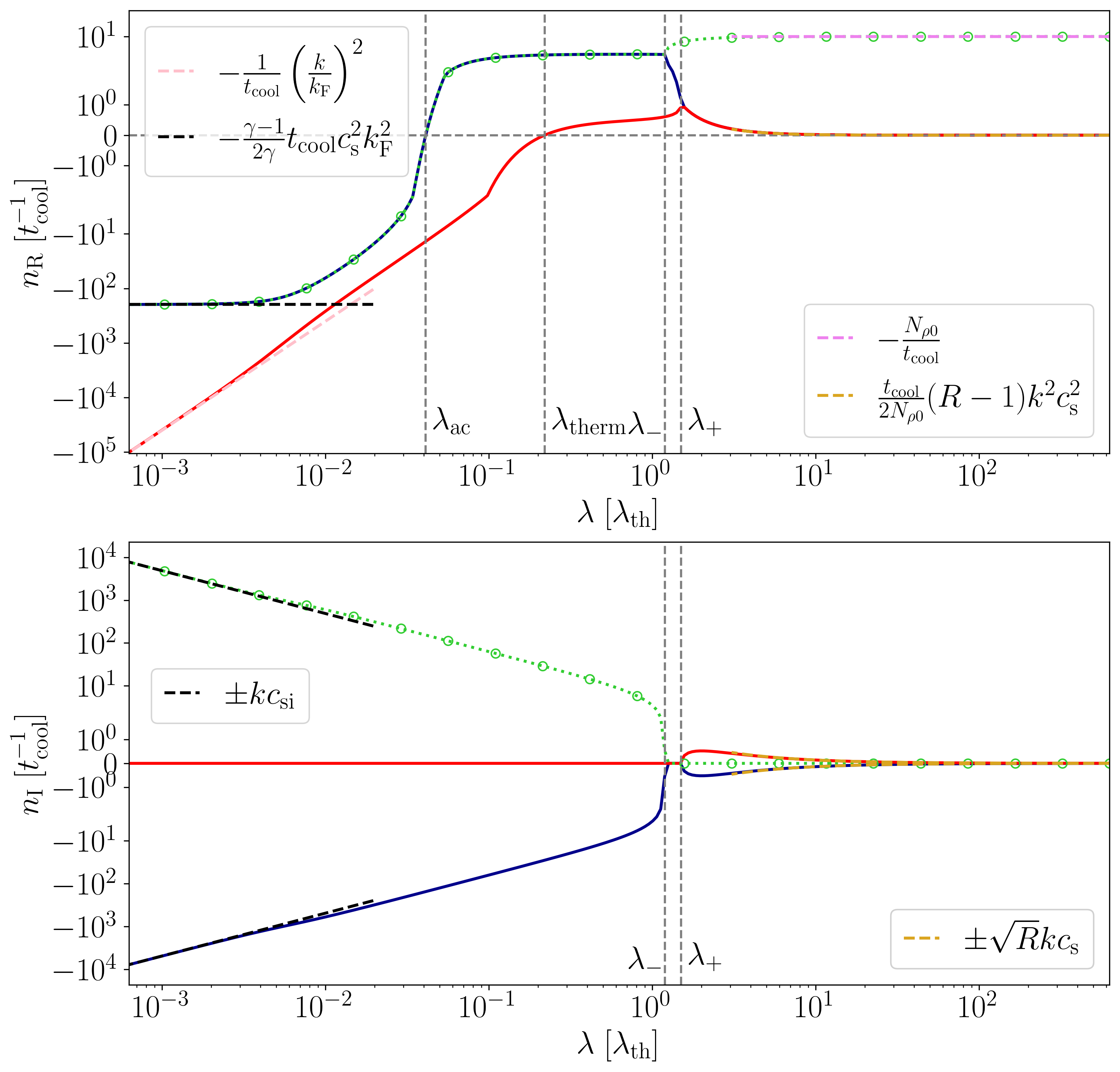}
    
    \caption{As in Figure~\ref{fig:row1} but for case $0 < R < 1/9$, $\Nrhoz, \Npz < 0$. The legends refer to the asymptotic curves in the limits $\lambda \rightarrow 0$ (dominant term of Equation~(\ref{eq:lambda0_condens_n}) and Equation~(\ref{eq:lambda0_acoustic_n})) and $\lambda \rightarrow \infty$ (dominant term of Equation~(\ref{eq:k0_condens_n}) and Equation~(\ref{eq:k0_acoustic_n})). The parameter values are $\Nrhoz = -10$, $R = 0.05$, and $\lambdaF = 0.2 \lambdath$.}
    \label{fig:row3}
    \end{minipage}\hfill
    \begin{minipage}[t]{0.48\textwidth}
        \centering
        \includegraphics[width=\linewidth]{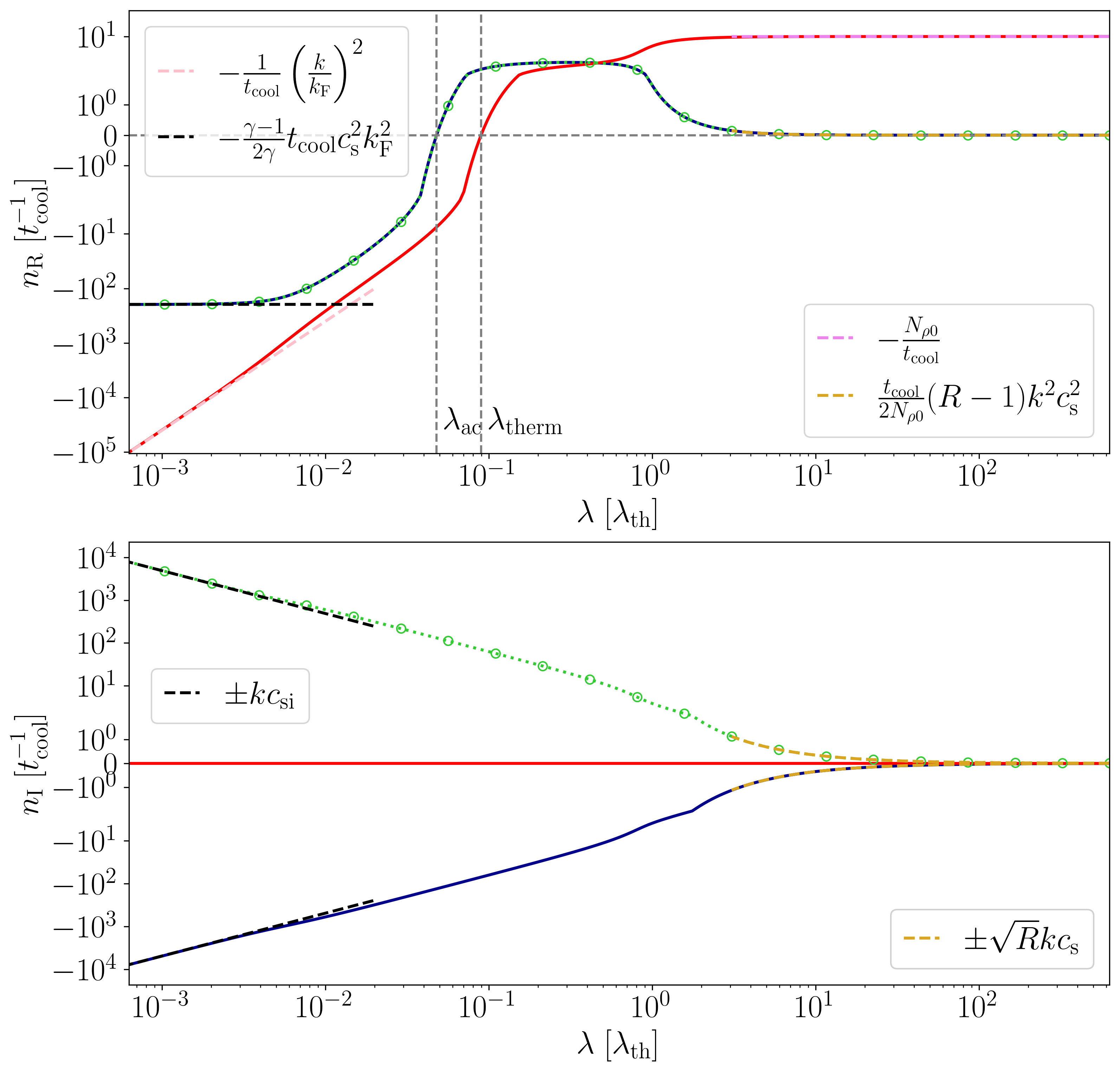}
    
    \caption{As in Figure~\ref{fig:row1} but for case $1/9 < R < 1/\gamma$, $\Nrhoz, \Npz < 0$. The legends refer to the asymptotic curves in the limits $\lambda \rightarrow 0$ (dominant term of Equation~(\ref{eq:lambda0_condens_n}) and Equation~(\ref{eq:lambda0_acoustic_n})) and $\lambda \rightarrow \infty$ (dominant term of Equation~(\ref{eq:k0_condens_n}) and Equation~(\ref{eq:k0_acoustic_n})). The parameter values are $\Nrhoz = -10$, $R = 0.3$, and $\lambdaF = 0.2 \lambdath$.}
    \label{fig:row4}
    \end{minipage}
\end{figure*}

We begin our discussion with a plasma configuration that is isobarically unstable and isochorically and isentropically\footnote{According to Equations~(\ref{eq:inst_criteria2}) and (\ref{eq:NSz}), isentropic instability requires $R-1$ and $\Nrhoz$ to have the same sign. This condition is met in rows 2--5 and 7 of Table~\ref{table:R_and_lambda}.} stable: see row~1 of Table~\ref{table:R_and_lambda} and Figure~\ref{fig:row1}; this figure is representative of the atomic ISM, the cool-core ICM and the X-ray photoionized outflow. Starting at small wavelengths, we find a characteristic that is common to all values of $R$: there is one thermal and two acoustic waves, all stable, whose growth rates are well reproduced by  Equations~(\ref{eq:lambda0_condens_n}) and (\ref{eq:lambda0_acoustic_n}). Since the system is isentropically stable, acoustic modes are stable for all wavelengths and this is the reason why $\lambdaacoustic$ (Equation~(\ref{eq:kacoustic})) does not appear in this row of Table~\ref{table:R_and_lambda}. On the other hand, the thermal mode becomes unstable at $\lambda = \lambdacond$ (see Equation~(\ref{eq:kcond})), after which $\nR$ reaches a maximum, at a wavelength that is denoted by~$\lambdam$. \citet{field1965thermal} gave an approximate expression for the corresponding wavenumber, $\km$, but here we prefer to compute it exactly. To do so, we take the derivative with respect to $k$ of Equation~(\ref{eq:dr_WP}) with $n$ substituted by $\nR$ and, upon imposing $d\nR/dk = 0$, we obtain

\begin{equation} \label{eq:km}
    \km^2 = -\frac{\kF^2}{2\cs^2} \left(\frac{\gamma \nR^2}{\kF^2} + \gamma \tc \cs^2 \nR + \Npz \cs^2\right), \hspchorhalf \lambdam = \frac{2\pi}{\km}.
\end{equation}

\noindent Now, this expression is inserted into Equation~(\ref{eq:dr_WP}) with $n$ substituted by $\nR$, which leads to a fourth-degree polynomial equation for $\nRmax$, the maximum growth rate. The wavenumber at which this maximum takes place is then given by Equation~(\ref{eq:km}).

At $\lambda = \lambdac$, the two acoustic modes transition to two stable thermal modes ($\nR < 0$, $\nI = 0$). Finally, in the large wavelength range, Equations~(\ref{eq:k0_condens_n}) and (\ref{eq:k0_condens_n_Rlt0}) show an excellent agreement with the growth rates of the three thermal modes.

\begin{figure*}
    \centering
    \includegraphics[width=0.495\linewidth]{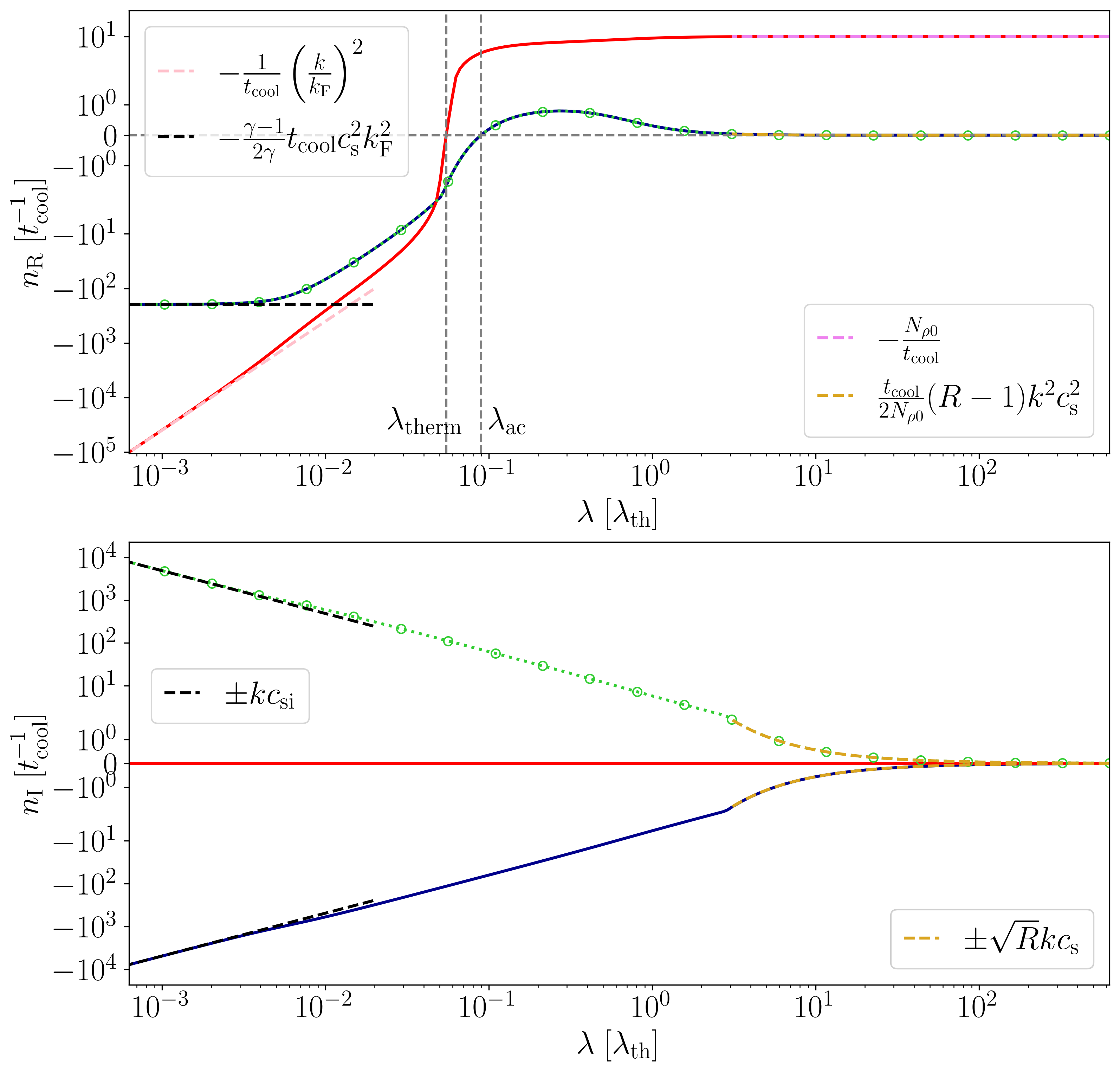}
    
    \caption{As in Figure~\ref{fig:row1} but for case $1/\gamma < R < 1$, $\Nrhoz, \Npz < 0$. The legends refer to the asymptotic curves in the limits $\lambda \rightarrow 0$ (dominant term of Equation~(\ref{eq:lambda0_condens_n}) and Equation~(\ref{eq:lambda0_acoustic_n})) and $\lambda \rightarrow \infty$ (dominant term of Equation~(\ref{eq:k0_condens_n}) and Equation~(\ref{eq:k0_acoustic_n})). The parameter values are $\Nrhoz = -10$, $R = 0.8$, and $\lambdaF = 0.2 \lambdath$.}
    \label{fig:row5}
\end{figure*}

We next move to row~2 of Table~\ref{table:R_and_lambda} and Figure~\ref{fig:row2}, which correspond to a plasma that is isochorically and isentropically unstable and isobarically stable. The main differences from Figure~\ref{fig:row1} are that the thermal mode is stable for all wavelengths, while the two acoustic modes first become overstable at $\lambda = \lambdaacoustic$ and later switch to two unstable thermal modes at $\lambda = \lambdac$. These two modes remain unstable as $\lambda \rightarrow \infty$, their growth rates given by Equations~(\ref{eq:k0_condens_n}) and (\ref{eq:k0_condens_n_Rlt0}).

Figure~\ref{fig:row3} and the third row of Table~\ref{table:R_and_lambda} present us with the most complex case in terms of wavelength ranges. Now all three stability criteria are met and, moreover, there are two critical wavelengths, $\lambda_\mp$, at which the two (overstable) acoustic modes transition to two (unstable) thermal modes and then back to two (overstable) acoustic modes. At wavelengths $\lambdaacoustic$ and $\lambdacond$, smaller than $\lambda_-$, we find that the modes shift from damped to undamped, as discussed above. And at wavelengths larger than $\lambda_+$ the growth rates behave as predicted by Equations~(\ref{eq:k0_condens_n}) and (\ref{eq:k0_acoustic_n}). The thermal mode is given by the red curve for $\lambda < \lambda_+$ and by the green curve with circles for $\lambda > \lambda_+$. A detailed view of the mode transitions is displayed in the right panels of Figure~\ref{fig:lambda_pm}.

Next, we discuss together rows 4 and 5 of Table~\ref{table:R_and_lambda} and their corresponding Figures~\ref{fig:row4} and~\ref{fig:row5} because they only differ in the ordering of $\lambdacond$ and $\lambdaacoustic$. In both cases we find isobaric, isochoric, and isentropic instability, and for this reason both the thermal mode and the pair of acoustic modes become unstable/overstable at sufficiently large wavelengths. Given that there are no critical wavenumbers, we find no mode conversions between the two mode types, and both modes have growth rates that tend to their asymptotic Equations~(\ref{eq:k0_condens_n}) and (\ref{eq:k0_acoustic_n}).  The active-region coronal loop sits in row~5 of Table~\ref{table:R_and_lambda} if a constant heating per unit mass is assumed instead of the commonly used constant heating per unit volume.

Row~6 of Table~\ref{table:R_and_lambda} and Figure~\ref{fig:row6} show a thermal mode that becomes unstable at $\lambda = \lambdacond$, because both the isobaric and isochoric criteria are satisfied. This mode reaches a maximum value of $\nR$ before tending to the value given by Equation~(\ref{eq:k0_condens_n}). With regard to the acoustic modes, they are stable for all wavelengths because the system is isentropically stable. Figure~\ref{fig:row6} is representative of both coronal loops and the CGM.

Row~7 of Table~\ref{table:R_and_lambda} is analogous to row~6, but now the roles of the thermal and the acoustic modes are interchanged: the former is stable for all $\lambda$ and the latter are overstable for $\lambda > \lambdaacoustic$. See Figure~\ref{fig:row7}.

\onecolumngrid
\begin{center}\figurehere
    \begin{minipage}[t]{0.48\textwidth}
        \centering
        \includegraphics[width=\linewidth]{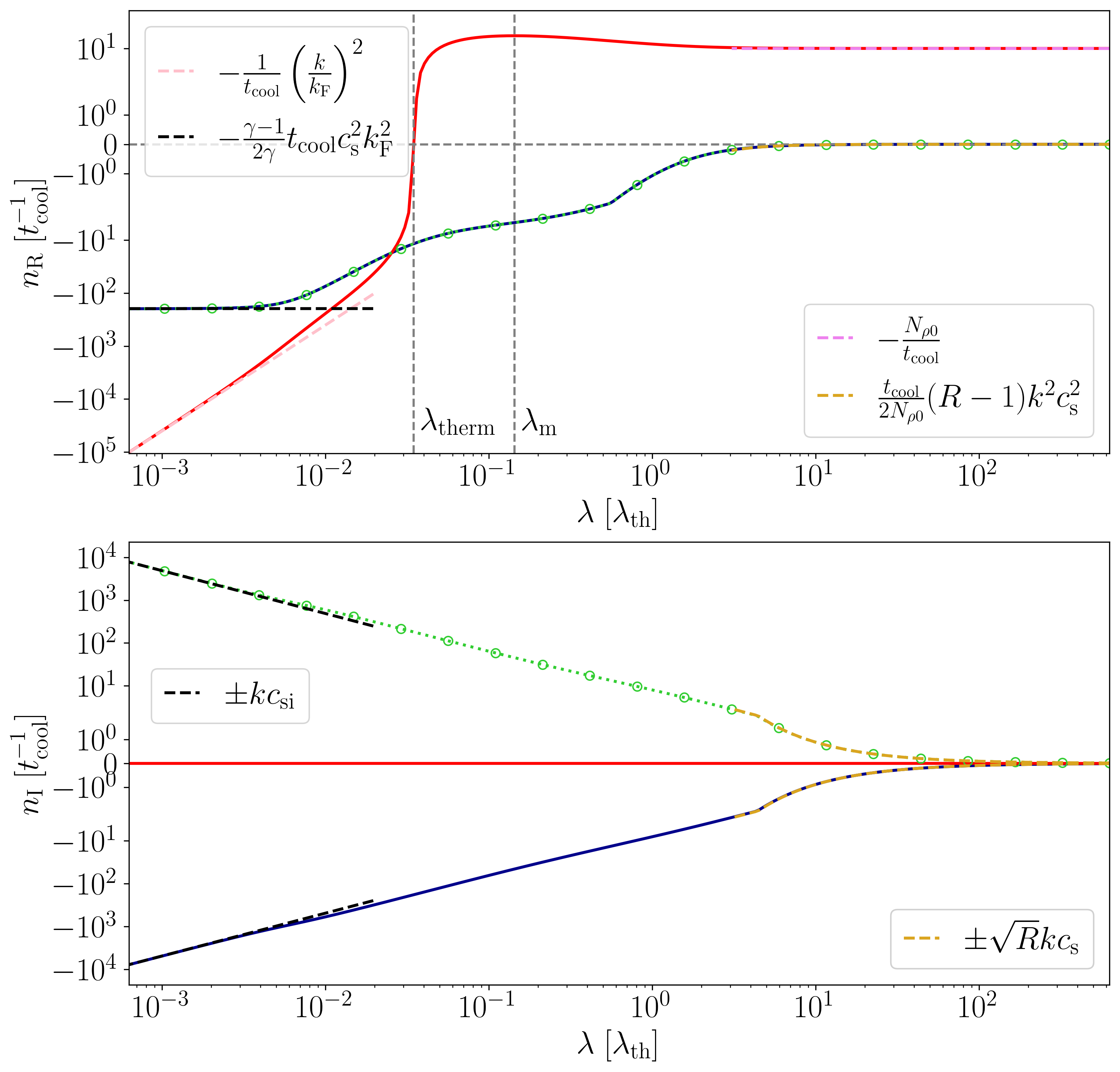}
        \caption{As in Figure~\ref{fig:row1} but for case $R > 1$, $\Nrhoz, \Npz < 0$. The legends refer to the asymptotic curves in the limits $\lambda \rightarrow 0$ (dominant term of Equation~(\ref{eq:lambda0_condens_n}) and Equation~(\ref{eq:lambda0_acoustic_n})) and $\lambda \rightarrow \infty$ (dominant term of Equation~(\ref{eq:k0_condens_n}) and Equation~(\ref{eq:k0_acoustic_n})). The parameter values are $\Nrhoz = -10$, $R = 2$, and $\lambdaF = 0.2 \lambdath$.}
        \label{fig:row6}
    \end{minipage}\hfill
    \begin{minipage}[t]{0.48\textwidth}
        \centering
        \includegraphics[width=\linewidth]{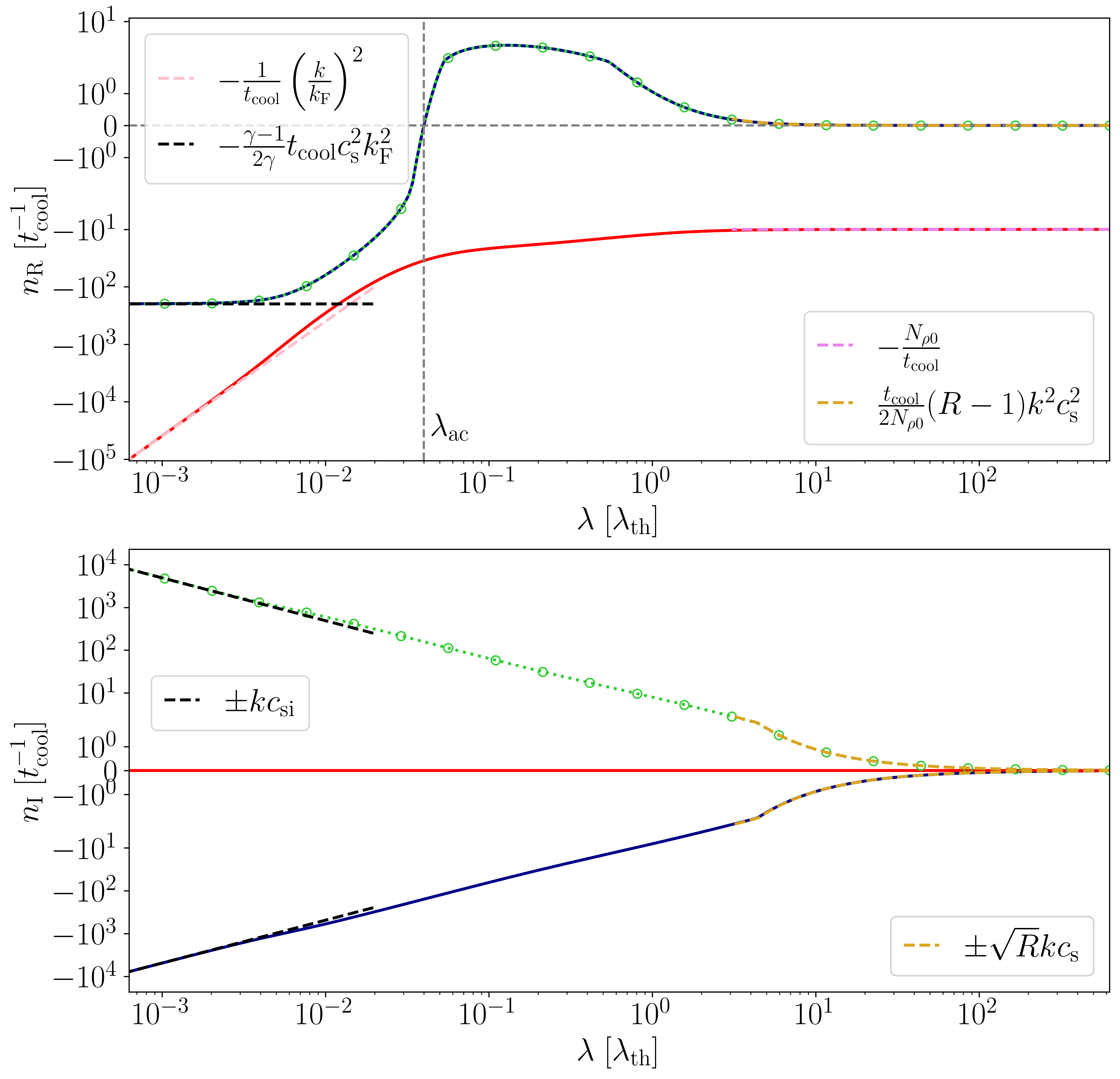}
        \caption{As in Figure~\ref{fig:row1} but for case $R > 1$, $\Nrhoz, \Npz > 0$. The legends refer to the asymptotic curves in the limits $\lambda \rightarrow 0$ (dominant term of Equation~(\ref{eq:lambda0_condens_n}) and Equation~(\ref{eq:lambda0_acoustic_n})) and $\lambda \rightarrow \infty$ (dominant term of Equation~(\ref{eq:k0_condens_n}) and Equation~(\ref{eq:k0_acoustic_n})). The parameter values are $\Nrhoz = 10$, $R = 2$, and $\lambdaF = 0.2 \lambdath$.}
        \label{fig:row7}
    \end{minipage}
\end{center}
\twocolumngrid

\section{Isobaric and isochoric perturbations at intermediate wavelengths} \label{sect:isobaric_perturbations}

In Sections~\ref{sect:short_wavelength} and~\ref{sect:long_wavelength} we have seen that, under certain circumstances, truly isochoric perturbations are allowed: this happens for the thermal mode when either $\lambda \rightarrow 0$ or $\lambda \rightarrow \infty$. This leads us to wonder whether purely isobaric perturbations exist. According to Equation~(\ref{eq:perturb_p}), this requires $n/k = 0$. Of course, whenever $k$ is finite and the growth rate of the thermal mode vanishes, the perturbations $\delta p$ and $\delta v$ also vanish, and the system has sinusoidal density and temperature perturbations that do not change in time. This takes place at the wavelength $\lambdacond$, given by Equation~(\ref{eq:kcond}), at which the thermal mode changes from being damped to being amplified.

Apart from this trivial solution, all other wavelengths produce pressure perturbations. Nevertheless, a truly isobaric thermal mode exists in the absence of thermal conduction; this is a situation that has not been considered in this paper and is only analyzed in this section and in the limit of very short wavelength. If $\kappa = 0$ and $\lambda \rightarrow 0$, then the results of Section~\ref{sect:short_condensation} are not valid. Now, the thermal mode has the growth rate

\begin{equation} \label{eq:n_condens_noconduct}
    \nR = -\frac{\Npz}{\gamma \tc}, \hspchor \nI = 0.
\end{equation}

\noindent Given that $n/k \rightarrow 0$ as $\lambda \rightarrow 0$, Equations~(\ref{eq:perturb_rho})--(\ref{eq:perturb_p}) and (\ref{eq:perturb_T}) tell us that $\delta v, \delta p \rightarrow 0$ and, consequently, this mode only gives rise to density and temperature perturbations. In the presence of thermal conduction (Section~\ref{sect:short_condensation}) the thermal mode is isochoric and always stable. Now we find that in the absence of thermal conduction, the thermal mode is unstable if the system is isobarically unstable ($\Npz < 0$), otherwise it is damped. The solution of Equation~(\ref{eq:n_condens_noconduct}) corresponds to Equation~(31) of \citet{field1965thermal} and has also been studied by, for example, \citet{Das2021}. For completeness, we also give the growth rate of acoustic modes for $\kappa = 0$ and $\lambda \rightarrow 0$,

\begin{equation} \label{eq:n_acoustic_noconduct}
    \nR = \frac{\Nrhoz}{2 \tc} (R-1), \hspchor \nI = \pm k \cs,
\end{equation}

\noindent which corresponds to Equation~(32) of \citet{field1965thermal}.

Our discussion of Section~\ref{sect:timescales} suggests that a uniform and infinite plasma with thermal conduction can support a thermal mode with an {\it almost isobaric} perturbation, i.e., whose pressure perturbation normalized to $\pz$ is much smaller than its density perturbation normalized to $\rhoz$, or an {\it almost isochoric} behavior, for which the normalized density perturbation is negligible compared with the normalized pressure perturbation. We next examine the character of the thermal mode by computing the amplitude of the density and pressure perturbations associated to Figures~\ref{fig:row1}--\ref{fig:row7}, which have been computed with $\lambdaF = 0.2 \lambdath$. The analysis of time scales carried out in Section~\ref{sect:timescales} indicates that, since $\lambdaF < \lambdath$, the thermal mode is isochoric for $\lambda \ll \lambdaisoch \equiv 0.04 \lambdath$ and for $\lambda \gg \lambdath$, while it is isobaric in the range $\lambdaisoch \equiv 0.04 \lambdath \ll \lambda \ll \lambdath$. Here is a summary of the results: (i) In all cases of Table~\ref{table:R_and_lambda}, we find that the thermal mode is essentially isochoric for $\lambda \lesssim 10^{-3} \lambdath$; this is consistent with the prediction $\lambda \ll 0.04 \lambdath$ from the time-scale analysis. (ii) When $\lambda$ is increased beyond this value, the thermal mode transitions to an almost isobaric character. Isobaric behavior is found in the range $0.01 \lambdath \lesssim \lambda \lesssim 0.1 \lambdath$ for rows 1, 5, and 6 of Table~\ref{table:R_and_lambda}, in the range $0.01 \lambdath \lesssim \lambda \lesssim \lambda_+ \simeq \lambdath$ for row 3, and in the range $0.01 \lambdath \lesssim \lambda \lesssim \lambdath$ for row 4. Rows 2 and 7 display the smallest wavelength range in which the thermal mode is isobaric: this only happens for $\lambda \simeq 0.04 \lambdath$, the reason being that $\Npz > 0$ in these two cases (and only in these two). We thus see that the time-scale prediction that the isobaric character is found for wavelengths satisfying $0.04 \lambdath \ll \lambda \ll \lambdath$ fails at the lower limit of these inequalities in rows 1 and 3--6 and is satisfied at the upper limit in all rows except 3 and 4. (iii) For even longer wavelengths, we abandon the isobaric regime. The properties of thermal modes again depend on the instability domain of Table~\ref{table:R_and_lambda}. In rows 1 and 2, the thermal mode is neither isochoric nor isobaric and it becomes one of the two asymptotic solutions of Equation~(\ref{eq:k0_condens_n_Rlt0}). But there are two more thermal modes that come from the conversion of acoustic modes at $\lambda = \lambdac \simeq 3 \lambdath$: one of them is neither isochoric nor isobaric and for $\lambda \rightarrow \infty$ becomes the other asymptotic solution of Equation~(\ref{eq:k0_condens_n_Rlt0}). The second of these thermal modes is isochoric for $\lambda > \lambdac \simeq 3 \lambdath$ and its asymptotic behavior is given by Equation~(\ref{eq:k0_condens_n}). On the other hand, in rows 3--7 the thermal mode becomes isochoric for $\lambda \gtrsim \lambdath$. Therefore, in all cases of Table~\ref{table:R_and_lambda} there is an isochoric thermal mode for $\lambda \gtrsim \lambdath$, which is in rough agreement with the time-scale prediction $\lambda \gg \lambdath$.

All astrophysical environments of Section~\ref{sect:environments} have $\lambdaF < \lambdath$ and therefore possess a range of wavelengths in which the thermal mode is isobaric, sandwiched between the lower and higher wavelength ranges, in which it is isochoric. The smaller the ratio $\lambdaF / \lambdath$, the larger the width of the isobaric range. The narrowest case among the environments considered is the TNE loop, with $\lambdaF/\lambdath = 0.56$, for which the thermal mode is still isobaric in $0.06 \lesssim \lambda / \lambdath \lesssim 0.15$.

We have also computed the perturbation amplitudes as a function of wavelength for each of the seven rows of Table~\ref{table:R_and_lambda} with $\lambdaF=5\lambdath$ instead of $\lambdaF=0.2\lambdath$; these plots are not shown. Now the conditions of Section~\ref{sect:timescales} for the thermal mode to be isochoric are $\lambda \ll \lambdaisoch = 25 \lambdath$ and $\lambda \gg \lambdath$. In agreement with these inequalities, in all seven rows the thermal mode is isochoric for almost all wavelengths and only behaves isobarically in a very narrow range of wavelengths around $\lambdath$ in rows 1, 3--5, 7.


\section{{Discussion and} conclusions} \label{sect:conclusions}

\subsection{Characteristic time scales and wavelengths} \label{sect:scales_summary}

The analysis carried out in this work involves a number of time scales and wavelengths, listed in Table~\ref{table:scales}, each of them defined either by the balance of two physical processes or by a change in the character of a mode. Having used them throughout, we can now say how they fit together.

Three time scales suffice to organize the problem, and their different dependence on the wavelength is what makes the spectrum rich.  The cooling time $\tc$ does not depend on $\lambda$ at all; the sound crossing time $t_{\rm snd}$ is proportional to $\lambda$; and the conduction time $t_{\rm cond}$ is proportional to $\lambda^2$.  Equating them in pairs produces the three wavelengths $\lambdath$, $\lambdaF$, and $\lambdaisoch$, which between them fix the ranges in which the thermal mode is isobaric or isochoric (Sections~\ref{sect:timescales} and~\ref{sect:isobaric_perturbations}).

The remaining wavelengths of Table~\ref{table:scales} are defined instead by a change in the character of a mode.  Two of them, $\lambdacond$ and $\lambdaacoustic$, which mark the change of the thermal and the acoustic modes from damped to growing, still follow in closed form from the equilibrium parameters, through Equations~(\ref{eq:kcond}) and (\ref{eq:kacoustic}). The others have no such expression and must be obtained by solving Equations~(\ref{eq:WP2019_eq13}) and (\ref{eq:km}): $\lambda_{\rm c}$ and $\lambda_\pm$, which determine the conversion between the two types of mode, and $\lambda_{\rm m}$, the wavelength of fastest growth.

A distinction that deserves emphasis, because the two are often used interchangeably, is that between $\lambdaF$ and $\lambdacond$.  The Field length $\lambdaF$ is the wavelength at which conduction and cooling act on the same time scale; it is a property of the equilibrium alone and does not know whether the plasma is unstable.  The wavelength at which the thermal mode actually changes from damped to growing is $\lambdacond = \lambdaF / \sqrt{-\Npz}$, which is shorter than $\lambdaF$ when $-\Npz > 1$ and longer when $-\Npz < 1$.  For the equilibrium coronal loop of Section~\ref{sect:environments}, for instance, $-\Npz = 4.1$ and the two differ by a factor of two.  Finally, none of these wavelengths is meaningful unless it is compared with the size $L$ of the system, since the uniform medium of our analysis is an idealization of a finite object; this comparison is made in Section~\ref{sect:fit_inside}.

\subsection{The isobaric/isochoric character of the thermal mode} \label{sect:discussion_character}

In this subsection we discuss our results on the isobaric and isochoric character of the thermal mode; for completeness, the character of the acoustic modes in the limits $\lambda \rightarrow 0$ and $\lambda \rightarrow \infty$ is also reviewed. We have presented a detailed analysis of the modes of the homogeneous hydrodynamical system studied by \citet{field1965thermal} and \citet{waters2019non} for very short and very long wavelengths. We have also shown that, to study the perturbations caused by modes with an isochoric evolution, it is essential to express all perturbations in terms of the amplitude of the pressure perturbation instead of the amplitude of the density perturbation. When $\lambda \rightarrow 0$,  the thermal mode becomes isochoric. It has no velocity perturbation, and the temperature and pressure perturbations damp by heat exchange via thermal conduction. Moreover, there is a pair of damped isothermal acoustic modes that propagate in opposite directions at the isothermal sound speed. This clarifies the physical nature of the thermal and acoustic modes at short wavelengths and shows that the commonly employed assumption of an isobaric thermal mode in the $\lambda\rightarrow0$ limit is not valid when thermal conduction is included.

Equation~(16) of \citet{seno_etal2026} provides an approximation to the asymptotic behavior of the growth rate of the thermal mode for small wavelength. In terms of the parameters used in this paper, this equation can be written as

\begin{equation} \label{eq:lambda0_seno_etal2026}
    \nR = -\frac{\Nrhoz}{\gamma \tc} + \frac{\cs}{\gamma} - \frac{1}{\gamma \tc} \left(\frac{k}{\kF}\right)^2.
\end{equation}

\noindent There is a clear discrepancy between this expression and our Equation~(\ref{eq:lambda0_condens_n}), the reason being that, to derive it, \cite{seno_etal2026} incorrectly assumed that the thermal mode is isobaric when $\lambda \rightarrow 0$. We refer to the discussion in Section~\ref{sect:isobaric_perturbations}, where we show that for $\lambda \lesssim 10^{-3} \lambdath$ the thermal mode is isobaric (isochoric) in the absence (presence) of thermal conduction. A comparison of Equation~(\ref{eq:lambda0_seno_etal2026}) with the growth rate of the thermal mode for the cases of Table~\ref{table:R_and_lambda} (not shown in Figures~\ref{fig:row1}--\ref{fig:row7}) reveals that \citeauthor{seno_etal2026}'s approximation is not too far from the actual value of $\nR$ for $\lambda \simeq 10^{-2} \lambdath$, but that it departs from it at smaller wavelengths. Hence, Equation~(\ref{eq:lambda0_seno_etal2026}) does not provide a valid estimate of $\nR$ for $\lambda \rightarrow 0$, while Equation~(\ref{eq:lambda0_condens_n}) does.

When $\lambda \rightarrow \infty$ another isochoric thermal mode is present, its stability being governed by the isochoric stability criterion. Thermal conduction plays no role in the exponential damping/growth of this mode. When the parameter $R$ is negative, this thermal mode is accompanied by two additional thermal modes with growth rate $\pm \sqrt{-R}k \cs$ \citep[see][]{waters2019non}. When $R > 0$, these two thermal modes are replaced by a pair of acoustic solutions with phase speed $\pm \sqrt{R} \cs$ that are overstable if $R-1$ and $\Nrhoz$ have the same sign, and damped otherwise; among the cases collected in Table~\ref{table:R_and_lambda}, they are damped only in the sixth row.

The amplitudes of the perturbed variables in the linear regime (see Equations~(\ref{eq:perturb_rho})--(\ref{eq:perturb_T})) are such that, when thermal conduction is included, perturbations of the medium always produce pressure and density changes. In other words, truly isobaric and truly isochoric perturbations are not allowed, and this is true regardless of whether the medium is isobarically or isochorically unstable for $\lambda \rightarrow \infty$ ($\Npz < 0$ or $\Nrhoz < 0$) or not. There are two exceptions to this, since the medium supports a thermal mode with isochoric perturbation both for $\lambda \rightarrow 0$ and $\lambda \rightarrow \infty$. Isochoric perturbations were discarded by \citet{field1965thermal} on the basis that ``pressure variations accompanying isochoric temperature variations will in general drive motions which destroy the constancy of density''. This sentence is concluded as follows, ``so that this criterion is incompatible with the force equation.'' We have shown that this incompatibility does not exist for very small and very long wavelengths. Furthermore, if we relax the condition on the thermal mode to be {\it purely} isobaric or isochoric, we find that it is {\it almost} isochoric for very small or very large wavelengths and is {\it almost} isobaric for intermediate wavelengths. For example, for $\lambdaF = 0.2 \lambdath$, the thermal mode is almost isochoric for $\lambda \lesssim 10^{-3} \lambdath$ and for $\lambda \gtrsim \lambdath$ and almost isobaric for $0.01 \lambdath \lesssim \lambda \lesssim 0.1 \lambdath$. We also find that increasing $\lambdaF$ results in a reduction of the range of $\lambda$ for which the thermal mode is almost isobaric, together with a shift to higher wavelengths. For example, for $\lambdaF = \lambdath$, this range becomes $0.1 \lambdath \lesssim \lambda \lesssim \lambdath$ for all rows of Table~\ref{table:R_and_lambda}, except rows 2 and 7, for which the thermal mode behaves almost isochorically for all wavelengths.

\subsection{Coronal rain and the catastrophic cooling instability} \label{sect:discussion_CR_CC}

\begin{figure*}
    \centering
    \includegraphics[width=0.58\linewidth]{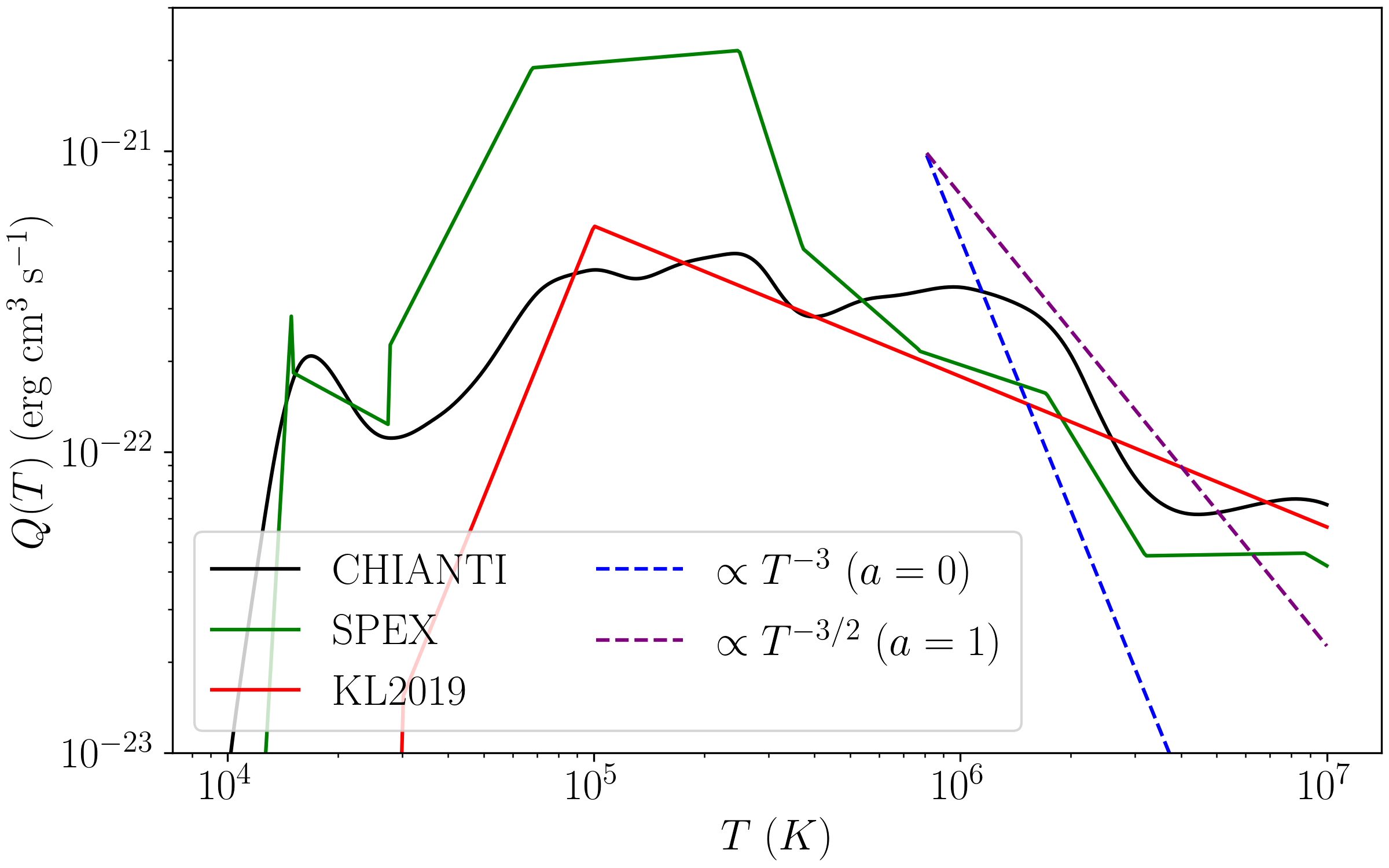}
    \includegraphics[width=0.58\linewidth]{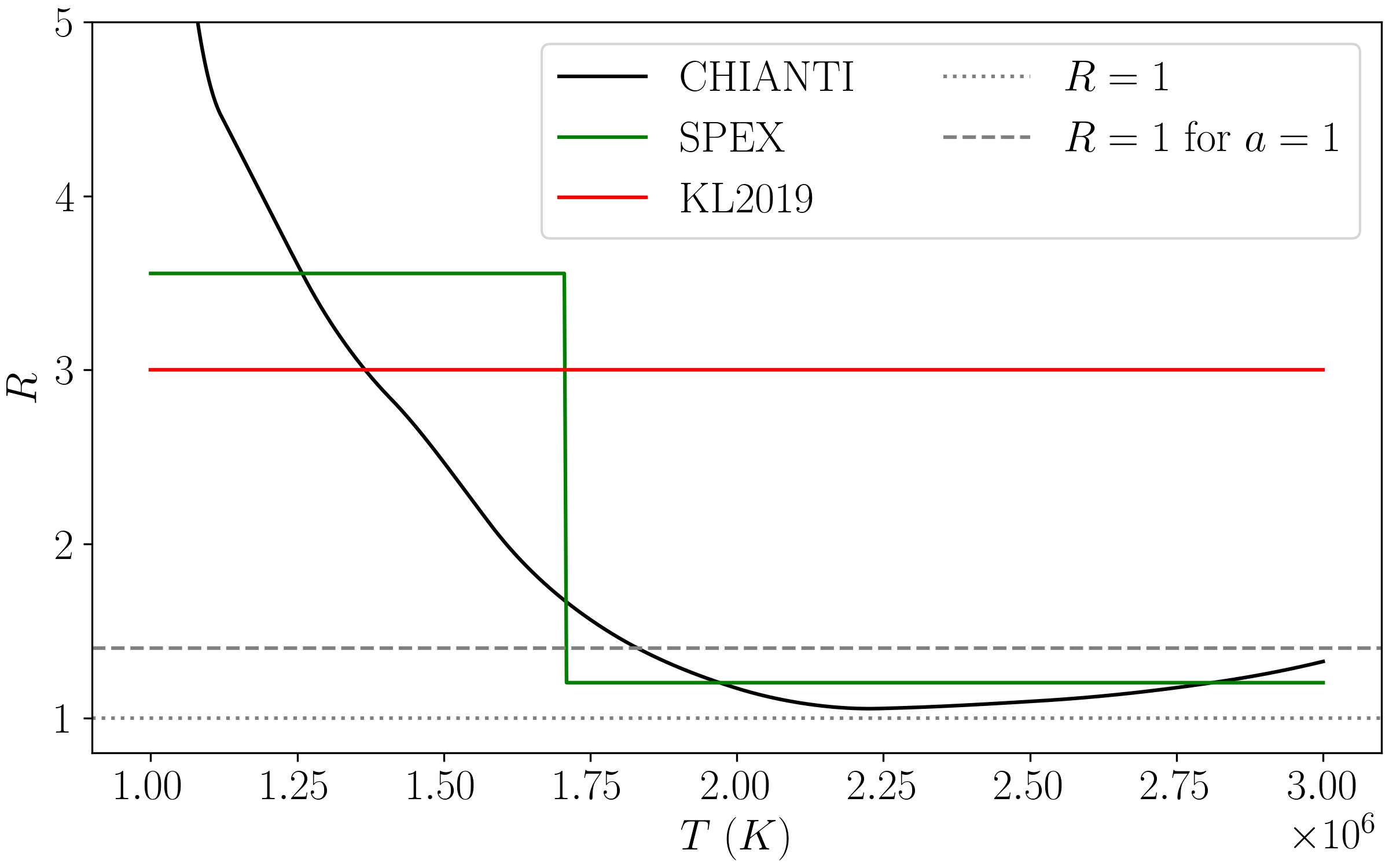}

    \caption{Top: three optically thin radiative loss functions for the solar corona, from the CHIANTI atomic database \citep{Dere_etal1997,Dufresne_etal2024}, from the piecewise power-law fit of \citet{HermansKeppens2021} to the SPEX cooling function of \citet{Schure_etal2009}, and from \citet{KlimchukLuna2019}. The two dashed lines have power-law slopes $-3$ and $-3/2$, which are the critical slopes for which $R = 1$ when $a = 0$ and $a = 1$, respectively; see Equation~(\ref{eq:appcrit}). A loss function steeper than the relevant line gives $R < 1$. The three functions are not defined for the same chemical composition, the CHIANTI values being those for coronal abundances, the SPEX fit those for solar abundances, and the expression of \citet{KlimchukLuna2019} that for a fully ionized hydrogen plasma; this shifts the curves vertically with respect to one another but has no effect on their logarithmic slopes and hence none on the bottom panel. Bottom: the corresponding values of $R$, obtained from Equation~(\ref{eq:appR}) with $b = 0$ and a volumetric heating rate independent of density, $a = 0$. The horizontal dotted line corresponds to $R = 1$. The horizontal dashed line, at $R = (2\gamma - 1)/\gamma = 7/5$, marks the value of $R$ at which the same plasma would have $R = 1$ if the heating rate per unit mass were independent of density ($a = 1$); the curves lying below it therefore correspond to $R < 1$ under that alternative prescription.}
    \label{fig:corona}
\end{figure*}

 Next, we discuss the catastrophic cooling (CC) instability recently proposed by \citet{waters_stricklan2025} as the agent causing coronal rain plasma condensations. We have already made clear that the CC solution coincides with the thermal mode in the $\lambda \rightarrow \infty$ limit. This solution is thus the $k = 0$ member of the thermal branch, for which $\delta\rho = \delta v = 0$ identically and the temperature perturbation is uniform in space (see Equation~(\ref{eq:k0_condens_perturb})); it therefore describes the plasma of the loop cooling as a whole, and cannot by itself redistribute mass so as to produce a localized condensation. A perturbation that does so must have $k \neq 0$, and in a loop of length $L$ the available wavenumbers satisfy $k \gtrsim \pi/L$, so that whether such a perturbation still evolves isochorically is decided by the comparison of $\lambdath$ with $L$ carried out in Section~\ref{sect:fit_inside}. Quite apart from this, a comparison between linear thermal instability in a homogeneous medium and coronal rain formation must be taken with a grain of salt because thermal conduction plays a major role in the unperturbed, non-uniform coronal loop (equilibrium or out-of-equilibrium) and this can affect perturbations in ways that are not captured by the Field length.
 
\citet{waters_stricklan2025} argued that for $0 < R < 1$ the CC mode is the fastest growing linear mode and that, as a consequence, it will be the first to reach non-linearity and will then dominate the evolution of the system. This is in line with Figures~\ref{fig:row3}--\ref{fig:row5}, which show that the highest growth rate of the thermal mode occurs as $\lambda \rightarrow \infty$, exclusively when $0 < R < 1$. Whether coronal loops satisfy this condition is therefore a quantitative question, and it is the one we address here.

Equation~(\ref{eq:appcrit}) --- with $b=0$ for a heating independent of temperature --- provides the answer in a compact form: for power-law heating and cooling, the value $R = 1$ is attained when the logarithmic slope of the radiative loss function satisfies $\beta = (2 - a)/(\gamma - 1)$, so that $R < 1$ requires $\beta > 3$ if the heating rate per unit volume is independent of density ($a = 0$) and $\beta > 3/2$ if it is the heating rate per unit mass that is independent of density ($a = 1$). The first of these is the prescription adopted in Section~\ref{sect:environments} and in the loop models that motivate the present comparison, and we take it as our reference; we return below to the consequences of the other choice.

The top panel of Figure~\ref{fig:corona} presents three optically thin radiative loss functions in use for the solar corona: the tabulated values of the CHIANTI atomic database \citep{Dere_etal1997,Dufresne_etal2024}, the piecewise power-law fit of \citet{HermansKeppens2021} to the SPEX cooling function of \citet{Schure_etal2009}, and the simplified expression employed by \citet{KlimchukLuna2019} in their loop simulations, which modifies the piecewise fit of \citet{Klimchuk_etal2008}.\footnote{These three functions are not independent determinations. CHIANTI and SPEX are computed from atomic databases under the assumption of collisional ionization equilibrium and differ mainly through atomic completeness and the adopted ionization balance, as does the independent calculation of \citet{Colgan_etal2008}; the expression of \citet{KlimchukLuna2019}, by contrast, is a deliberate simplification, chosen for its analytical convenience in loop modeling rather than fitted to a spectral calculation. All of them share the limitation that ionization equilibrium is assumed, which is questionable precisely for the rapidly cooling plasma whose stability is at issue here; non-equilibrium calculations such as those of \citet{GnatSternberg2007} give appreciably different losses in that regime.
} None of them is anywhere near as steep as the reference slope $\beta = 3$ drawn in the same panel: over $1$--$3$~MK the CHIANTI slope reaches an absolute value of about $2.65$ at most, the SPEX fit gives $\beta = 0.41$ below $1.7$~MK and $\beta = 1.99$ above it, and \citet{KlimchukLuna2019} adopt $\beta = 1/2$ throughout. The corresponding values of $R$, computed from Equation~(\ref{eq:appR}) with $a = 0$ and $b = 0$, are shown in the bottom panel of Figure~\ref{fig:corona}. All three curves remain above $R = 1$ at every temperature in the range $\simeq$~1--3~MK, typical of active-region coronal loops hosting coronal rain, the smallest values being $R \simeq 1.05$ for CHIANTI and $R \simeq 1.20$ for SPEX. The conclusion is therefore that the coronal plasma is better represented by $R > 1$, that is, by the sixth row of Table~\ref{table:R_and_lambda}, whose associated Figure~\ref{fig:row6} implies that the thermal mode achieves its maximum growth rate at a finite wavelength $\lambda = \lambdam$ rather than for $\lambda \rightarrow \infty$. The CC instability solution is thus not the fastest growing mode in this regime.

This conclusion, however, is conditional on the assumed density scaling of the heating, and the condition is not a weak one. For $a = 1$ the critical slope drops from $3$ to $3/2$, which the coronal loss functions do exceed: $R$ then falls below unity above about $1.7$~MK for the SPEX fit and above roughly $1.8$~MK for CHIANTI, reaching $R \simeq 0.90$ and $R \simeq 0.83$ for SPEX and CHIANTI, respectively, and only the function of \citet{KlimchukLuna2019}, with its shallow $\beta = 1/2$, stays above $R = 1$ throughout. A coronal plasma with $a = 1$ would therefore belong to the fifth row of Table~\ref{table:R_and_lambda} over much of the relevant temperature range, and the argument of \citet{waters_stricklan2025} would apply to it. We adopt $a = 0$ because it is the prescription used in the loop models that motivate this comparison, and because the scaling laws proposed for coronal heating favor small values of $a$, but the heating of coronal loops is not known well enough for the alternative to be excluded\footnote{The mechanisms that have been proposed differ in the way their heating rate depends on density, though less than one might expect. Table~5 of \citet{Mandrini_etal2000} collects scaling laws for eighteen coronal heating models, retaining the explicit dependence on density and with no dependence on temperature, so that $b = 0$ throughout; of the twenty-two sets of exponents listed, eleven correspond to $a = 0$ exactly, all but one of the remainder to $a \lesssim 1/2$, and only one reaches $a = 1$ or $a = 3/2$. For a typical coronal slope $\beta = 2.1$, Equation~(\ref{eq:appcrit}) places the crossing of $R = 1$ at $a \simeq 0.6$, so that almost all of the tabulated models give $R > 1$. These scaling laws are, as those authors stress, highly idealized, and they are formulated for a structured and magnetized loop; we therefore quote them as an indication of the spread rather than adopting any particular value for our uniform, non-magnetic model.}, and the question of whether the CC mode is the fastest growing one in the corona ultimately rests on this choice rather than on the radiative losses alone. That the outcome should depend on this assumption is not a new observation: \citet{DahlburgMariska1988} already found the development of condensations in coronal loops to be sensitive to the way the heating rate scales with the local plasma properties.

A second consideration, of a different nature, follows from the comparison of $\lambdath$ with the size of the system carried out in Section~\ref{sect:fit_inside}. The CC solution is recovered in the limit $\lambda \gg \lambdath$, and that limit is physically available only in a system large enough to accommodate such wavelengths, that is, one for which $\lambdath \lesssim L$. Among the environments of Table~\ref{table:environments_output} this is comfortably satisfied by the X-ray photoionized outflow and by the coronal plasma in a TNE loop, for which $\lambdath \simeq 0.17\,L$, but not by the equilibrium loop, whose $\lambdath \simeq 8.2\,L$, nor by the cluster cool core. Whether the catastrophic cooling mode can be reached is thus a question about the size of the system as much as about the value of $R$, and in a coronal loop the answer changes during the evolution of the loop itself.

None of this implies that thermal instability is irrelevant for coronal loop conditions. Rather, the mode properties and their stability are systematically categorized in Table~\ref{table:R_and_lambda} and illustrated in the dispersion diagrams of Figures~\ref{fig:row5} and~\ref{fig:row6}.

\subsection{Summary} \label{sect:summary}

We have revisited and extended the analysis of the linear modes of an infinite and homogeneous hydrodynamic medium, investigated by \citet{field1965thermal} and \citet{waters2019non}. In this work we have imposed no particular form of the heating and radiative cooling processes, nor the thermal conductivity, in the general analysis of Sections~\ref{sect:main_equations}--\ref{sect:isobaric_perturbations}; specific forms are adopted only when the results are applied to particular astrophysical media. We find that the presentation of the different regimes of linear thermal instability given in Figure~2 of \citet{waters2019non} is easier to understand than in Figure~1 of \citet{field1965thermal}. Nevertheless, the former contains an error that has been corrected (namely the relevance of the value $R = 1/3$) and we extend the previous classification by incorporating the wavelengths $\lambdacond$ and $\lambdaacoustic$, as well as the effects of thermal conduction, into our Table~\ref{table:R_and_lambda}. This table provides a systematic synthesis of seven distinct instability domains, making clear which modes are present in different wavelength ranges and whether they are unstable or damped. The domains of instability are separated by critical values of $R$ that were derived by \citet{waters2019non} under the assumption of vanishing Field length ($\lambdaF \rightarrow 0$), or equivalently, in the limit of no thermal conduction. The physical meaning of $R$ comes from its definition as the quotient $\Npz / (\gamma \Nrhoz)$, where $\Npz < 0$ and $\Nrhoz < 0$ are a dimensionless form of the isobaric and isochoric instability criteria obtained by \citet{field1965thermal}; we stress that both parameters must be evaluated in the equilibrium state, so that $R$ is independent of the wavenumber.
 
Including thermal conduction in the analysis implies that the critical values of $R$ can shift slightly and we have shown that Table~\ref{table:R_and_lambda} and the auxiliary Figures~\ref{fig:row1}--\ref{fig:row7} remain an excellent guide to the domains of instability of a hydrodynamic plasma for which $\lambdaF \lesssim \lambdath$. For larger values of the Field length the instability domains have no simple description in terms of $R$, thereby establishing the limits of validity of the classification proposed by \citet{waters2019non}. This limit turns out to be a generous one. Equation~(\ref{eq:appratio}) shows that the ratio $\lambdaF/\lambdath$ is independent of the density of the plasma apart from a weak logarithmic dependence, and that it is a slowly varying function of its temperature, of the radiative loss function, and of the conductivity; the five environments examined in Section~\ref{sect:environments} all have $\lambdaF/\lambdath$ between $0.059$ and $0.56$, and the equality $\lambdaF = \lambdath$ is not reached until $T \simeq 1.3 \times 10^{9}$~K, a temperature at which the assumptions underlying that expression no longer hold. Any suppression of the thermal conductivity below its Spitzer value, such as is expected in the intracluster and circumgalactic media, only reinforces this conclusion, since $\lambdaF \propto f^{1/2}$. We are therefore not aware of an environment in which thermal instability operates and the Field length exceeds the thermal length: Table~\ref{table:R_and_lambda} is applicable wherever thermal instability is invoked, and the regime $\lambdaF \gtrsim \lambdath$ analyzed in Section~\ref{sect:arbitrary_lambdaF} is a remote corner of parameter space.
 
In Section~\ref{sect:environments} we have placed five astrophysical media in this classification: the two-phase atomic interstellar medium, a coronal loop in equilibrium and the same loop late in a thermal non-equilibrium cycle, the circumgalactic medium, the cool core of a galaxy cluster, and the X-ray photoionized outflow of an active galactic nucleus. For each of them we have adopted a heat-loss function, a thermal conductivity, and a heating prescription in common use in its field, and evaluated $\Npz$, $\Nrhoz$, $R$, and the characteristic scales. All of them fall in the first or the sixth row of Table~\ref{table:R_and_lambda}. This is not an accident: the parameterization of the heating rate introduced in Appendix~\ref{sect:app_powerlaw} shows that $\Nrhoz - \Npz = 2 - a$, so that a plasma whose heating rate per unit volume does not grow with density faster than the two-body radiative losses, $a \le 2$, is confined to the first, fifth, and sixth rows, and in particular can never produce the elaborate spectrum of the third row, with its two critical wavelengths and its double conversion between acoustic and thermal modes. Finally, we have compared the wavelengths of the classification with the size $L$ of each system, since the uniform medium of our analysis is an idealization of a finite object; this comparison determines whether an unstable wavelength fits inside the system at all, and whether the isochoric regime $\lambda \gg \lambdath$, in which the catastrophic cooling solution is recovered, is physically available.
 
Furthermore, we have presented a detailed analysis of the modes for very short and very long wavelengths, and of the isobaric or isochoric character of the thermal mode at intermediate ones, which is summarized in Section~\ref{sect:discussion_character}. When $\lambda \rightarrow 0$ the thermal mode is isochoric, not isobaric, whenever thermal conduction is present, and the commonly employed assumption to the contrary leads to incorrect results. Growth rate and dispersion diagrams are presented in Section~\ref{sect:nRnI_vs_lambda} for all the domains of instability. These diagrams provide a graphical counterpart to Table~\ref{table:R_and_lambda} and give evidence of how the conversion between thermal and acoustic modes takes place, how the modes behave for very small or very large wavelengths, in what wavelength ranges there is instability, etc. A discussion of the instability domains in terms of $\Npz$, $\Nrhoz$, and $\lambdaF/\lambdath$ is given in Section~\ref{sect:arbitrary_lambdaF}. Figure~\ref{fig:Nzeros} and its accompanying movie can be considered a simplification of Figure~1 of \citet{field1965thermal}: the former provides the number of critical wavenumbers in three dimensions (with $\lambdaF/\lambdath$ represented along the third axis) and, at the same time, does not involve the wavenumber, which is present in the latter.

\begin{acknowledgements}
This publication is part of the R+D+i project PID2023-147708NB-I00, financed by MCIN/AEI/10.13039/501100011033. This research was supported by the International Space Science Institute (ISSI) in Bern, through ISSI International Team project \#545. The work of James A Klimchuk was supported by the NASA HISFM competed grant program. This work was also supported by the U.S. Department of Energy through the Los Alamos National Laboratory. Los Alamos National Laboratory is operated by Triad National Security, LLC, for the National Nuclear Security Administration of U.S. Department of Energy (contract No. 89233218CNA000001). The authors thank Rony Keppens for useful discussions. CHIANTI is a collaborative project involving George Mason University, the University of Michigan (USA), University of Cambridge (UK) and NASA Goddard Space Flight Center (USA). We thank the referee for two careful and constructive reports, whose insistence on connecting the classification to real astrophysical systems has substantially improved the paper.
\end{acknowledgements}


\bibliographystyle{aasjournalv7}
\bibliography{ref}


\begin{appendix}

\section{Power-law heating, cooling, and thermal conduction} \label{sect:app_powerlaw}
 
In this appendix we collect closed-form expressions for the equilibrium parameters and characteristic scales of Sections~\ref{sect:dispersion_relation}, \ref{sect:instability_criteria}, and \ref{sect:timescales} for the case in which the radiative losses, the heating rate, and the thermal conductivity are all power laws of the temperature. These expressions are used in Section~\ref{sect:environments} to place representative astrophysical systems in the classification of Table~\ref{table:R_and_lambda}.

\subsection{Radiative losses and chemical composition} \label{app:losses}

For an optically thin medium, the radiative losses per unit volume are the result of two-body processes and are therefore proportional to the product of two number densities. We write them as

\begin{equation}
  \rho\Lambda = x\, n_{\rm H}^{2}\, Q(T), \qquad Q(T) = Q_0 T^{-\beta},
  \label{eq:appQ}
\end{equation}

\noindent where $n_{\rm H}$ is the number density of hydrogen nuclei, whether neutral or ionized, $Q(T)$ is the radiative loss function, $Q_0$ and $\beta$ are constants, and $x$ is the ratio of the product of the two densities appearing in the tabulation of $Q$ to $n_{\rm H}^{2}$. Loss functions for ionized plasmas are given either as $n_{\rm e}n_{\rm H}Q(T)$, for which $x = n_{\rm e}/n_{\rm H}$, or as $n_{\rm e}n_{\rm i}Q(T)$, for which $x = n_{\rm e}n_{\rm i}/n_{\rm H}^{2}$; those for neutral media are given as $n_{\rm H}^{2}Q(T)$, for which $x = 1$. In every case the contributions of helium and of the heavier elements are contained in $Q(T)$ itself, which is computed for a definite chemical composition. We characterize the composition by the two constants

\begin{equation}
  n_{\rm H} = \frac{\rho}{\mu_{\rm H} m_{\rm H}}, \qquad
  n = \frac{\rho}{\mu\, m_{\rm H}},
  \label{eq:appcomp}
\end{equation}

\noindent where $n$ is the total particle number density and $m_{\rm H}$ is the hydrogen mass. Thus, $\mu_{\rm H}$ is the plasma mass per hydrogen nucleus in units of $m_{\rm H}$ and accounts for the presence of helium and heavier elements, and $\mu$ is the mean molecular weight, which gives the gas pressure as $p = \rho k_{\rm B}T/(\mu m_{\rm H})$. These two are properties of the medium, their ratio being the number of particles per hydrogen nucleus. For a fully ionized hydrogen plasma $\mu_{\rm H} = 1$ and $\mu = 1/2$; for a fully ionized plasma of solar composition $\mu_{\rm H} \simeq 1.4$ and $\mu \simeq 0.6$, with $x \simeq 1.2$ for losses tabulated as $n_{\rm e}n_{\rm H}Q$ and $x \simeq 1.32$ for those tabulated as $n_{\rm e}n_{\rm i}Q$; and for the neutral atomic gas of the interstellar medium $\mu_{\rm H} \simeq 1.4$, $\mu \simeq 1.27$, and $x = 1$. No particular composition is assumed in the main text of this paper.

The heat losses per unit mass then read
\begin{equation}
  \Lambda = \frac{x}{(\mu_{\rm H}m_{\rm H})^{2}}\,\rho\, Q(T).
  \label{eq:appLambda}
\end{equation}
Because the composition enters Equation~(\ref{eq:appLambda}) only through a constant prefactor, it has no effect whatsoever on the dimensionless parameters of Section~\ref{app:params} ($\Npz$, $\Nrhoz$, $R$, \ldots); it only rescales the time and length scales of Section~\ref{app:scales}, and does so weakly (see Section~\ref{app:ratio}).
 
\subsection{Parameterized heating} \label{app:heating}
 
The heating rate is far less well constrained than the cooling rate in every environment of interest, and it is customary to parameterize it.  Following \citet[][their Equation~(2)]{stricklan_etal2025}, we write the heating rate \emph{per unit volume} as

\begin{equation}
  \rho\,H = H_\ast\,\rho^{\,a}\,T^{\,b},
  \label{eq:appH}
\end{equation}
with $H_\ast$, $a$, and $b$ constants. A parameterization of this kind has a long history in studies of the thermal instability. \citet{DahlburgMariska1988} examined precisely how the assumed dependence of the heating rate on density and temperature alters the onset of condensations in the solar corona, and \citet{AntiochosNoci1986} adopted a similar prescription with $b = 0$ and $a$ unrestricted in their coronal loop calculations; this is the earliest reference we are aware of in the coronal literature for the form of Equation~(\ref{eq:appH}), although those authors were concerned with low-lying, cool loops in which thermal conduction plays a minor role in the equilibrium, unlike the hot loops considered here. The equilibrium condition $\calL=0$ fixes $H_\ast$ in terms of the equilibrium state,
\begin{equation}
  H_\ast = \frac{x}{(\mu_{\rm H}m_{\rm H})^{2}}\,
              \rhoz^{\,2-a}\,Q(\Tz)\,\Tz^{-b}.
  \label{eq:appCH}
\end{equation}
Three choices of $a$ cover most of the literature:
\begin{itemize}
  \item $a=0$: the heating rate per unit volume is independent of density.  This is the prescription adopted in essentially all global coronal loop simulations that produce thermal non-equilibrium solutions \citep[see the list of references in][]{stricklan_etal2025}.
  \item $a=1$: the heating rate per unit mass is independent of density, i.e., the volumetric rate scales as $n$.  This is the physically motivated choice for the neutral interstellar medium, where photoelectric heating on dust grains and cosmic-ray heating proceed at a rate per particle \citep{KoyamaInutsuka2002,vazquez-semadeni_etal2007}.
  \item $a=2$: the heating shares the density scaling of the cooling.  As shown below, in this case the isobaric and isochoric instability criteria coincide and $R=1/\gamma$ identically, whatever the value of $\beta$.
\end{itemize}
The exponent $b$ vanishes for all rows of Table~\ref{table:environments_input} except the last one, in which the cooling and heating functions are not power laws of $T$ and $\rho$.

\subsection{Instability parameters} \label{app:params}
 
Inserting Equations~(\ref{eq:appLambda}) and (\ref{eq:appH}) into Equations~(\ref{eq:Nrho0_Np0}), (\ref{eq:R}), and (\ref{eq:NSz}) gives, exactly,
\begin{equation}
  \Nrhoz = -(\beta+b), \qquad
  \Npz    = -(\beta+b+2-a),
  \label{eq:appN}
\end{equation}
\begin{equation}
  \NSz = \frac{1}{\gamma-1}\left[\,(2-a) - (\gamma-1)(\beta+b)\,\right],
  \qquad
  R = \frac{\beta+b+2-a}{\gamma\,(\beta+b)}.
  \label{eq:appR}
\end{equation}
All these quantities are unaffected by the chemical composition, as long as $x$ and $\mu$ can be treated as constants; see Sections~\ref{app:losses} and~\ref{app:ratio}.

It is also worth noting that the power-law description is less restrictive than it may appear. The parameters $\Nrhoz$, $\Npz$, and $\NSz$ are defined in terms of the first derivatives of the heat-loss function at the equilibrium state, so that the exponents appearing in Equations~(\ref{eq:appN}) and (\ref{eq:appR}) are simply the local logarithmic slopes of the cooling and heating rates:

\begin{equation}
  \beta  = -\frac{\Tz}{\Lambda_0}\,\Lambda_T, \qquad
  a = 1 + \frac{\rhoz}{\Lambda_0}\,H_\rho, \qquad
  b = \frac{\Tz}{\Lambda_0}\,H_T,
  \label{eq:appslopes}
\end{equation}
 
\noindent where the $T$- and $\rho$-derivatives are evaluated in the equilibrium state. Any sufficiently smooth $Q(T)$ and $H(\rho, T)$ are power laws in the neighborhood of the equilibrium, which is all that linear theory samples. The one caveat concerns the piecewise power-law fits in common use, whose slopes change discontinuously from one segment to the next: a plasma whose equilibrium temperature lies close to such a boundary is assigned an ambiguous value of $\beta$, and the ambiguity is then a property of the fit rather than of the plasma.

Several remarks follow.
 
First, the isochoric and isobaric parameters differ by a quantity that depends on the density scaling of the heating alone,
\begin{equation}
  \Nrhoz - \Npz = 2-a,
  \label{eq:appsep}
\end{equation}
independently of $\beta$ and $b$.  In the $(\Npz,\Nrhoz)$ plane of Figure~\ref{fig:Nzeros}, therefore, all systems sharing a given heating exponent $a$ lie on the straight line $\Nrhoz=\Npz+(2-a)$, and their position along that line is set by $\beta+b$.  In other words, real systems do not populate the plane of Figure~\ref{fig:Nzeros} at random: they form a one-parameter family along a line of unit slope.
 
Second, the boundaries between the instability domains of Table~\ref{table:R_and_lambda} correspond to particular values of $\beta+b$:
\begin{equation}
  R = 0 \Leftrightarrow \beta+b = a-2, \qquad
  R = \frac{1}{9} \Leftrightarrow \beta+b = -\frac{9\,(2-a)}{9-\gamma}, \qquad
  R = \frac{1}{\gamma} \Leftrightarrow a = 2, \qquad
  R = 1 \Leftrightarrow \beta+b = \frac{2-a}{\gamma-1}.
  \label{eq:appcrit}
\end{equation}
The last of these deserves emphasis, since $R = 1$ is equivalent to $\NSz = 0$, the marginal case of the isentropic criterion, and it controls whether the thermal mode attains its largest growth rate at a finite wavelength or as $\lambda \rightarrow \infty$. For $\gamma=5/3$ the value of $\beta + b$ at which $R = 1$ --- that is, of the combined logarithmic slopes of the cooling and heating rates with temperature --- is 3 when $a = 0$ but only $3/2$ when $a = 1$; a plasma whose slopes sum to more than this has $R < 1$ and belongs to the fifth row of Table~\ref{table:R_and_lambda} rather than the sixth. Radiative loss functions for the solar corona have $\beta \simeq 0.4$--$2.65$ in the range 1--3~MK, so that some of them fall between these two thresholds; the value of $R$ for coronal conditions is therefore sensitive to the assumed density scaling of the heating, and we return to this point in Section~\ref{sect:discussion_CR_CC}.

Third, Equations~(\ref{eq:appN})--(\ref{eq:appcrit}) amount to a complete classification of the plasmas that admit a power-law description, and it is convenient to express it in terms of $\Npz$ and $\Nrhoz$, the variables of Figure~\ref{fig:Nzeros}. A plasma moves \emph{down} the line of Equation~(\ref{eq:appsep}) as the heat-loss function $\calL$ becomes more sensitive to temperature, since $\Nrhoz = -(\beta+b)$ by Equations~(\ref{eq:appN}), and in doing so it visits the instability domains of Table~\ref{table:R_and_lambda} in a fixed order. For $a < 2$, we find
 
\begin{equation}
\label{eq:appclassification}
\begin{aligned}
  \Nrhoz &> 2-a:                       &&\Npz>0, && 0<R<1/\gamma
    &&\text{(all modes damped at all wavelengths)},\\
  2-a &> \Nrhoz > 0:                   &&\Npz<0, && R<0
    &&\text{(first row of Table~\ref{table:R_and_lambda})},\\
  0 &> \Nrhoz > -\frac{2-a}{\gamma-1}: &&\Npz<0, && R>1
    &&\text{(sixth row)},\\
  \Nrhoz &< -\frac{2-a}{\gamma-1}:     &&\Npz<0, && 1/\gamma<R<1
    &&\text{(fifth row)},
\end{aligned}
\end{equation}
 
\noindent where the three boundaries are, in order, $\Npz = 0$, $\Nrhoz = 0$, and $R = 1$; the last of them is the condition $R = 1$ of Equation~(\ref{eq:appcrit}) written in terms of $\Nrhoz$ instead of $\beta+b$.  There is no fifth case, because Equation~(\ref{eq:appsep}) with $a < 2$ forbids $\Nrhoz < 0 < \Npz$: a plasma with $a < 2$ that is isochorically unstable is necessarily isobarically unstable as well.  Read on the plane of Figure~\ref{fig:Nzeros}, the condition $a < 2$ places a plasma above the diagonal $\Nrhoz = \Npz$ and therefore empties three regions of the plane: the fourth quadrant, which contains the second row of Table~\ref{table:R_and_lambda}; the part of the third quadrant that lies below that diagonal, which contains the third and fourth rows; and the part of the first quadrant below the same diagonal, which contains the seventh row.  The exclusion of the third row for $a < 2$ is the most consequential, since it is the only row whose spectrum contains the pair of critical wavelengths $\lambda_\pm$ at which \emph{overstable} acoustic modes are converted into \emph{unstable} thermal modes and back again.\footnote{Even when $a < 2$, a conversion between acoustic and thermal modes is not by itself forbidden.  The value $R = 1/9$, whose position is given by Equation~(\ref{eq:appcrit}), lies in the first range of Equation~(\ref{eq:appclassification}), and more precisely in the interval $2-a < \Nrhoz < 9(2-a)/(9-\gamma)$, so that a plasma there does possess two critical wavelengths.  All of its modes are damped, however, and the conversion takes place between a pair of damped acoustic modes and two damped thermal modes; $R = 1/9$ therefore plays no role in the stability of such a system.  This is the situation encountered in the first quadrant of Figure~\ref{fig:Nzeros} and discussed in Section~\ref{sect:arbitrary_lambdaF}.}

The case $a = 2$ is degenerate. Setting $a = 2$, the second and third ranges of Equation~(\ref{eq:appclassification}) become empty, only the first and the fourth survive, and in them the inequalities satisfied by $R$ collapse onto the single value $R = 1/\gamma$, whatever the value of $\Nrhoz$. Such a plasma therefore does not belong to any single row of Table~\ref{table:R_and_lambda}: if $\Nrhoz > 0$ all its modes are damped at all wavelengths, whereas if $\Nrhoz < 0$ it lies exactly on the boundary between the fourth and fifth rows, where $\lambdaacoustic = \lambdacond$ and the thermal and the two acoustic modes change their stability at one and the same wavelength (see Section~\ref{sect:WP_fig2_kF}). This is no accident of the algebra: since $R = 1/\gamma$ is equivalent to $\Npz = \Nrhoz$, the line $R = 1/\gamma$ of Figure~\ref{fig:Nzeros} is at once the $a = 2$ member of the family of diagonals $\Nrhoz = \Npz + (2-a)$, and the family approaches it from above as $a \to 2$.
 
The remaining rows of Table~\ref{table:R_and_lambda}, namely the second, third, fourth, and seventh, all require $a > 2$: a heating rate whose density dependence is steeper than that of the two-body radiative losses. Every row of Table~\ref{table:R_and_lambda} is thus reachable by some power-law plasma, but no single value of the heating exponent reaches more than four of them.
 
Finally, the classification above also accounts for the position of the astrophysical environments considered in Section~\ref{sect:environments}, which fall in the first and the sixth rows of Table~\ref{table:R_and_lambda}, the fifth being reached only by the solar corona under the alternative heating prescription ($a = 1$).

Although Equation~(\ref{eq:appclassification}) has been obtained for power-law heating and cooling, its content is more general. Equations~(\ref{eq:Nrho0_Np0}) give, without any assumption on the form of $\calL$, $\Nrhoz - \Npz = (\rhoz/\Lambda_0)\,\calL_\rho$. By analogy with Equation~(\ref{eq:appsep}), of which this is the general form, we define the effective heating exponent

\begin{equation}
  \aeff \equiv 2 - \frac{\rhoz}{\Lambda_0}\,\calL_\rho ,
  \label{eq:aeff}
\end{equation}

\noindent which is the value that $a$ would take if the radiative losses were the result of two-body processes, and which reduces to $a$ for the power laws (\ref{eq:appQ}) and (\ref{eq:appH}).\footnote{More generally, if the radiative losses are not proportional to $n_{\rm H}^2$, as happens when they depend on the ionization parameter, they may be written locally as $\rho\Lambda \propto \rho^{\,\alpha}T^{-\beta}$. Equations~(\ref{eq:appN}) then read $\Nrhoz = -(\beta+b)$ and $\Nrhoz - \Npz = \alpha - a$, so that of the four exponents the linear analysis determines only the two combinations $\beta+b$ and $\alpha-a$, and the condition $a < 2$ that underlies Equation~(\ref{eq:appclassification}) becomes $a < \alpha$: the heating rate per unit volume must not grow with density faster than the losses do. Equation~(\ref{eq:aeff}) absorbs this into a single number, which is the form we use.} Whenever $\aeff \le 2$, that is, whenever the heating rate per unit volume does not grow with density faster than the two-body radiative losses, only the first, fifth, and sixth rows of Table~\ref{table:R_and_lambda} are accessible. In Section~\ref{sect:environments} we check that this is so for all the environments considered, including the X-ray photoionized plasma, whose heating and cooling rates are controlled by the ionization parameter and are not power laws of $\rho$ and $T$. For that medium the radiative losses are not proportional to $n_{\rm H}^2$ either, because the ionization parameter varies as $\rho^{-1}$ at a fixed distance from the source; they are nevertheless proportional to $n_{\rm H}^2$ at a fixed value of that parameter, so that Equation~(\ref{eq:appQ}) and the scales derived from it in Sections~\ref{app:scales} and \ref{app:ratio} apply to it unchanged.

\subsection{Characteristic time and length scales} \label{app:scales}
 
We take the thermal conductivity to be a power law of the temperature,
\begin{equation}
  \kappa = f\,\kappa_0\,T^{\,\nu},
  \label{eq:appkappa}
\end{equation}
with $\nu=5/2$ for the Spitzer conductivity of a fully ionized plasma (field-aligned in the magnetized case) and $\nu=1/2$ for conduction by neutral atoms; $f\le1$ is a suppression factor, discussed in Section~\ref{sect:environments}.  Then the cooling time~(\ref{eq:tcool}), the thermal length~(\ref{eq:lambdath}), and the Field length~(\ref{eq:lambdaF}) are
\begin{equation}
  \tc = \frac{\mu_{\rm H}}{\mu}\,\frac{k_{\rm B}\,\Tz}{(\gamma-1)\,x\,\nHz\,Q(\Tz)},
  \label{eq:apptcool}
\end{equation}
\begin{equation}
  \lambdath = \cs\tc
   = \left(\frac{\gamma k_{\rm B}}{\mu m_{\rm H}}\right)^{1/2}
     \frac{\mu_{\rm H}}{\mu}\,
     \frac{k_{\rm B}\,\Tz^{\,3/2}}{(\gamma-1)\,x\,\nHz\,Q(\Tz)},  \label{eq:applamth}
\end{equation}
\begin{equation}
  \lambdaF = \frac{2\pi}{\nHz}
   \left(\frac{f\,\kappa_0}{x\,Q(\Tz)}\right)^{1/2}
   \Tz^{\,(\nu+1)/2}.
  \label{eq:applamF}
\end{equation}

\noindent Note that $\tc$, $\lambdath$, and $\lambdaF$ do not depend explicitly on $\beta$ when written in this form; the exponent enters the remaining wavelengths defined in the main text only through $\Npz$ and $\NSz$:
\begin{equation}
  \lambdacond = \frac{\lambdaF}{\sqrt{-\Npz}},\qquad
  \lambdaacoustic    = \frac{\lambdaF}{\sqrt{-\NSz}},\qquad
  \lambdaisoch = \frac{\lambdaF^{2}}{\lambdath} .
  \label{eq:applams}
\end{equation}

\subsection{The ratio $\lambda_{\rm F}/\lambda_{\rm th}$} \label{app:ratio}

Combining Equations~(\ref{eq:applamth}) and (\ref{eq:applamF}), the explicit density dependence cancels\footnote{Nevertheless, the Spitzer conductivity retains a weak dependence on density through the Coulomb logarithm, which amounts to a few per cent per decade in density.} and we obtain the general result
\begin{equation}
  \frac{\lambdaF}{\lambdath}
  = \frac{2\pi(\gamma-1)}{\sqrt{\gamma}}
    \left[\frac{\mu^{3}x}{\mu_{\rm H}^{2}}\,
          \frac{m_{\rm H}\,\kappa(\Tz)\,Q(\Tz)}{k_{\rm B}^{3}\,\Tz^{2}}
    \right]^{1/2} \;
  \label{eq:appratio}
\end{equation}
which for the power laws (\ref{eq:appQ}) and (\ref{eq:appkappa}) scales as $T_0^{(\nu-\beta-2)/2}$.  Equation~(\ref{eq:appratio}) is one of the more useful results of this work: whether the classification of Table~\ref{table:R_and_lambda} applies to a given plasma does not depend on its density, but only on its temperature, on the shape of the radiative loss function, and on the conductivity.  We stress that, although it has been derived in this appendix, Equation~(\ref{eq:appratio}) does not rest on the power-law assumption: it involves the values $\kappa(\Tz)$ and $Q(\Tz)$ rather than the coefficients $\kappa_0$ and $Q_0$, and the heating does not enter at all, since neither $\lambdath$ nor  $\lambdaF$ depends on it. The power-law forms are required only for the temperature scaling quoted above. Note that the derivation assumes two-body cooling, Equation~(\ref{eq:appQ}); it does not hold when Compton processes, line trapping, or collisional de-excitation above the critical density make the volumetric cooling rate depart from an $n^{2}$ scaling.

Equation~(\ref{eq:appratio}) also quantifies the (small) influence of the chemical composition.  Passing from a fully ionized hydrogen plasma to one of solar composition whose losses are tabulated as $n_{\rm e}n_{\rm H}Q$, so that $x = 1.2$, changes the prefactor $(\mu^{3}x/\mu_{\rm H}^{2})^{1/2}$ from $0.354$ to $0.364$, that is, it changes $\lambda_{\rm F}/\lambda_{\rm th}$ by $3\%$.  At fixed $\nHz$ and $\Tz$ the individual scales change by $3\%$ ($t_{\rm cool}$), $11\%$ ($\lambda_{\rm th}$), and $9\%$ ($\lambda_{\rm F}$).  These differences are negligible compared with the factor of two to three that separates the radiative loss functions available in the literature for the same environment, and with the effect of metallicity in the intracluster and circumgalactic media. Finally, we stress that Equations~(\ref{eq:appN}) and (\ref{eq:appR}) are unaffected by the composition only as long as $x$, $\mu_{\rm H}$, and $\mu$ can be treated as constants. This ceases to be true in temperature ranges over which hydrogen or helium ionize: the mean molecular weight then varies with temperature, which modifies the derivative taken at constant pressure and hence $\Npz$. The accompanying variation of the electron-to-hydrogen density ratio is a separate matter, and is already contained in any $Q(T)$ tabulated under the assumption of collisional ionization equilibrium.

\section{Heating, cooling, and thermal conduction in the astrophysical environments} \label{sect:app_environments}
 
This appendix collects the prescriptions adopted for each of the five environments considered in Section~\ref{sect:environments}, together with the reasons for adopting them and the fiducial values of the equilibrium temperature and density. In every case we have taken a heat-loss function, a thermal conductivity, and a heating prescription that are in common use in the corresponding field, rather than constructing our own. The resulting inputs are summarized in Table~\ref{table:environments_input}, and the parameters and characteristic scales derived from them in Table~\ref{table:environments_output}; the quantities in this table are obtained from Equations~(\ref{eq:appN}), (\ref{eq:appR}), and (\ref{eq:apptcool})--(\ref{eq:appratio}), except for the X-ray photoionized outflow, whose heating is not a power law and for which the quantities are evaluated directly from their equations in Section~\ref{sect:main_equations}.

\subsubsection*{Two-phase atomic interstellar medium (ISM)}

We use the heating and cooling rates of \citet{KoyamaInutsuka2002}, which are the standard choice in simulations of the two-phase interstellar medium. Their radiative loss function is given by their Equation~(4), which contains two typographical errors and should be taken in the corrected form given in Equation~(4) of \citet{vazquez-semadeni_etal2007}. Since the losses are given per $n_{\rm H}^{2}$, this function is $Q(T)$ in the notation of Equation~(\ref{eq:appQ}) and $x = 1$. The two terms of the fitting function $Q(T)$ represent the Lyman~$\alpha$ line of hydrogen and the C\,{\sc ii} fine-structure line, which dominate the cooling of the warm and of the cold phase respectively; the calculation on which the fit is based includes several further coolants. The heating is $n_{\rm H}\Gamma$, with $\Gamma = 2\times10^{-26}$~erg~s$^{-1}$ a rate per hydrogen nucleus, so that $a = 1$ and $b = 0$ in Equation~(\ref{eq:appH}). \citet{KoyamaInutsuka2002} include several heating processes, of which photoelectric emission from small grains and polycyclic aromatic hydrocarbons dominates at the temperatures of interest here, with contributions from cosmic rays and X-rays; that the total rate may be taken as constant per hydrogen nucleus is their simplification, and it is what fixes the value of $a$.

These authors write the two rates as $(\rho/m_{\rm H})\Gamma$ and $(\rho/m_{\rm H})^{2}Q(T)$, with $\rho$ the mass density of a gas of pure hydrogen. We retain their $\Gamma$ and $Q(T)$ but identify $\rho/m_{\rm H}$ with the hydrogen number density of Equation~(\ref{eq:appcomp}), $n_{\rm H} = \rho/(\mu_{\rm H}m_{\rm H})$ with $\mu_{\rm H} = 1.4$, since their fitting function was obtained for a gas that does contain helium; with helium at ten per cent of hydrogen by number and the gas neutral, the same counting gives $\mu = 1.27$. The values quoted below therefore differ slightly from those that would be obtained by using their equations verbatim.

Thermal conduction is carried by neutral atoms, for which $\kappa = 2.5\times10^{3}\,T^{1/2}$ in cgs units \citep{parker1953instability, KoyamaInutsuka2004}, so that $\nu=1/2$; being isotropic, it does not require the caveats that apply to the magnetized cases below. The equilibrium condition $n_{\rm H}Q(T)=\Gamma$ fixes the density once the temperature is chosen. We adopt $\Tz = 10^{3}$~K, which lies on the thermally unstable branch that joins the cold and the warm neutral phases, and obtain $\nHz = 2.5$~cm$^{-3}$ and a thermal pressure $p/k_{\rm B} = 2.7\times10^{3}$~K~cm$^{-3}$, in the middle of the range over which the two stable phases coexist.

\subsubsection*{Solar corona}

Optically thin radiative losses are taken from the CHIANTI atomic database \citep{Dere_etal1997, Dufresne_etal2024}.  Thermal conduction along the magnetic field is Spitzer's, $\kappa_\parallel = \kappa_0\,T^{5/2}$ \citep{Spitzer1962}, so $\nu = 5/2$ and $f=1$; the wavelengths quoted below therefore refer to perturbations propagating along the field.  For the heating we take $a=b=0$, i.e., a volumetric heating rate independent of density and temperature, which is the prescription used in essentially all global loop models that produce thermal non-equilibrium \citep[see][]{stricklan_etal2025}.  Two equilibria are considered: an active-region loop in equilibrium, with $\Tz = 2$~MK and $\nHz = 10^{9}$~cm$^{-3}$, and the denser and cooler plasma of a loop late in a TNE cycle, with $\Tz = 10^{6}$~K and $\nHz = 10^{10}$~cm$^{-3}$.  The characteristic size is the loop length, $L \simeq 100$~Mm.
 
\subsubsection*{Circumgalactic medium (CGM) and cool-core intracluster medium (ICM)}
 
For both we use the collisional-ionization-equilibrium cooling function of \citet{SutherlandDopita1993} at $Z = 0.3\,Z_\odot$, in the analytic parameterization of \citet{TozziNorman2001}.  The logarithmic slope differs greatly between the two: at $\Tz = 10^{6}$~K, typical of a hot galactic halo, $\beta = 1.47$, whereas at $k_{\rm B}\Tz = 2$~keV, typical of a cluster cool core, line cooling has given way to bremsstrahlung and $\beta = -0.25$, so that the medium is isochorically \emph{stable} while remaining isobarically unstable. We note that this is a collisional-ionization-equilibrium calculation, whereas at the low densities of the circumgalactic medium photoionization by the ultraviolet background modifies the losses; tabulations that include it are available \citep[e.g.,][]{Wiersma_etal2009}, although the effect is modest at the temperature adopted here. The heating is attributed to the ultraviolet background in the first case and to active-galactic-nucleus feedback in the second, and in both it is imposed rather than derived: the usual procedure is to add whatever heating is required for the volume-averaged or shell-averaged rates to balance the losses. Both prescriptions of Section~\ref{app:heating} are in use. \citet{sharma_etal2010} and \citet{Sharma_etal2012} adopt a heating rate per unit volume that is independent of density, $a = 0$, while \citet{mccourt_etal2012} run both that case and the mass-weighted one, $a = 1$, and find the evolution to be very similar once the time scales are rescaled; \citet{sharma_etal2010} report the same insensitivity. We therefore give both values for these two media in Tables~\ref{table:environments_input} and~\ref{table:environments_output}, with $b = 0$ throughout. Thermal conduction is Spitzer's, reduced by a suppression factor $f$ that accounts for the tangling of the magnetic field; we take $f = 0.1$, within the range $f \simeq 0.1-0.3$ usually quoted \citep[e.g.,][]{NarayanMedvedev2001}, and comment on this choice in Section~\ref{sect:environments}.  We take $n_{\rm H0} = 10^{-3}$~cm$^{-3}$ and $L \simeq 10$~kpc for the circumgalactic medium and $n_{\rm H0} = 4.2\times10^{-2}$~cm$^{-3}$ and $L \simeq 50$~kpc for the cool core.

\subsubsection*{X-ray photoionized outflow}
 
Our last environment is the X-ray irradiated gas of active galactic nuclei and X-ray binaries, for which we use the heating and cooling rates of \citet{Blondin1994}. Since the radiative losses are given per $n_{\rm H}^2$, in the notation of Equation~(\ref{eq:appQ}) we have $x = 1$.  Compton heating and cooling, photoionization heating, bremsstrahlung, and line cooling are all functions of the temperature and of the ionization parameter $\xi = L_X/(n_{\rm H} r^2)$.  Because $\xi$ varies as $n_{\rm H}^{-1}$ at a fixed distance from the source, the density dependence of the heating is determined by the physics rather than assumed, and the rates are not power laws of $\rho$ and $T$; we therefore evaluate the quantities in Table~\ref{table:environments_output} from their equations in Section~\ref{sect:main_equations}.  We adopt $\xi = 300$ and $T_X = 1.16\times10^{8}$~K, for which the equilibrium temperature is $\Tz = 4.5\times10^{5}$~K on the isobarically unstable branch of the thermal-equilibrium curve, and $\nHz = 10^{9}$~cm$^{-3}$.  For a source luminosity $L_X \sim 10^{44}$~erg~s$^{-1}$ this corresponds to a distance $r \simeq 2\times10^{16}$~cm, which we take as the characteristic size. Thermal conduction is usually neglected in this environment, but we retain it here for consistency with the other four: we take the Spitzer formula with $f = 1$ and $\nu = 5/2$.

\end{appendix}
\end{document}